\documentclass[pra,preprint,preprintnumbers,amsmath,amssymb]{revtex4}
\usepackage{graphicx}
\usepackage{dcolumn}
\usepackage{epic}%
\usepackage{eepic}%
\usepackage{bm}

\begin{document}

\preprint{APS/123-QED}

\newcommand{\scscs}{\scriptscriptstyle}

\title{Developments for Reference--State One--Particle Density--Matrix Theory}

\author{James P. Finley}

\affiliation{
Department of Physical Sciences,
Eastern New~Mexico University,
Station \#33, Portales, NM 88130}
\email{james.finley@enmu.edu}

\affiliation{Department of Applied Chemistry, Graduate School of Engineering,
The University of Tokyo, Tokyo, Japan 113-8656}

\date{\today}

\begin{abstract}
Brueckner orbitals, and the density of the Brueckner reference-state, are shown to
satify the same cusp condition -- involving the nuclear charges -- as natural- and
Hartree--Fock-orbitals. Using the cusp condition, the density of a determinantal
state can be used to determine the external potential, if the determinantal state
is from either Hartee--Fock or Brueckner-orbital theory, as well as, determinant
states obtained by many other formalisms that are defined by a one-body operator,
if a portion of the one-body operator -- the portion not associated with the
kinetic energy or external potential -- generates a well behaved function when
acting on an occupied orbital. Using this relationship involving a determinant and
its external potential, a variation of Reference--State One--Particle
Density--Matrix Theory [arXiv:physics/0308056] is formulated, where the trial
wavefunctions are universal, in the Kohn-Sham sense, since they do not depend on
the external potential. The resulting correlation-energy functionals, are also,
universal, except for a relatively small term involving the portion of the
expectation value of the external potential with the trial wavefunctions that
appears beyond the first order. The same approximate energy functionals that were
shown to be valid for the previous $v$-dependent, Reference--State One--Particle
Density--Matrix Theory [arXiv:physics/0308084], are shown to be valid for the
current approach, except that the use of the LYP and Colle--Salvetti functional
appear more natural within the current approach, since these functionals are
universal ones. And since the BLYP and B3LYP functionals contain the LYP
functional, these approaches are also better suited with the current approach.
\end{abstract}

\maketitle

\section{External potential determined by the one-particle density matrix and the 
particle density} \label{v.det.rho} 

There is a one-to-one correspondence between determinant states and their
one-particle density-matrices \cite{Blaizot:86,Parr:89}. Because of this
correspondence, it is convenient to denote a determinantal state that is
determined by a one-particle density-matrix, say $\gamma$, simply by
$|\gamma\rangle$. In addition, any function, say $G$, that depends on $\gamma$,
can be written as $G(\gamma)$; this same notation also indicates that $G$ is
determined by the corresponding determinant, $|\gamma\rangle$.

Consider the following noninteracting Hamiltonian:
\begin{eqnarray} \label{H.nonint} 
H_s^\gamma=\sum_i^n
\left(
-\mbox{$\frac12$}\nabla^2_{\mathbf{r}_i}
+
 v(\mathbf{r}_i)
 +
 \hat{w}^{\scscs \gamma}(\mathbf{x}_i)
\right),
\end{eqnarray}
where the external potential $v$ is given by a fixed set of point charges
\begin{eqnarray} \label{potential} 
v(\mathbf{r}) = 
-\!\!\!\!\!\sum_{m\in\{\mathbf{R}_{\text{nuc}}\}} \frac{Z_m}{|\mathbf{R}_m - \mathbf{r}|},
\end{eqnarray}
and the summation is over the coordinates of the nuclear point charges, denoted by
$\{\mathbf{R}_{\text{nuc}}\}$; furthermore, $\hat{w}^{\scscs \gamma}$ may be
non-local and this operator can depend on the spin-coordinate $\omega$, where the
spatial and spin coordinates are denoted collectively by $\mathbf{x}$;
in addition, the $\gamma$ superscript appended to $w$ indicates that this operator
may also depend on $\gamma$ (or equivalently $|\gamma\rangle$).

Consider a determinantal state, say $|\gamma\rangle$, that satisfies the
Schr\"odinger Eq:
\begin{eqnarray} \label{sch} 
H_s^\gamma|\gamma\rangle = E_\gamma|\gamma\rangle, 
\end{eqnarray}
where the Hamiltonian $H_s$ is given by Eq.~(\ref{H.nonint}); furthermore, the
noninteracting eigenstate, $|\gamma\rangle$, can be expressed by a unique set of
occupied orbitals, denoted by $\{\psi_o \leftarrow \gamma, \mbox{\small
$\hat{f}_{\gamma}$}\}$; each of these orbitals satisfy the following one-particle
Schr\"odinger Eq:
\begin{eqnarray} \label{one.sch} 
\hat{f}_\gamma\psi_{x\sigma}^{\scscs \gamma}(\mathbf{x}) =
\epsilon_{i\sigma}^\gamma \psi_{x\sigma}^{\scscs \gamma}(\mathbf{x}),
\;\;\; \sigma = \alpha,\beta,
\;\;\; \psi_{x\sigma}^{\scscs \gamma} \in \{\psi_o \leftarrow \gamma, \mbox{\small
$\hat{f}_{\gamma}$}\},
\end{eqnarray}
where the one-body operator is given by
\begin{eqnarray} \label{one.op} 
\hat{f}_\gamma =
-\mbox{$\frac12$}\nabla^2_{\!\mathbf{r}} + v(\mathbf{r}) +
w^{\scscs \gamma}(\mathbf{x}).
\end{eqnarray}
In addition, we require the operator $w^{\scscs \gamma}$ to be Hermitian and
satisfy
\begin{eqnarray} \label{w.propZ} 
\lim_{|\mathbf{R}-\mathbf{r}|\rightarrow 0} 
|\mathbf{R}-\mathbf{r}| w^{\scscs \gamma}(\mathbf{r},\omega) 
\psi_{x\sigma}^{\scscs \gamma}(\mathbf{r},\omega)
= 0,
\;\;\; \mbox{for all $\mathbf{R}$}.
\end{eqnarray}
In order to emphasize an exclusive dependence upon $\mathbf{R}$, we modify the limit
in this Eq., giving
\begin{eqnarray} \label{w.prop} 
\lim_{\mathbf{r}\rightarrow\mathbf{R}} 
|\mathbf{r}-\mathbf{R}| w^{\scscs \gamma}(\mathbf{r},\omega) 
\psi_{x\sigma}^{\scscs \gamma}(\mathbf{r},\omega)
= 0.
\;\;\; \mbox{for all $\mathbf{R}$}.
\end{eqnarray}

Let us also mention that the set of unoccupied orbitals -- orthogonal to $\{\psi_o
\leftarrow \gamma, \mbox{\small $\hat{f}_{\gamma}$}\}$ -- is denoted by $\{\psi_u
\leftarrow \gamma, \mbox{\small $\hat{f}_{\gamma}$}\}$ and, in addition, all of
our spin-orbitals $\psi_{i\sigma}(\mathbf{x})$ have the following form:
\begin{equation} \label{spinorb} 
\psi_{i\sigma}(\mathbf{x})=
\chi_{i\sigma}(\mathbf{r})\sigma(\omega), 
\;\; \sigma=\alpha \mbox{ or } \beta,
\end{equation}
where the spatial and spin portions are given by $\chi_{i\sigma}(\mathbf{r})$ and
$\sigma(\omega)$, respectively, and the spatial functions
$\chi_{i\sigma}(\mathbf{r})$ are permitted to be unrestricted -- two spin orbitals
do not, in general, share the same spatial function, i.e., ($\chi_{i\alpha}\ne
\chi_{i\beta}$) is permitted.

Multiplying Eq.~(\ref{one.sch}) by $|\mathbf{r}-\mathbf{R}|$ followed by taking a
limit of this term vanishing, gives
\begin{eqnarray} \label{also.nat} 
\lim_{|\mathbf{r}-\mathbf{R}|\rightarrow 0} 
|\mathbf{R}-\mathbf{r}|
\left( -\mbox{$\frac12$}\nabla^2_{\!\mathbf{r}} 
-\sum_{m\in\{\mathbf{R}_{\text{nuc}}\}} \frac{Z_m}{|\mathbf{R}_m - \mathbf{r}|} 
\right) \psi_{x\sigma}^{\scscs \gamma}(\mathbf{x}) =0,
\end{eqnarray}
where we have used Eqs.~(\ref{potential}), (\ref{one.op}), and
(\ref{w.propZ}). (Note that this Eq.\ is also the cusp condition
\cite{Poling:71,Davidson:76}, however, in that case, $\psi_{x\sigma}^{\scscs
\gamma}$ is a natural orbital and $\gamma$ is the one-particle density-matrix of
an interacting target-state, say $\Psi$.)

In order to obtain an exclusive dependence upon $\mathbf{R}$, we, again, modify
the limit in this Eq, giving
\begin{eqnarray} 
\lim_{\mathbf{r}\rightarrow\mathbf{R}} 
|\mathbf{r}-\mathbf{R}|
\left( -\mbox{$\frac12$}\nabla^2_{\!\mathbf{r}} 
-\sum_{m\in\{\mathbf{R}_{\text{nuc}}\}} \frac{Z_m}{|\mathbf{R}_m - \mathbf{r}|} 
\right) \psi_{x\sigma}^{\scscs \gamma}(\mathbf{x}) =0.
\end{eqnarray}
Since the second term vanishes unless
($\mathbf{R}\in\{\mathbf{R}_{\text{nuc}}\}$), we have
\begin{eqnarray} \label{interx} 
\lim_{\mathbf{r}\rightarrow\mathbf{R}}  
|\mathbf{r}-\mathbf{R}|
\left( 
\left( -\mbox{$\frac12$}\nabla^2_{\!\mathbf{r}}\right) - 
\sum_{m\in\{\mathbf{R}_{\text{nuc}}\}}
\mbox{\large $\delta$}_{\!\mbox{\tiny $\mathbf{R}\mathbf{R}_m$}}  Z_m 
\right) \psi_{x\sigma}^{\scscs \gamma}(\mathbf{x}) =0,
\end{eqnarray}
which can be written as
\begin{eqnarray} 
\lim_{\mathbf{r} \rightarrow \mathbf{R}} 
|\mathbf{r}-\mathbf{R}|
\psi_{x\sigma}^{\scscs \gamma}(\mathbf{x})^{-1}
\left(-\mbox{$\frac12$}\nabla^2_{\!\mathbf{r}}\right) 
\psi_{x\sigma}^{\scscs \gamma}(\mathbf{x})
= 
\sum_{m\in\{\mathbf{R}_{\text{nuc}}\}}
\mbox{\large $\delta$}_{\!\mbox{\tiny $\mathbf{R}\mathbf{R}_m$}}  Z_m.
\end{eqnarray}
Defining the left side by
\begin{eqnarray} 
T(\psi_{x\sigma}^{\scscs \gamma},\mathbf{R})=
\lim_{\mathbf{r}\rightarrow\mathbf{R}} 
|\mathbf{r}-\mathbf{R}|
\psi_{x\sigma}^{\scscs \gamma}(\mathbf{x})^{-1}
\left(-\mbox{$\frac12$}\nabla^2_{\!\mathbf{r}}\right) 
\psi_{x\sigma}^{\scscs \gamma}(\mathbf{x}),
\end{eqnarray}
we can write
\begin{eqnarray} 
T(\psi_{x\sigma}^{\scscs \gamma},\mathbf{R})= 
\sum_{m\in\{\mathbf{R}_{\text{nuc}}\}}
\mbox{\large $\delta$}_{\!\mbox{\tiny $\mathbf{R}\mathbf{R}_m$}}  Z_m.
\end{eqnarray}

For a set of spatially restricted orbitals:
\begin{eqnarray} \label{res.orbs} 
\psi_{x\sigma}^{\scscs \gamma}(\mathbf{x}) =
\chi_{x}^{\scscs \gamma}(\mathbf{r}) \sigma(\omega),
\end{eqnarray}
it is readily proven that we have
\begin{eqnarray} 
T(\chi_{x}^{\scscs \gamma},\mathbf{R})= 
\sum_{m\in\{\mathbf{R}_{\text{nuc}}\}}
\mbox{\large $\delta$}_{\!\mbox{\tiny $\mathbf{R}\mathbf{R}_m$}}  Z_m.
\end{eqnarray}

Multiplying Eq.~(\ref{interx}) by $\left(\psi_{x\sigma}^{\scscs
\gamma}(\mathbf{x}^\prime)\right)^*$, and summing over all occupied orbitals from
the set $\{\psi_o \leftarrow \gamma, \mbox{\small $\hat{f}_{\gamma}$}\}$, gives
\begin{eqnarray} \label{intery} 
\lim_{\mathbf{r}\rightarrow\mathbf{R}} 
|\mathbf{r}-\mathbf{R}|
\left( 
\left( -\mbox{$\frac12$}\nabla^2_{\!\mathbf{r}}\right) - 
\sum_{m\in\{\mathbf{R}_{\text{nuc}}\}}
\mbox{\large $\delta$}_{\!\mbox{\tiny $\mathbf{R}\mathbf{R}_m$}}  Z_m 
\right) \gamma(\mathbf{x},\mathbf{x}^\prime) =0,
\end{eqnarray}
where the one-particle density matrix is given by
\begin{eqnarray} \label{opdm} 
\gamma(\mathbf{x},\mathbf{x}^\prime)
= \sum_{x\sigma\in\{\psi_o \leftarrow \gamma, 
\hat{f}_{\gamma}\}}
\psi_{x\sigma}^{\scscs \gamma}(\mathbf{x}) 
\left(\psi_{x\sigma}^{\scscs \gamma}(\mathbf{x}^\prime)\right)^*
\end{eqnarray}
and it is readily proven that we have
\begin{eqnarray} \label{T.gamma1} 
T(\gamma,\mathbf{R})= 
\sum_{m\in\{\mathbf{R}_{\text{nuc}}\}}
\mbox{\large $\delta$}_{\!\mbox{\tiny $\mathbf{R}\mathbf{R}_m$}}  Z_m,
\end{eqnarray}
where
\begin{eqnarray} 
T(\gamma,\mathbf{R})=
\lim_{\mathbf{r}\rightarrow\mathbf{R}} 
|\mathbf{r}-\mathbf{R}|
\gamma(\mathbf{x},\mathbf{x}^\prime)^{-1}
\left(-\mbox{$\frac12$}\nabla^2_{\!\mathbf{r}}\right)
\gamma(\mathbf{x},\mathbf{x}^\prime).
\end{eqnarray}
Since this expression is invariant to the variable $\mathbf{x}^\prime$, we can
choose ($\mathbf{x}^\prime=\mathbf{x}$), yielding
\begin{eqnarray} \label{Tdef.gamma1} 
T(\gamma,\mathbf{R})=
\lim_{\mathbf{r}\rightarrow\mathbf{R}} 
|\mathbf{r}-\mathbf{R}|
\gamma(\mathbf{x},\mathbf{x})^{-1}
\left(
\left(-\mbox{$\frac12$}\nabla^2_{\!\mathbf{r}}\right)
\gamma(\mathbf{x},\mathbf{x}^\prime)
\right)_{\mathbf{x}^\prime=\mathbf{x}}.
\end{eqnarray}
Since Eq.~(\ref{interx}) is also satisfied by the complex conjugate orbital,
$\psi_{x\sigma}^{\scscs \gamma}(\mathbf{x})^*$, it is readily shown that we have
\begin{eqnarray} \label{T.gamma1cc} 
T(\gamma^*,\mathbf{R})= 
\sum_{m\in\{\mathbf{R}_{\text{nuc}}\}}
\mbox{\large $\delta$}_{\!\mbox{\tiny $\mathbf{R}\mathbf{R}_m$}}  Z_m,
\end{eqnarray}
where
\begin{eqnarray} \label{Tdef.gamma1cc} 
T(\gamma^*,\mathbf{R})=
\lim_{\mathbf{r}\rightarrow\mathbf{R}} 
|\mathbf{r}-\mathbf{R}|
\gamma(\mathbf{x},\mathbf{x})^{-1}
\left(
\left(-\mbox{$\frac12$}\nabla^2_{\!\mathbf{r}}\right)
\gamma(\mathbf{x}^\prime,\mathbf{x})
\right)_{\mathbf{x}^\prime=\mathbf{x}}.
\end{eqnarray}

Adding together Eq.~(\ref{T.gamma1}) and (\ref{T.gamma1cc}), using
(\ref{Tdef.gamma1}) and (\ref{Tdef.gamma1cc}), and using the following identity:
\begin{eqnarray} 
\left(-\mbox{$\frac12$}\nabla^2_{\!\mathbf{r}}\right)
\gamma(\mathbf{x},\mathbf{x})
=\left(
\left(-\mbox{$\frac12$}\nabla^2_{\!\mathbf{r}}\right)
\gamma(\mathbf{x},\mathbf{x}^\prime)
\right)_{\mathbf{x}^\prime=\mathbf{x}}
+
\left(
\left(-\mbox{$\frac12$}\nabla^2_{\!\mathbf{r}}\right)
\gamma(\mathbf{x}^\prime,\mathbf{x})
\right)_{\mathbf{x}^\prime=\mathbf{x}},
\end{eqnarray}
we get
\begin{eqnarray} \label{T.spind} 
T(\rho_\gamma^{s},\mathbf{R})= \;
2\!\!\!\!\sum_{m\in\{\mathbf{R}_{\text{nuc}}\}}
\mbox{\large $\delta$}_{\!\mbox{\tiny $\mathbf{R}\mathbf{R}_m$}}  Z_m,
\end{eqnarray}
where
\begin{eqnarray} \label{Tdef.gamma} 
T(\rho_\gamma^{s},\mathbf{R})= 
\lim_{\mathbf{r}\rightarrow\mathbf{R}} 
|\mathbf{r}-\mathbf{R}|
\rho_\gamma^{s}(\mathbf{x})^{-1}
\left(-\mbox{$\frac12$}\nabla^2_{\!\mathbf{r}}\right)
\rho_\gamma^{s}(\mathbf{x}),
\end{eqnarray}
and $\rho_\gamma^{s}(\mathbf{x})$ is the spin density,
i.e.,
\begin{eqnarray}
\rho_\gamma^{s}(\mathbf{x}) = \gamma(\mathbf{x},\mathbf{x}).
\end{eqnarray}

Eq.~(\ref{intery}) is also valid for
$\gamma(\mathbf{r},\omega;\mathbf{r}^\prime,\omega)$ replacing
$\gamma(\mathbf{x},\mathbf{x}^\prime)$; making this substitution and summing over
the spin-variable $\omega$ we obtain the same expression, as Eq.~(\ref{intery}),
except that it involves the spinless density matrix $\rho_1$, given by
\begin{eqnarray} 
\rho_1(\mathbf{r},\mathbf{r}^\prime)
= \sum_\omega
\gamma(\mathbf{r},\omega;\mathbf{r}^\prime,\omega),
\end{eqnarray}
and it is readily proven that we have
\begin{eqnarray} \label{T.rho} 
T(\rho_1,\mathbf{R})= 
\frac12 T(\rho,\mathbf{R})= 
\sum_{m\in\{\mathbf{R}_{\text{nuc}}\}}
\mbox{\large $\delta$}_{\!\mbox{\tiny $\mathbf{R}\mathbf{R}_m$}}  Z_m,
\end{eqnarray}
where $T(\rho_1,\mathbf{R})$ and $T(\rho^{\scscs\gamma},\mathbf{R})$
are defined by Eqs.~(\ref{Tdef.gamma1}) and (\ref{Tdef.gamma}), respectively;
$\rho$ is the electron density, i.e., 
\begin{eqnarray}
\rho(\mathbf{r}) = \rho_1(\mathbf{r},\mathbf{r}).
\end{eqnarray}

Consider the set of (ground and excited) determinantal states, denoted
$\{|\gamma\rangle_{v}\}$, that are eigenfunctions of $H_s$, given by
Eq.~(\ref{H.nonint}), where the states from the set, $\{|\gamma\rangle_{v}\}$,
are obtained from all $w^{\scscs \gamma}$ that satisfy Eq.~(\ref{w.prop}), and
from all Coulombic external-potentials $v$, given by Eq.~(\ref{potential}). From
the density of any one of theses states, say $\rho$, we can determine its
Coulombic external-potential $v$ by using Eqs.~(\ref{T.rho}) and
(\ref{potential}). Hence, $v$ is a unique function of the density. In other words,
we have $v(\rho)$, and this function is defined for all densities that are from
this set of determinantal states, $\{|\gamma\rangle_{v}\}$.

\section{Invariance of occupied-orbital transformation} 
\label{INV} 

We now partition the operator $w^{\scscs \gamma}$ into the following four
components:
\begin{eqnarray} \label{w.one.part} 
w^{\scscs \gamma}
= 
w^{\scscs \gamma}_{\text{ex}} +
w^{\scscs \gamma}_{\text{de}} +
w^{\scscs \gamma}_{\text{oc}} +
w^{\scscs \gamma}_{\text{un}},
\end{eqnarray}
where the excitation (ex), de-excitation (de), occupied (oc), and unoccupied (un)
parts are given by the following expressions:
\begin{eqnarray}  \label{w.ex} 
w^{\scscs \gamma}_{\text{ex}}
&=& 
\sum_{w\sigma r\sigma^\prime} w^{r\sigma}_{w\sigma^\prime} 
a_{r\sigma}^\dagger a_{w\sigma^\prime},
\\ 
w^{\scscs \gamma}_{\text{de}}
&=& 
\sum_{r\sigma w\sigma^\prime} w^{w\sigma}_{r\sigma^\prime} 
a_{w\sigma}^\dagger a_{r\sigma^\prime},
\\
w^{\scscs \gamma}_{\text{oc}}
&=& 
\sum_{w\sigma x\sigma^\prime} w^{w\sigma}_{x\sigma^\prime} 
a_{w\sigma}^\dagger a_{x\sigma^\prime},
\\ \label{w.un} 
w^{\scscs \gamma}_{\text{un}}
&=& 
\sum_{r\sigma s\sigma^\prime} w^{r\sigma}_{s\sigma^\prime} 
a_{r\sigma}^\dagger a_{s\sigma^\prime},
\end{eqnarray}
and the occupied- and unoccupied-orbitals are, respectively, given by
\begin{eqnarray} \label{occ.orbs} 
\psi_{w\sigma}^{\scscs \gamma},\psi_{x\sigma}^{\scscs \gamma} & \in& 
\{\psi_o \leftarrow \gamma, \mbox{\small $\hat{f}_{\gamma}$}\}, \\
\label{unocc.orbs} 
\psi_{r\sigma}^{\scscs \gamma}, \psi_{s\sigma}^{\scscs \gamma} & \in&  
\{\psi_u \leftarrow \gamma, \mbox{\small $\hat{f}_{\gamma}$}\}.
\end{eqnarray}

The results from the previous Sec.\ indicate that $v$ is a function of $\rho$ for
any $\rho$ determined from $H_s^\gamma$, given by Eq.~(\ref{H.nonint}) -- or,
equivalently, any $\rho$ determined from the one-body operator $\hat{f}_\gamma$
given by (\ref{one.op}) -- when the operator $w^{\scscs \gamma}$ satisfies
Eq.~(\ref{w.prop}). Using the partitioning method given above, Eq.~(\ref{w.prop})
becomes
\begin{eqnarray} \label{w.prop2} 
\lim_{|\mathbf{r}-\mathbf{R}|\rightarrow 0} 
|\mathbf{r}-\mathbf{R}| 
\left[
w^{\scscs \gamma}_{\text{oc}}(\mathbf{r},\omega) + 
w^{\scscs \gamma}_{\text{ex}}(\mathbf{r},\omega)
\right]
\psi_{w\sigma}^{\scscs \gamma}(\mathbf{r},\omega)
= 0,
\;\;\; \mbox{for all $\mathbf{R}$},
\end{eqnarray}
indicating that Eq.~(\ref{w.prop}) can be satisfied with any choice of $w^{\scscs
\gamma}_{\text{de}}$ and $w^{\scscs \gamma}_{\text{un}}$. 

The above relation is satisfied when we have
\begin{subequations} \label{w.propT} 
\begin{eqnarray}
\label{w.prop.oc} 
\lim_{\mathbf{r}\rightarrow\mathbf{R}}
|\mathbf{r}-\mathbf{R}| 
w^{\scscs \gamma}_{\text{oc}}(\mathbf{r},\omega)
\psi_{w\sigma}^{\scscs \gamma}(\mathbf{r},\omega)&=&0, \\
\label{w.prop.ex} 
\lim_{\mathbf{r}\rightarrow\mathbf{R}}
|\mathbf{r}-\mathbf{R}| 
w^{\scscs \gamma}_{\text{ex}}(\mathbf{r},\omega)
\psi_{w\sigma}^{\scscs \gamma}(\mathbf{r},\omega)&=&0.
\end{eqnarray}
\end{subequations}

It is easily proven that a determinantal state $|\gamma\rangle$ that satisfies
Eq.~(\ref{sch}) -- and the corresponding density $\rho$ from $|\gamma\rangle$ --
does not depend on $w^{\scscs \gamma}_{\text{oc}}$; so, when considering the
statements appearing in the last paragraph of the previous section, we can relax
the requirement that Eq.~(\ref{w.prop}) be satisfied, and only require
Eq.~(\ref{w.prop.ex}) to be satisfied. In other words, the density of a
determinantal state, that satisfies Eq.~(\ref{sch}), can be used to determine its
external potential, given by Eq.~(\ref{potential}), by using Eq.~(\ref{T.rho}), if
(\ref{w.prop.ex}) is satisfied. An equivalent statement refers to the one-body
operator $\hat{f}_\gamma$: The density of a determinantal state can be used to
determine its external potential, given by Eq.~(\ref{potential}), by using
Eq.~(\ref{T.rho}), if (\ref{w.prop.ex}) is satisfied, where the orbitals defining
the determinantal state $|\gamma\rangle$ are the occupied eigenfunctions of
$\hat{f}_\gamma$, defined by Eq.~(\ref{one.op}). Note that the $w^{\scscs
\gamma}_{\text{oc}}$ and $w^{\scscs \gamma}_{\text{un}}$ portions of the operator
$w^{\scscs \gamma}$ are at our disposal, since the determinantal state does not
depend on these components; $w^{\scscs \gamma}_{\text{de}}$ is determined by
$w^{\scscs \gamma}_{\text{ex}}$, since $w^{\scscs \gamma}$ is required to be
Hermitian. (The Hermitian requirement can be dropped by using a biorthogonal basis
set.)

\section{Hartree--Fock Determinantal states}

We now show that the set of Hartree--Fock determinantal states, say
$\{|\tilde{\tau}\}$, are members of $\{|\gamma\rangle_{v}\}$, indicating that
their Coulombic external-potentials $v$ can be uniquely determined by their
electron density, i.e., $v(\rho)$, by using Eq.~(\ref{T.rho}).

The occupied, canonical Hartree--Fock orbitals satisfy the following single
particle Eq:
\begin{eqnarray} \label{HF.eq} 
\hat{F}_{\tilde{\tau}}\psi_{x\sigma}^{\scscs \tilde{\tau}}(\mathbf{x}) =
\epsilon_{x\sigma}^{\tilde{\tau}} \psi_{x\sigma}^{\scscs \tilde{\tau}}(\mathbf{x}), 
\;\;\; \sigma = \alpha,\beta,
\;\;\; \psi_{x\sigma}^{\scscs \tilde{\tau}} \in 
\{\psi_o \leftarrow \tilde{\tau}, \mbox{\small
$\hat{F}_{\tilde{\tau}}$}\}
\end{eqnarray}
where the Fock operator is given by
\begin{eqnarray} \label{Fockop} 
\hat{F}_{\tilde{\tau}}=
-\mbox{$\frac12$}\nabla^2_{\!\mathbf{r}}
+ v(\mathbf{r})
+
\int 
r_{12}^{-1}
\tilde{\tau}(\mathbf{x}_2,\mathbf{x}_2) 
\,d\mathbf{x}_2
+ \hat{v}_{\mathrm{x}}^{\scscs \tilde{\tau}}(\mathbf{x}),
\end{eqnarray}
and the one-particle density-matrix for the Hartree--Fock reference-state has the
following form:
\begin{eqnarray} \label{onepart.hf} 
\tilde{\tau}(\mathbf{x},\mathbf{x}^\prime)
= \sum_{x\sigma\in\{\psi_o \leftarrow \tilde{\tau}, 
\hat{F}_{\tilde{\tau}}\}}
\psi_{x\sigma}^{\scscs \tilde{\tau}}(\mathbf{x}) 
\left(\psi_{x\sigma}^{\scscs \tilde{\tau}}(\mathbf{x}^\prime)\right)^*;
\end{eqnarray}
furthermore, the exchange operator, $\hat{v}_{\mathrm{x}}^{\scscs \gamma}$, is a
non-local operator that is defined by its kernel, $-r_{12}^{-1}
\gamma$. Therefore, for an arbitrary function, say $\psi$, we have
\begin{eqnarray} \label{exchop.hf} 
\hat{v}_{\mathrm{x}}^{\scscs \tilde{\tau}}(\mathbf{x}_1) \psi(\mathbf{x}_1)
= - \int d\mathbf{x}_2 \, r_{12}^{-1} \tilde{\tau}(\mathbf{x}_1,\mathbf{x}_2)
\psi(\mathbf{x}_2).
\end{eqnarray}

Equating Eqs.~(\ref{HF.eq}) and (\ref{Fockop}) with (\ref{one.sch}) and
(\ref{one.op}), for ($\gamma = \tilde{\tau}$), we have
\begin{eqnarray} 
w^{\scscs \tilde{\tau}}(\mathbf{x}) =
\int 
r_{12}^{-1}
\tilde{\tau}(\mathbf{x}_2,\mathbf{x}_2) 
\,d\mathbf{x}_2
+ \hat{v}_{\mathrm{x}}^{\scscs \tilde{\tau}}(\mathbf{x}),
\end{eqnarray}
and it is easily seen that Eq.~(\ref{w.prop}) is satisfied; so, the Hartree--Fock
states are members of $\{|\gamma\rangle_{v}\}$, and we have $v(\tilde{\varrho})$ where
$\tilde{\varrho}$ is the Hartree--Fock density:
\begin{eqnarray} 
\tilde{\varrho}(\mathbf{r}) = 
\sum_\omega \tilde{\tau}(\mathbf{r},\omega;\mathbf{r},\omega). 
\end{eqnarray}

The Hartree--Fock Eqs.\ are usually solved using a iterative, self consistent
field (SCF) approach, where the ($m-1$)th iteration is given by
\begin{eqnarray} 
\hat{F}_{\tilde{\tau}_{m-1}}\psi_{x\sigma}^{\scscs \tilde{\tau}_m}(\mathbf{x}) =
\epsilon_{x\sigma}^{\tilde{\tau}_m} \psi_{x\sigma}^{\scscs \tilde{\tau}_m}(\mathbf{x}), 
\;\;\; \sigma = \alpha,\beta,
\;\;\; \psi_{x\sigma}^{\scscs \tilde{\tau}_m} \in 
\{\psi_o \leftarrow \tilde{\tau}, \mbox{\small
$\hat{F}_{\tilde{\tau}_m-1}$}\}
\end{eqnarray}
and its easily seen that Eq.~(\ref{w.prop}) is satisfied for $w^{\scscs
\tilde{\tau}_{m-1}}$, so all determinantal states determined during the SCF
approach are also members of $\{|\gamma\rangle_{v}\}$.

Consider another Hermitian Fock-type operator, say $\hat{F}^\prime_{\gamma}$,
that is given by
\begin{eqnarray} \label{F.prime} 
\hat{F}^\prime_{\gamma}=
\hat{F}_{\gamma} +
\hat{v}_{\mathrm{x}\mbox{\tiny $+$}}^{\scscs \gamma}
\end{eqnarray}
where the excitation (ex) portion of the additional exchange-operator
$\hat{v}_{\mathrm{x}\mbox{\tiny$+$}}^{\scscs \gamma}$ is zero:
\begin{eqnarray}
\left(\hat{v}_{\mathrm{x}\mbox{\tiny$+$}}^{\scscs \gamma}\right)_{\mathrm{ex}}=0.
\end{eqnarray}
Hence, according to Sec.~\ref{INV}, Eq.~(\ref{w.prop.ex}) remains satisfied and,
in addition, the determinantal state defined by the occupied orbitals, from
$\hat{F}^\prime_{\gamma}$, is the as same the determinantal state from
$\hat{F}_{\gamma}$; it is a member of $\{|\gamma\rangle_{v}\}$; so, again, the
density of this determinantal state can be used to determine the external
potential $v$, Eq.~(\ref{potential}), by using Eq.~(\ref{T.rho}).  Furthermore,
since the occupied eigenfunctions from $\hat{F}^\prime_{\tilde{\tau}}$, given by
\begin{eqnarray} \label{HF.eq2} 
\hat{F}^\prime_{\tilde{\tau}}\psi_{x\sigma}^{\scscs \prime\tilde{\tau}}(\mathbf{x}) =
\epsilon_{x\sigma}^{\tilde{\tau}} \psi_{x\sigma}^{\scscs \prime\tilde{\tau}}(\mathbf{x}), 
\;\;\; \sigma = \alpha,\beta,
\;\;\; \psi_{x\sigma}^{\scscs \prime\tilde{\tau}} \in 
\{\psi_o \leftarrow \tilde{\tau}, \mbox{\small
$\hat{F}^\prime_{\tilde{\tau}}$}\}
\end{eqnarray}
differ only by a unitary transformation from the $\hat{F}_{\tilde{\tau}}$ occupied
orbitals, $\{\psi_o \leftarrow \tilde{\tau}, \mbox{\small
$\hat{F}_{\tilde{\tau}}$}\}$. Therefore, and of course, the one-particle
density-matrix obtained from these occupied orbitals are equivalent:
\begin{eqnarray} 
\tilde{\tau}(\mathbf{x},\mathbf{x}^\prime)
= \sum_{x\sigma\in\{\psi_o \leftarrow \tilde{\tau}, 
\hat{F}^\prime_{\tilde{\tau}}\}}
\psi_{x\sigma}^{\scscs \prime\tilde{\tau}}(\mathbf{x}) 
\left(\psi_{x\sigma}^{\scscs \prime\tilde{\tau}}(\mathbf{x}^\prime)\right)^*,
\end{eqnarray}
where this one-particle density-matrix $\tilde{\tau}$ is the same one appearing in
Eq.~(\ref{onepart.hf}).

\section{Generalized Fock operator} \label{GFO} 

Consider a generalized Fock-operator ${\cal \hat{F}}_{\gamma}$, where its occupied
orbitals satisfy
\begin{eqnarray} \label{EHF.eq} 
{\cal \hat{F}}_{\gamma}\psi_{x\sigma}^{\scscs \gamma}(\mathbf{x}) =
\varepsilon_{x\sigma}^\gamma \psi_{x\sigma}^{\scscs \gamma}(\mathbf{x}), 
\;\;\; \sigma = \alpha,\beta,
\;\;\; \psi_{x\sigma}^{\scscs \gamma} \in 
\{\psi_o \leftarrow \gamma, \mbox{\small
${\cal \hat{F}}_{\gamma}$}\}
\end{eqnarray}
and ${\cal \hat{F}}_{\gamma}$ is given by
\begin{eqnarray} \label{EFockop} 
{\cal \hat{F}}_{\gamma}=
-\mbox{$\frac12$}\nabla^2_{\!\mathbf{r}}
+ v(\mathbf{r})
+
\int 
r_{12}^{-1}
\gamma(\mathbf{x}_2,\mathbf{x}_2) 
\,d\mathbf{x}_2
+ \hat{v}_{\mathrm{x}}^{\scscs \gamma}(\mathbf{x})
+ \hat{v}_{\mathrm{co}}^{\scscs \gamma}(\mathbf{x}).
\end{eqnarray}
Comparing the above two Eqs.\ with (\ref{one.sch}) and (\ref{one.op}), we obtain a
new definition for $w^{\scscs \gamma}$:
\begin{eqnarray} \label{w.newdef} 
w^{\scscs \gamma}(\mathbf{x}) =
\int 
r_{12}^{-1}
\gamma(\mathbf{x}_2,\mathbf{x}_2) 
\,d\mathbf{x}_2
+ \hat{v}_{\mathrm{x}}^{\scscs \gamma}(\mathbf{x}),
+ \hat{v}_{\mathrm{co}}^{\scscs \gamma}(\mathbf{x}).
\end{eqnarray} 
and substituting this expression into Eq.~(\ref{w.prop.ex}), gives
\begin{eqnarray} \label{wehf.prop} 
\lim_{\mathbf{r}\rightarrow\mathbf{R}} 
|\mathbf{r}-\mathbf{R}| 
\left[
\hat{v}_{\mathrm{co}}^{\scscs \gamma}
(\mathbf{r},\omega)
\right]_{\mathrm{ex}} 
\psi_{w\sigma}^{\scscs \gamma}(\mathbf{r},\omega)
= 0,
\;\;\; \mbox{for all $\mathbf{R}$}.
\end{eqnarray}
Hence, if this relation is satisfied, the determinantal state defined by the
occupied orbitals, from ${\cal \hat{F}}_{\gamma}$, is a member of
$\{|\gamma\rangle_{v}\}$; so, again, the density of this determinantal state can
be used to determine the external potential $v$, Eq.~(\ref{potential}), by using
Eq.~(\ref{T.rho}).

\section{Brueckner Determinantal states}

We seek solutions of the time-independent Schr\"odinger equation,
\begin{eqnarray} \label{SE} 
H_v|\Psi\rangle ={\cal E}|\Psi\rangle,
\end{eqnarray}
where $H_v$ denotes the Hamiltonian operator defined by the external potential
$v$, where the Hamiltonian is independent of the number of electrons when it is
expressed in second quantization:
\begin{eqnarray} \label{H} 
H_v = \sum_{i\sigma j\sigma} 
(i\sigma|\hat{h}|j\sigma)
a_{i\sigma}^\dagger a_{j\sigma} +  
\frac{1}{2} 
\sum_{i\sigma j\sigma}
\sum_{k\sigma^\prime l\sigma^\prime}
(i\sigma j\sigma|k\sigma^\prime l\sigma^\prime)
a_{i\sigma}^\dagger 
a_{k\sigma^\prime}^\dagger a_{l\sigma^\prime}  
a_{j\sigma} 
\end{eqnarray}
where our Hamiltonian is spin-free; the spin-free integrals are written using
chemist's notation \cite{Szabo:82}:
\begin{eqnarray} \label{h1} 
(i\sigma|\hat{h}|j\sigma)&=& \int \
\chi_{i\sigma}^*(\mathbf{r})
\left({-}\mbox{\small$\frac{1}{2}$}\nabla^2_\mathbf{r} + v(\mathbf{r}) \right)
\chi_{j\sigma}(\mathbf{r})
d \mathbf{r},
\\
\label{chemist} 
(i\sigma j\sigma|k\sigma^\prime l\sigma^\prime)
&=&\int 
\chi_{i\sigma}^*(\mathbf{r}_1)
\chi_{j\sigma}(\mathbf{r}_1) 
r_{12}^{-1}
\chi_{k\sigma^\prime}^*(\mathbf{r}_2) 
\chi_{l\sigma^\prime}(\mathbf{r}_2) \; d \mathbf{r}_1 d \mathbf{r}_{2},
\end{eqnarray}
and the creation and annihilation operators, $a_{i\sigma}^\dagger$ and
$a_{i\sigma}$, correspond to the unrestricted spin-orbitals, $\psi_{i\sigma}$,
defined by Eq.~(\ref{spinorb}).

The wavefunction of interest $|\Psi\rangle$ can be generated by a wave operator
$\Omega_\gamma$:
\begin{eqnarray} \label{WO} 
\Omega_\gamma |\gamma\rangle = 
(1 + \chi_\gamma) |\gamma\rangle =
|\Psi\rangle,
\end{eqnarray}
and the second relation defines the correlation operator, $\chi_\gamma$;
furthermore, $|\gamma\rangle$ is any determinantal reference-state that overlaps
with the target state: ($\langle \gamma | \Psi\rangle \ne 0$).

Brueckner orbital theory
\cite{Brueckner:54,Nesbet:58,Brenig:61,Lowdin:62,Kutzelnigg:64,Cizek:80,Chiles:81,Stolarczyk:84,Handy:85,Handy:89,Raghavachari:90,Hirao:90,Stanton:92,Hampel:92,Scuseria:94,Lindgren:02}
is a generalization of Hartree--Fock theory that utilizes a single-determinantal
state that has the maximum overlap with the target state \cite{Kobe:71,Shafer:71}.
By definition, if $|\tau\rangle$ is a Brueckner reference-state, then the target
state, $|\Psi\rangle$, contains no singly-excited states from $|\tau\rangle$:
\begin{eqnarray} \label{no.sing} 
\langle\tau_{w\sigma}^{r\sigma^\prime}|\Psi\rangle = 0,
\end{eqnarray}
and the singly-excited states are given by
\begin{eqnarray} \label{sing} 
|\tau_{w\sigma}^{r\sigma^\prime}\rangle = 
a_{r\sigma^\prime}^\dagger a_{w\sigma}
|\tau\rangle,
\end{eqnarray}
where the Brueckner-state occupied- and unoccupied-orbitals are, respectively, given by
\begin{eqnarray} \label{Bocc.orbs} 
\psi_{w\sigma}^{\scscs \tau},\psi_{x\sigma}^{\scscs \tau} \in 
\{\psi_o \rightarrow \tau\}, \\
\label{Bunocc.orbs} 
\psi_{r\sigma^\prime}^{\scscs \tau}, \psi_{s\sigma^\prime}^{\scscs \tau} \in 
\{\psi_u \rightarrow \tau\},
\end{eqnarray}
and this notation indicates that the occupied orbitals determine $\tau$;
furthermore, the unoccupied orbitals also determine $\tau$ since the union of the
two orthogonal sets (of orbitals) is a complete set. 

Note that, unlike the orbitals that are defined by Eq.~(\ref{occ.orbs}), the
occupied orbitals that satisfy Eq.~(\ref{Bocc.orbs}) are not completely defined;
they are invariant to a unitary transformation; similarly, the unoccupied orbitals
that satisfy Eq.~(\ref{Bunocc.orbs}) are also invariant to a unitary
transformation. Using a set of these orbitals, the Brueckner one-particle
density-matrix is given by
\begin{eqnarray} \label{onepart.br} 
\tau(\mathbf{x},\mathbf{x}^\prime)
= \sum_{x\sigma\in\{\psi_o \rightarrow \tau\}}
\psi_{x\sigma}^{\scscs \tau}(\mathbf{x}) 
\left(\psi_{x\sigma}^{\scscs \tau}(\mathbf{x}^\prime)\right)^*,
\end{eqnarray}
and, for future use, we mention that the virtual orbitals define the following
two-body function:
\begin{eqnarray}  \label{orthpart.br} 
\kappa_\tau(\mathbf{x},\mathbf{x}^\prime)
= \sum_{r\sigma\in\{\psi_u \rightarrow \tau\}}
\psi_{r\sigma}^{\scscs \tau}(\mathbf{x}) 
\left(\psi_{r\sigma}^{\scscs \tau}(\mathbf{x}^\prime)\right)^*,
\end{eqnarray}
where, for a complete set of one-particle functions, the sum of the two gives the
Dirac delta function:
\begin{eqnarray}
\delta(\mathbf{x},\mathbf{x}^\prime)
=
\kappa_\tau(\mathbf{x},\mathbf{x}^\prime) +
\tau(\mathbf{x},\mathbf{x}^\prime).
\end{eqnarray}

Since our Hamiltonian, given by Eq.~(\ref{H}), is spin-free, it is easily
demonstrated that we have \cite{Paldus:75}
\begin{eqnarray}
\label{any.sing} 
\langle\gamma_{w\sigma}^{r\sigma^\prime}|\Psi\rangle = 0, 
\;\; \mbox{for $\sigma\ne\sigma^\prime$ and $\langle\gamma|\Psi\rangle \ne 0$ };
\end{eqnarray}
hence, we can modify the definition for a Brueckner reference-state, given by
Eq.~(\ref{no.sing}), and only consider the spin-conserving matrix-elements:
\begin{eqnarray} \label{no.sing.cons} 
\langle\tau_{w\sigma}^{r\sigma}|\Psi\rangle = 0.
\end{eqnarray}
Because of spin symmetry, Eq.~(\ref{any.sing}) certainly holds when
$|\gamma\rangle$ is a determinantal state that is an eigenfunction of the total
spin angular-momentum operator, $\hat{S}^2$, e.g., a closed-shell ground-states
with spatially restricted spin orbitals. However, this identity should also hold
in more general cases, since, diagrammatically speaking, the spin state -- either
$\alpha$ or $\beta$ -- must be conserved along an oriented path \cite{Paldus:75},
and $w\sigma$ and $r\sigma^\prime$ are on the same oriented path. In order to
simplify our discussions, henceforth, we only consider cases where
Eq.~(\ref{any.sing}) holds; however, the result are easily generalized to the more
general case, e.g., when the Hamiltonian is spin-dependent.

Substituting Eqs.~(\ref{SE}) and (\ref{WO}) into (\ref{no.sing.cons}), sequentially, we
obtain
\begin{eqnarray} \label{bbcond} 
0= \langle\tau_{w\sigma}^{r\sigma}|\Psi\rangle =
\langle\tau_{w\sigma}^{r\sigma}|H_v|\Psi\rangle =
\langle\tau_{w\sigma}^{r\sigma}|H_v\Omega_\tau|\tau\rangle =
\langle\tau_{w\sigma}^{r\sigma}|H_v|\tau\rangle +
\langle\tau_{w\sigma}^{r\sigma}|H_v\chi_\tau|\tau\rangle,
\end{eqnarray}
and the vanishing of the above matrix elements involving $H_v$ is know as the
Brillouin--Brueckner condition
\cite{Brenig:57,Lowdin:62,Nesbet:58,Lowdin:62,Kobe:71,Schafer:71}. Writing the
operator-product $H_v\chi_\tau$ in normal-ordered form
\cite{Cizek:66,Cizek:69,Lindgren:86,Paldus:75} with respect to the reference state
$|\tau\rangle$, the last matrix element of the above Eq.\ becomes
\begin{eqnarray} \label{lastme} 
\langle\tau_{w\sigma}^{r\sigma}|H_v\chi_\tau|\tau\rangle =
\langle\tau_{w\sigma}^{r\sigma}|
\left( H_v\chi_\tau \right)_1
|\tau\rangle,
\end{eqnarray}
where the the one-body portion, $\left( H_v\chi_\tau \right)_1$, can be partitioned
in the following manner:
\begin{eqnarray} \label{Hchi.1} 
\left( H_v\chi_\tau \right)_1 =
\left[\left( H_v\chi_\tau \right)_1\right]_{\mathrm{op}} +
\left[\left( H_v\chi_\tau \right)_1\right]_{\mathrm{re}}, 
\end{eqnarray}
and where the open (op) portion and remaining (re) portions have the
following explicit forms \cite{Finley:bdmt.arxiv,Finley:temp}:
\begin{eqnarray} \label{Hchi.1.op} 
\left[\left( H_v\chi_\tau \right)_1\right]_{\mathrm{op}}&=& 
{\sum_{w\sigma r\sigma}}^{\!\tau} U_{w\sigma}^{r\sigma} 
a_{r\sigma}^\dagger a_{w\sigma}, \\
\label{Hchi.1.re} 
\left[\left( H_v\chi_\tau \right)_1\right]_{\mathrm{re}}&=& 
{\sum_{r\sigma w\sigma}}^{\!\tau} U_{r\sigma}^{w\sigma} 
a_{w\sigma}^\dagger a_{r\sigma}
+
{\sum_{r\sigma s\sigma}}^{\!\tau} U_{r\sigma}^{s\sigma} 
a_{s\sigma}^\dagger a_{r\sigma} -
{\sum_{w\sigma x\sigma}}^{\!\tau} U_{x\sigma}^{w\sigma}
a_{x\sigma} a_{w\sigma}^\dagger;
\end{eqnarray}
furthermore, the one-body matrix-elements are defined by
\begin{eqnarray} 
U_{i\sigma}^{j\sigma} =
\langle \psi_{j\sigma}^{\scscs \tau} 
|\left( H_v\chi_\tau \right)_1 
|\psi_{i\sigma}^{\scscs \tau}\rangle,
\end{eqnarray}
and the orbital indices are given by the right side of Eqs.~(\ref{Bocc.orbs}) and
(\ref{Bunocc.orbs}); this choice is indicated by the $\tau$ superscripts appended
to the summations, i.e., ${\sum}^\tau$. (Note that the definition of an open
operator given above differs from the definition used by other authors
\cite{Lindgren:85,Lindgren:86,Lindgren:87}.)

In the above matrix elements, the ones that do not preserve the spin, i.e.,
($U_{i\sigma}^{j\sigma^\prime}$ for $\sigma\ne\sigma^\prime$), are omitted, since
they can easily be shown to vanish for a spin-free Hamiltonian. (The vanishing of
these matrix elements occurs, diagrammatically speaking, since the spin state --
either $\alpha$ or $\beta$ -- must be conserved along an oriented path
\cite{Paldus:75}, and $i\sigma$ and $j\sigma^\prime$ are on the same oriented
path.)

Substituting Eq.~(\ref{Hchi.1}) into (\ref{lastme}) and using (\ref{Hchi.1.op})
and (\ref{Hchi.1.re}),  gives
\begin{eqnarray} \label{chi1.sopA} 
\langle\tau_{w\sigma}^{r\sigma}|
\left( H_v\chi_\tau \right)_1|\tau\rangle =
\langle\tau_{w\sigma}^{r\sigma}|
\left[\left( H_v\chi_\tau \right)_1\right]_{\mathrm{op}}
|\tau\rangle.
\end{eqnarray}
Since the one-body operator-product $\left[\left( H_v\chi_\tau
\right)_1\right]_{\mathrm{op}}$ can also act within the one-body sector of the
Hilbert space, we have the following identity:
\begin{eqnarray} \label{chi1.sop} 
\langle\tau_{w\sigma}^{r\sigma}|
\left[\left( H_v\chi_\tau \right)_1\right]_{\mathrm{op}}
|\tau\rangle
=
\langle \psi_{r\sigma}^{\scscs \tau}|
\left[\left( H_v\chi_\tau \right)_1\right]_{\mathrm{op}}
|\psi_{w\sigma}^{\scscs \tau}\rangle.
\end{eqnarray}
Substituting Eq.~(\ref{lastme}) into the Brillouin--Brueckner condition,
Eq.~(\ref{bbcond}), and using Eq.~(\ref{chi1.sopA}) and (\ref{chi1.sop}), and also
the following identity:
\begin{eqnarray} \label{bcond} 
\langle \psi_{r\sigma}^{\scscs \tau}|
(\hat{F}_{\tau})_{\mathrm{ex}}
|\psi_{w\sigma}^{\scscs \tau}\rangle =
\langle\tau_{w\sigma}^{r\sigma}|H_v|\Psi\rangle,
\end{eqnarray}
involving the Fock operator, Eq.~(\ref{Fockop}), yields
\begin{eqnarray} \label{bbc.form} 
\langle \psi_{r\sigma}^{\scscs \tau}|
(\hat{F}_{\tau})_{\mathrm{ex}}
|\psi_{w\sigma}^{\scscs \tau}\rangle
+
\langle \psi_{r\sigma}^{\scscs \tau}|
\left(\hat{v}_{\mathrm{co}}^{\scscs \tau}\right)_{\mathrm{ex}}
|\psi_{w\sigma}^{\scscs \tau}\rangle
=
0,
\end{eqnarray}
where the introduced correlation potential $\hat{v}_{\mathrm{co}}^{\scscs \tau}$,
by definition, satisfies
\begin{eqnarray} \label{vco.def} 
\left(\hat{v}_{\mathrm{co}}^{\scscs \tau}\right)_{\mathrm{ex}}=
\left[\left( H_v\chi_\tau \right)_1\right]_{\mathrm{op}},
\end{eqnarray}
and the operators, $(\hat{F}_{\tau})_{\mathrm{ex}}$ and
$(\hat{v}_{\mathrm{co}}^{\scscs \tau})_{\mathrm{ex}}$, are defined in an analogous
way as $w^{\scscs \gamma}_{\text{ex}}$, as indicated by Eqs.~(\ref{w.one.part})
through (\ref{w.un}).

Since the above form of the Brillouin--Brueckner condition, given by
Eq.~(\ref{bbc.form}), is satisfied by all pairs of orbitals involving one
unoccupied-orbital and one occupied-orbital, we have
\begin{eqnarray} 
\left({\cal \hat{F}_{\tau}}\right)_{\mathrm{ex}}
= 0,
\end{eqnarray}
where the generalized, or exact, Fock operator is defined by
\begin{eqnarray} \label{gen.Fockop} 
{\cal \hat{F}_{\tau}} =
\hat{F}_{\tau} +  \hat{v}_{\mathrm{co}}^{\scscs \tau}.
\end{eqnarray}
Comparing this definition of ${\cal \hat{F}_{\tau}}$ with the one given by
Sec.~\ref{GFO}, Eq.~(\ref{EFockop}), and using Eq.~(\ref{Fockop}), we see that,
for ($\gamma = \tau$), the two definitions are equivalent, except that in this
Sec.\ we require the excitation (ex) portion of the correlation potential
$\hat{v}_{\mathrm{co}}^{\scscs \tau}$ to satisfy Eq.~(\ref{vco.def}); by
arbitrarily defining the other portions of $\hat{v}_{\mathrm{co}}^{\scscs \tau}$
we can diagonalize ${\cal \hat{F}_{\tau}}$, and this eigenvalue Eq.\ is given by
Eq.~(\ref{EHF.eq}) for ($\gamma = \tau$):
\begin{eqnarray} \label{EHF.eq2} 
{\cal \hat{F}}_{\tau}\psi_{x\sigma}^{\scscs \tau}(\mathbf{x}) =
\varepsilon_{x\sigma}^\tau \psi_{x\sigma}^{\scscs \tau}(\mathbf{x}), 
\;\;\; \sigma = \alpha,\beta,
\;\;\; \psi_{x\sigma}^{\scscs \tau} \in 
\{\psi_o \leftarrow \tau, \mbox{\small
${\cal \hat{F}}_{\tau}$}\},
\end{eqnarray}
and, furthermore, Eq.~(\ref{wehf.prop}) becomes
\begin{eqnarray} \label{wehf.prop2} 
\lim_{\mathbf{r}\rightarrow\mathbf{R}} 
|\mathbf{r}-\mathbf{R}| 
\left[\hat{v}_{\mathrm{co}}^{\scscs \tau}(\mathbf{r},\omega)\right]_{\mathrm{ex}}
\psi_{w\sigma}^{\scscs \tau}(\mathbf{r},\omega)
= 0,
\;\;\; \mbox{for all $\mathbf{R}$},
\end{eqnarray}
where $\left(\hat{v}_{\mathrm{co}}^{\scscs \tau}\right)_{\mathrm{ex}}$ is given by
Eq.~(\ref{vco.def}).  Hence, if this relation is satisfied, Bruckner determinantal
states $\{|\tau\rangle\}$ are member of $\{|\gamma\rangle_{v}\}$; so, again, the
density of a Brueckner determinantal state can be used to determine its external
potential $v$, Eq.~(\ref{potential}), by using Eq.~(\ref{T.rho}). 

Using the results from appendix~(\ref{prooft}), we have
\begin{eqnarray} \label{vcopsi} 
\lim_{\mathbf{r}_1\rightarrow\mathbf{R}} 
|\mathbf{r}_1-\mathbf{R}|
\left[\hat{v}_{\mathrm{co}}^{\scscs\tau} (\mathbf{x}_1)
\right]_{\mathrm{ex}}
\psi_{w\sigma}^{\scscs \tau}(\mathbf{x}_1) 
=
\lim_{\mathbf{r}_1\rightarrow\mathbf{R}} 
|\mathbf{r}_1-\mathbf{R}|\left(
C_{w\sigma}^{x\sigma}
\hat{h}_{v \mbox{\tiny $1$}}
\psi_{x\sigma}^{\scscs \tau}(\mathbf{x}_1) 
+
D_{w\sigma}^{r\sigma}
\hat{h}_{v \mbox{\tiny $1$}}
\psi_{r\sigma}^{\scscs \tau}(\mathbf{x}_1) 
\right),
\end{eqnarray}
where
\begin{eqnarray} \label{hv1} 
\hat{h}_{v \mbox{\tiny $1$}}=
-\mbox{$\frac12$}\nabla^2_{\!\mathbf{r}_1}
+ v(\mathbf{r}_1),
\end{eqnarray}
and there are summations over the repeated indices $x\sigma$ and $r\sigma$ for the
orbital sets $\{\psi_o \leftarrow \tau, \hat{F}_{\tau}\}$ and $\{\psi_u \leftarrow
\tau, \hat{F}_{\tau}\}$. (The coefficients $C_{w\sigma}^{x\sigma}$
and $D_{w\sigma}^{r\sigma}$ are defined by Eqs.~(\ref{Cwx}) and (\ref{Dwr}).)

Unfortunately we have been unable to prove that Eq.~(\ref{wehf.prop2}) is an
identity by using Eq.~(\ref{vcopsi}). So, as an alternative approach, consider the
case where the above identity, given by Eq.~(\ref{wehf.prop2}), is not necessarily
satisfied. As in the derivation Eq.~(\ref{also.nat}), by multiplying
Eq.~(\ref{EHF.eq2}) by $|\mathbf{r}_1-\mathbf{R}|$ followed by taking a limit of
this term vanishing, gives the following identity that {\em must} be satisfied:
\begin{eqnarray} \label{also.nat2} 
\lim_{\mathbf{r}_1\rightarrow\mathbf{R}} 
|\mathbf{R}-\mathbf{r}_1|
\hat{h}_{v \mbox{\tiny $1$}}
\psi_{w\sigma}^{\scscs \tau}(\mathbf{x}_1) 
+
\lim_{\mathbf{r}_1\rightarrow\mathbf{R}} 
|\mathbf{r}_1-\mathbf{R}| 
\left[\hat{v}_{\mathrm{co}}^{\scscs \tau}(\mathbf{r}_1,\omega)\right]_{\mathrm{ex}}
\psi_{w\sigma}^{\scscs \tau}(\mathbf{r}_1,\omega)
=0,
\end{eqnarray}
where we have used Eqs.~(\ref{gen.Fockop}), (\ref{hv1}), and (\ref{Fockop}) and, also, omitted
the Coulomb and exchange terms, since these terms vanish; furthermore, we have
used the decomposition of $\hat{v}_{\mathrm{co}}^{\scscs \tau}$ as defined by
Eq.~(\ref{w.one.part}), and have chosen $(\hat{v}_{\mathrm{co}}^{\scscs
\tau})_{\mathrm{oc}}$ to be zero, since, according to the discussion within
Sec.~\ref{INV}, this portion is at our disposal; the one-particle density-matrix
$\tau$ is invariant to this choice.

Substituting Eq.~(\ref{vcopsi}) into (\ref{also.nat2}), gives
\begin{eqnarray} 
\lim_{\mathbf{r}_1\rightarrow\mathbf{R}} 
|\mathbf{r}_1-\mathbf{R}|\left(
\tilde{C}_{w\sigma}^{x\sigma}
\hat{h}_{v \mbox{\tiny $1$}}
\psi_{x\sigma}^{\scscs \tau}(\mathbf{x}) +
D_{w\sigma}^{r\sigma}
\hat{h}_{v \mbox{\tiny $1$}}
\psi_{r\sigma}^{\scscs \tau}(\mathbf{x}) 
\right) = 0,
\end{eqnarray}
where
\begin{eqnarray}
\tilde{C}_{w\sigma}^{x\sigma} = \delta_{w\sigma\!,x\sigma} + C_{w\sigma}^{x\sigma}.
\end{eqnarray}
Since both terms from the above identity are independent, apparently, we must have
\begin{eqnarray} \label{also.nat3} 
\lim_{|\mathbf{r}-\mathbf{R}|\rightarrow 0} 
|\mathbf{R}-\mathbf{r}|
\hat{h}_{v \mbox{\tiny $1$}}\psi_{x\sigma}^{\scscs \tau}(\mathbf{x})&=&0, \\
\lim_{|\mathbf{r}-\mathbf{R}|\rightarrow 0} 
|\mathbf{R}-\mathbf{r}|
\hat{h}_{v \mbox{\tiny $1$}}\psi_{r\sigma}^{\scscs \tau}(\mathbf{x})&=&0.
\end{eqnarray}
Substituting these relations into Eq.~(\ref{vcopsi}) proves that
Eq.~(\ref{wehf.prop2}) is an identity.  Hence, the density of the
Brueckner-determinantal state, $|\tau\rangle$, can be used to determine the
external potential $v$, Eq.~(\ref{potential}), by using Eq.~(\ref{T.rho}).  Note
that Eq.~(\ref{also.nat3}) is identical to Eq.~(\ref{also.nat}), the cusp
condition, except that the orbitals are now Brueckner; Eq.~(\ref{also.nat3}) can
also be used to prove all relations within Sec.\ref{v.det.rho} that appear after
Eq.~(\ref{also.nat}), including the one above that states that the density of the
Brueckner-determinantal state can be used to determine the external potential.

\section{One-particle density-matrix theory}

\subsection{Variational Brueckner orbital theory} \label{VBOT} 

Reference-state one-particle density-matrix theory
\cite{Finley:bdmt.arxiv,Finley:temp,Finley:bdft.arxiv} is based on Brueckner
orbital theory
\cite{Brueckner:54,Nesbet:58,Brenig:61,Lowdin:62,Kutzelnigg:64,Cizek:80,Chiles:81,Stolarczyk:84,Handy:85,Handy:89,Raghavachari:90,Hirao:90,Stanton:92,Hampel:92,Scuseria:94,Lindgren:02}.
Unlike many other density functional formalisms based on variants of the
Konh--Sham method, the correlation operator for this approach is non-local.  The
approach also uses an energy functional that depends on the one-particle
density-matrix of a reference determinantal-state, and not the exact one from the
target state, where the energy functional is partitioned into the exact
exchange-energy and a correlation-energy functional that is non-universal, since
this functional depends on the external potential~$v$.  We now modify this
formalism to remove -- for the most part -- the dependence of the correlation
energy-functional upon the external potential.  However, an additional term that
describes a portion of the potential energy is also included that does not have an
analog in Kohn--Sham approaches. However, this term can be easily treated once the
kinetic energy functional is known or, in many cases, this term can be neglected,
since it is probably quite small. For convenience, we refer to the previous works
\cite{Finley:bdmt.arxiv,Finley:temp,Finley:bdft.arxiv} as being $v$-dependent, even though we
still retain some $v$-dependence in the correlation-energy functionals for the
current approach under consideration.

In this previous work \cite{Finley:bdmt.arxiv,Finley:temp,Finley:bdft.arxiv}, we
introduced four $v$-dependent trial wavefunctions -- say $|\Psi_{\gamma
v}^{\scscs(\eta)}\rangle$, where ($\eta= \mbox{{\small I, II, III}, and {\small
IV}}$) -- that are defined with respect to an external potential $v$ and a
one-particle density-matrix, where $\gamma$ is from a single-determinantal
reference-state, $|\gamma\rangle$.

The first trial-wavefunction $|\Psi_{\gamma v}^{\scscs (\mathrm{I})}\rangle$ is
simply the target state of interest, say $|\Psi_{\!\mbox{\tiny $N$}v}\rangle$,
with the single excitations removed:
\begin{equation} \label{first.tr} 
|\Psi_{\gamma v}^{\scscs (\mathrm{I})}\rangle = 
\left(1-P_{11}^{\gamma}\right)|\Psi_{\!\mbox{\tiny $N$}v}\rangle,
\end{equation}
where $|\Psi_{\!\mbox{\tiny $N$}v}\rangle$ is a ground-state determined by the
external potential $v$ and the number of electrons $N$ and, furthermore, the
spin-conserved projector for the singly-excited states is
\begin{equation} \label{P11} 
P_{11}^{\gamma} = 
\sum_{w\sigma\in \{\psi_o\rightarrow \gamma\}} 
\sum_{r\sigma\in \{\psi_u\rightarrow \gamma\}}
|\gamma_{w\sigma}^{r\sigma}\rangle\langle\gamma_{w\sigma}^{r\sigma}|,
\end{equation}
where we assume that Eq.~(\ref{any.sing}) holds, but this requirement can easily
be dropped by appending the states $|\gamma_{w\sigma}^{r\sigma^\prime}\rangle$ the
the right side of Eq.~(\ref{P11}).  Note that the $P_{11}^{\gamma}$ subspace is
completely determined by $|\gamma\rangle$; $P_{11}^{\gamma}$ is also invariant to
a unitary transformation of occupied, or virtual, orbitals \cite{Stolarczyk:84}.

We now require the state $|\gamma\rangle$ to be a member of $\{|\gamma\rangle_{v}\}$, 
so Eq.~(\ref{first.tr}) becomes
\begin{equation} \label{first.trb} 
|\Psi_{\gamma v^\prime}^{\scscs (\mathrm{I})}\rangle = 
\left(1-P_{11}^{\gamma}\right)|\Psi_{\!\mbox{\tiny $N$}v^\prime}\rangle,
\;\;\; \gamma \in  \{|\gamma\rangle_{v}\}, \;\; \gamma \rightarrow v,
\;\; \gamma \rightarrow N,
\end{equation}
where the prime superscripts appended to the external potentials, i.e.,
$v^\prime$, emphasizes that the potential defining the states, $|\Psi_{\gamma
v^\prime}^{\scscs (\mathrm{I})}\rangle$ and $|\Psi_{\!\mbox{\tiny
$N$}v^\prime}\rangle$, may differ from the one determined by $\gamma$ (indicated
by $\gamma \rightarrow v$). However, we now restrict these potentials to be
equivalent, i.e., ($v^\prime=v$), and generate the target state using a wave
operator:
\begin{equation} \label{first.trc} 
|\Psi_{\gamma v}^{\scscs (\mathrm{I})}\rangle = 
\left(1-P_{11}^{\gamma}\right)|\Psi_{\!\mbox{\tiny $N$}v}\rangle
=
\left(1-P_{11}^{\gamma}\right)\Omega_{\gamma v}|\gamma\rangle,
\;\;\; \gamma \in  \{|\gamma\rangle_{v}\},
\;\; \gamma \rightarrow v,
\end{equation}
and since the external potential is a unique function of $\gamma$ (or the density
$\rho$ of $\gamma$), we can use the function $v(\gamma)$ to express the wave
operator as a unique function of $\gamma$:
\begin{equation} \label{omega.vnot} 
\Omega_{\gamma v} = \Omega_{\gamma} 
\;\;\; \gamma \in  \{|\gamma\rangle_{v}\},
\;\; \gamma \rightarrow v,
\;\; \gamma \rightarrow N,
\end{equation}
and for are target state, we have
\begin{equation} \label{wf.vnot} 
|\Psi_{\gamma}\rangle = |\Psi_{\!\mbox{\tiny $N$}v}\rangle,\;\;
\gamma \rightarrow v,
\;\; \gamma \rightarrow N.
\end{equation}
Therefore, we are assuming that the target state is completely determined by
$\gamma$. This is a reasonable assumption, since the Hamiltonian is completely
determined by $\gamma$, since $\gamma$ gives the number of electrons, $N$, and the
external potential $v$.  In addition, however, we must also make the assumption
that the wave operator is, or -- at least, in principle -- can be uniquely defined
so that it generates only one exact eigenstate -- the ground state -- from all
$\gamma$ that have ($\gamma \rightarrow v$). This implies, however, two
wavefunctions can differ only by a constant, say $c$, if they are obtained from
density-matrices that determine the same external potential:
\begin{equation}
|\Psi_{\gamma}\rangle =
c|\Psi_{\gamma^\prime}\rangle, 
 \;\;\;
\mbox{if $\gamma \rightarrow v$, 
$\gamma^\prime \rightarrow v^\prime$, and $v=v^\prime$},
\end{equation}
where the $\gamma$ subscript also indicates the normalization of the target state:
\begin{equation}
\langle \gamma |\Psi_{\gamma}\rangle = 1.
\end{equation}

By substituting Eqs.~(\ref{omega.vnot}) and (\ref{wf.vnot}) into
(\ref{first.trc}), we obtain a trial wavefunction that can be assumed to be
determined by $\gamma$:
\begin{equation} 
|\Psi_{\gamma}^{\scscs (\mathrm{I})}\rangle=
\left(1-P_{11}^{\gamma}\right)\Omega_{\gamma}|\gamma\rangle
 = 
\left(1-P_{11}^{\gamma}\right)|\Psi_{\gamma}\rangle,
\;\;\; \gamma \in  \{|\gamma\rangle_{v}\},
\end{equation}
and we have
\begin{eqnarray}
|\Psi_{\gamma}^{\scscs (\mathrm{I})}\rangle &=&
c|\Psi_{\gamma^\prime}^{\scscs (\mathrm{I})}\rangle, 
 \;\;\;
\mbox{if $\gamma \rightarrow v$, $\gamma^\prime \rightarrow v^\prime$, and $v=v^\prime$},
\\
\langle \gamma |\Psi_{\gamma}^{\scscs (\mathrm{I})}\rangle&=&1.
\end{eqnarray}

The second trial-wavefunction $|\Psi_{\gamma}^{\scscs (\mathrm{II})}\rangle$ is
defined with respect to the target state expressed by an exponential ansatz:
($|\Psi\rangle= e^{S_{\gamma}}|\gamma\rangle$), where $|\Psi_{\gamma}^{\scscs
(\mathrm{II})}\rangle$ is generated by removing the single-excitation amplitudes
$S_1^{\gamma}$ from the cluster-operator $S$:
\begin{equation} \label{trial.IIb} 
|\Psi_{\gamma}^{\scscs (\mathrm{II})}\rangle=e^{(S_{\gamma}-S_1^{\gamma})}|\gamma\rangle,
\end{equation}
where, as in Eq.~(\ref{omega.vnot}), we have
\begin{equation} \label{omega.vnotb} 
S_{\gamma v} = S_{\gamma} 
\;\;\; \gamma \in  \{|\gamma\rangle_{v}\},
\;\; \gamma \rightarrow v,
\;\; \gamma \rightarrow N.
\end{equation}

The third trial-wavefunction $|\Psi_{\gamma}^{\scscs (\mathrm{III})}\rangle$ can be
generated by its wave-operator:
\begin{equation} \label{ccf.wo} 
\hat{\Omega}_\gamma|\gamma\rangle = |\Psi_{\gamma}^{\scscs (\mathrm{III})}\rangle,
\end{equation}
that can be expressed in an exponential form: ($\hat{\Omega}_\gamma=
e^{\hat{S}_{\gamma}}|\gamma\rangle$), where $\hat{S}_{\gamma}$ can be written as a sum
$n$-body excitations, with the exclusion of a one-body operator:
\begin{equation} \label{T.ccf} 
\hat{S}_{\gamma}= \hat{S}_2^{\gamma}+\hat{S}_3^{\gamma}+\cdots.
\end{equation}
The wave operator $\hat{\Omega}_{\gamma}$ is a solution to the coupled cluster
equations
\cite{Hubbard:57,Coester:58,Cizek:66,Cizek:69,Cizek:71,Lindgren:78,Bartlett:78,
Pople:78,Lindgren:86,Harris:92} with the single excitation portion removed:
\begin{equation} \label{lct.op.iv} 
\left(1-P_{11}^{\gamma}\right) \left(H_v\hat{\Omega}_\gamma \right)_{\text{op,cn}} = 0,
\;\; \gamma \rightarrow v,
\;\; \gamma \rightarrow N,
\end{equation}
where only the open (op) and connected (cn) portions enter into the relation. This
expression defines the trial functional $|\Psi_{\gamma}^{\scscs
(\mathrm{III})}\rangle$ using Eq.~(\ref{ccf.wo}) and, again,
Eq.~(\ref{omega.vnot}) is satisfied with $\hat{\Omega}_\gamma$ replacing
$\Omega_\gamma$.

The fourth trial wavefunctions $|\Psi_{\gamma}^{\scscs (\mathrm{IV})}\rangle$ is not
considered here, except to mention that it is obtained by solving the
configuration-interaction equations \cite{Schaefer:72,Szabo:82,Harris:92} in an
approximate way, i.e., by neglecting the single-excitation portion.

All of the trial states $|\Psi_{\gamma}^{\scscs (\eta)}\rangle$ share the property
that they contain no single excitations, i.e., ($P_{11}^{\gamma}|\Psi_{\gamma}^{\scscs
(\eta)}\rangle = 0$), and they generate the target state $|\Psi_\tau\rangle$ when their
reference state satisfies ($|\gamma\rangle=|\tau\rangle$), where $|\tau\rangle$
is the determinantal state constructed from occupied Bruckner orbitals. In other
words, we have
\begin{equation} \label{trwf=exact} 
|\Psi_\tau^{\scscs (\eta)}\rangle = |\Psi_\tau\rangle.
\end{equation}

Since the target state is a solution of the Schr\"odinger equation (\ref{SE}) with
an exact energy, say ${\cal E}_{\mbox{\tiny $N$}v}$, we have
\begin{eqnarray} \label{targetEa} 
{\cal E}_{\mbox{\tiny $N$}v} = 
\frac{\langle\Psi_\gamma|H_v|\Psi_{\gamma}\rangle}
{\langle\Psi_\gamma|\Psi_{\gamma}\rangle} =
E_1[\gamma,v] +
{\cal E}_{\mathrm{co}}[\gamma,v]=
\langle H_v \rangle_\gamma,
\end{eqnarray}
where the first-order energy $E_1$ is given by the expectation value of the
Hamiltonian involving the reference state, $\langle \gamma| H_v|\gamma\rangle$,
and the correlation energy ${\cal E}_{\mathrm{co}}$ is defined above as
(${\cal E}_{\mbox{\tiny $N$}v}~-~E_1$); furthermore, the introduced notation
$\langle H_v \rangle_\gamma$ indicates the expectation value of the Hamiltonian
involving the target state~$|\Psi_{\gamma}\rangle$.

Using the trial wavefunctions, we can define variational energy-functionals that
depend on the one-particle density-matrix:
\begin{eqnarray} \label{Efuncts.var} 
\bar{E}_\eta[\gamma,v] = 
\langle H_v\rangle_{\!\gamma\eta} =
E_1[\gamma,v] + \bar{E}_{\mathrm{co}}^{\scscs (\eta)}[\gamma,v],
\end{eqnarray}
where we use the notation for the expectation value of an operator, say $\hat{A}$,
given by
\begin{eqnarray} \label{expect.not} 
\langle \hat{A}\rangle_{\!\gamma\eta} =
\frac{\langle\Psi_\gamma^{\scscs (\eta)}|
\hat{A}|\Psi_{\gamma}^{\scscs (\eta)}\rangle}
{\langle\Psi_\gamma^{\scscs (\eta)}|\Psi_{\gamma}^{\scscs (\eta)}\rangle}
\end{eqnarray}
and the last relation within Eq.~(\ref{Efuncts.var}) defines the correlation-energy
functionals $\bar{E}_{\mathrm{co}}^{\scscs (\eta)}$ as
($\bar{E}_\eta-E_1$); furthermore, the first order energy is given
by
\begin{eqnarray} \label{first.eb} 
E_1[\gamma,v]
=
\int \left[{-}\mbox{\small$\frac{1}{2}$}\nabla_{\mathbf{r}}^2\,
\rho_1(\mathbf{r},\mathbf{r}^\prime) 
\right]_{\mbox{\tiny $\mathbf{r}^\prime\!\!=\!\!\mathbf{r}$}}
\!d\mathbf{r} + 
\int v(\mathbf{r}) 
\rho(\mathbf{r})
\,d\mathbf{r} 
+ E_{\mathrm{J}}[\rho] + E_{\mathrm{x}}[\rho_1^\sigma], 
\end{eqnarray}
where the Coulomb and exchange energies have their usual forms:
\begin{eqnarray} \label{coul.dm} 
E_{\mathrm{J}}[\rho]&=&
\frac12 \int \!\! \int \!
d\mathbf{r}_1
d\mathbf{r}_2 \,
r_{12}^{-1}
\rho(\mathbf{r}_1) 
\rho(\mathbf{r}_2),
\\ \label{exch.dm} 
\mbox{$-E_{\mathrm{x}}[\rho_1^\sigma]$} &=&
\frac12 \int \!\! \int \! 
d\mathbf{r}_1
d\mathbf{r}_2 \,
r_{12}^{-1}
\left(
\rho_1^\alpha(\mathbf{r}_1,\mathbf{r}_2) 
\rho_1^\alpha(\mathbf{r}_2,\mathbf{r}_1)
+
\rho_1^\beta(\mathbf{r}_1,\mathbf{r}_2) 
\rho_1^\beta(\mathbf{r}_2,\mathbf{r}_1)
\right),
\end{eqnarray}
and the spin-components of the one particle density matrix are given by
\begin{eqnarray} \label{opdm.sc} 
\rho_1^\alpha(\mathbf{r}_1,\mathbf{r}_2)
&=& \sum_{x\alpha\in\{\psi_o\rightarrow \gamma\}}
\chi_{x\alpha}^{\scscs \gamma}(\mathbf{r}_1) 
\left(\chi_{x\alpha}^{\scscs \gamma}(\mathbf{r}_2)\right)^*
=\gamma(\mathbf{r}_1,\mbox{\small $1$},\mathbf{r}_2,\mbox{\small $1$}), \\
\rho_1^\beta(\mathbf{r}_1,\mathbf{r}_2)
&=& \sum_{x\beta\in\{\psi_o\rightarrow \gamma\}}
\chi_{x\beta}^{\scscs \gamma}(\mathbf{r}_1) 
\left(\chi_{x\beta}^{\scscs \gamma}(\mathbf{r}_2)\right)^*
=\gamma(\mathbf{r}_1,\mbox{\small $-1$},\mathbf{r}_2,\mbox{\small $-1$}).
\end{eqnarray}
Substituting Eq.~(\ref{first.eb}) into (\ref{Efuncts.var}) gives the following:
\begin{eqnarray} 
\bar{E}_\eta[\gamma,v]
&=&
\int \left[{-}\mbox{\small$\frac{1}{2}$}\nabla_{\mathbf{r}}^2\,
\gamma(\mathbf{x},\mathbf{x}^\prime) 
\right]_{\mbox{\tiny $\mathbf{x}^\prime\!\!=\!\!\mathbf{x}$}}
\!d\mathbf{x} + 
\int v(\mathbf{r}) 
\rho(\mathbf{r})
\,d\mathbf{r} 
+ E_{\mathrm{J}}[\rho] + 
\bar{E}_{\mathrm{xc}}^{\scscs (\eta)}[\gamma,v],
\end{eqnarray}
where the exchange-correlation energy-functionals are defined by
\begin{eqnarray} 
\label{Excorr.v} 
\bar{E}_{\mathrm{xc}}^{\scscs (\eta)}[\gamma,v] = 
E_{\mathrm{x}}[\gamma]
+
\bar{E}_{\mathrm{co}}^{\scscs (\eta)}[\gamma,v] .
\end{eqnarray}

Returning to our energy functionals, Eq.~(\ref{Efuncts.var}), let the functional
derivative of these functionals yield two-body functions that serve as the kernels
of exact Fock operators:
\begin{eqnarray} \label{kernEfock} 
{\cal \zeta}_{\gamma v}^{\scscs (\eta)}(\mathbf{x}_1,\mathbf{x}_2)&=&
\frac{\bar{E}_\eta[\gamma,v]}
{\delta \gamma(\mathbf{x}_2,\mathbf{x}_1)} 
\\ \nonumber 
&=&
\delta(\mathbf{x}_2-\mathbf{x}_1) \left(
-\mbox{\small$\frac{1}{2}$}
\nabla_{\mbox{\tiny $2$}}^2
+ v(\mathbf{r}_2)
+
\int 
r_{23}^{-1}
\gamma(\mathbf{x}_3,\mathbf{x}_3) 
\,d\mathbf{x}_3
\right)
+
\nu_{\mathrm{xc}}^{\scscs\!\gamma\eta v}(\mathbf{x}_1,\mathbf{x}_2),
\end{eqnarray}
where the kernels of the exchange-correlation operators,
$\nu_{\mathrm{xc}}^{\scscs \!\gamma\eta v}(\mathbf{x}_1,\mathbf{x}_2)$, are obtained
from the exchange-correlation energy-functionals:
\begin{eqnarray} \label{kernexchcorr} 
\nu_{\mathrm{xc}}^{\scscs \!\gamma\eta v}(\mathbf{x}_1,\mathbf{x}_2)
=
\frac{\delta \bar{E}_{\mathrm{xc}}^{\scscs (\eta)}[\gamma,v]} 
{\delta \gamma(\mathbf{x}_2,\mathbf{x}_1)} 
=
\frac{\delta \bar{E}_{\mathrm{co}}^{\scscs (\eta)}[\gamma,v]} 
{\delta \gamma(\mathbf{x}_2,\mathbf{x}_1)}
-r_{12}^{-1}
\gamma(\mathbf{x}_1,\mathbf{x}_2),
\end{eqnarray}
where the last relation uses Eqs.~(\ref{Excorr.v}) and the identity:
\begin{eqnarray} \label{kernexch} 
\frac{\delta E_{\mathrm{x}}[\gamma]} 
{\delta \gamma(\mathbf{x}_2,\mathbf{x}_1)} 
=
-r_{12}^{-1}
\gamma(\mathbf{x}_1,\mathbf{x}_2)
=v_{\mathrm{x}}^{\scscs \gamma}(\mathbf{x}_1,\mathbf{x}_2),
\end{eqnarray}
and the function $v_{\mathrm{x}}^{\scscs \gamma}(\mathbf{x}_1,\mathbf{x}_2)$ is
the kernel of the exchange operator, denoted by $\hat{v}_{\mathrm{x}}^{\scscs
\gamma}$.

Using the variation theorem, and by noting the identity given by
Eq.~(\ref{trwf=exact}), it becomes obvious -- as in our previous $v$-dependent
approach \cite{Finley:bdmt.arxiv,Finley:temp} -- that the minimizing of the functionals
$\bar{E}_\eta[\gamma,v]$, subject to the constraint that the one-particle
density-matrix comes from a single-determinantal state, yields
\begin{eqnarray} \label{trf=exact} 
{\cal E}_{\mbox{\tiny $N$}v}&=&\bar{E}_\eta[\tau,v], \\
 \label{co=cof} 
{\cal E}_{\mathrm{co}}[\tau,v]&=&\bar{E}_{\mathrm{co}}^{\scscs (\eta)}[\tau,v].
\end{eqnarray}
where ${\cal E}_{\mbox{\tiny $N$}v}$ and ${\cal E}_{\mathrm{co}}$ the electronic-energy and
correlation energy arising from the target state, defined by
Eqs.~(\ref{targetEa}); furthermore, $\tau$ is the one-particle density-matrix of
the Brueckner reference-state $|\tau\rangle$ that determines $v$:
\begin{equation} \label{bdm} 
\tau(\mathbf{x},\mathbf{x}^\prime) 
= \sum_{w\in \{\psi_o\rightarrow \tau\}} 
\psi_{w}(\mathbf{x}) \psi_{w}^*(\mathbf{x}^\prime),
\;\; \tau \rightarrow v,
\end{equation} 
and the Brueckner orbitals satisfy the following equivalent conditions:
\begin{subequations}
\label{eF.noncan} 
\begin{eqnarray} \label{eF.noncana} 
\langle \psi_r
|{\cal \hat{\zeta}}_{\tau}^{\scscs (\eta)}|
\psi_w\rangle&=&0; \;\; \psi_w\in \{\psi_o\rightarrow \tau\}, \;\;
\psi_r\in \{\psi_u\rightarrow \tau\}, \;\;
\tau \rightarrow v,
\\
\label{eF.noncanbX} 
\left(\mbox{\small $\hat{1}$}-\tau\right)
{\cal \hat{\zeta}}_{\tau}^{\scscs (\eta)}
\tau&=&0,
\end{eqnarray}
\end{subequations}
where these orbitals do not depend of $\eta$ -- any trial wavefunction gives the
same results.

A unique set of occupied and unoccupied orbitals is obtained by requiring the
occupied and unoccupied blocks of ${\cal \hat{\zeta}}_{\tau}^{\scscs (\eta)}$ to
be diagonal:
\begin{subequations}
\label{eF.can} 
\begin{eqnarray} 
{\cal \hat{\zeta}}_{\tau}^{\scscs (\eta)}
\psi_w^{\tau}(\mathbf{x})&=&
\xi_{w}^{\mbox{\tiny $\tau$}}
\psi^{\tau}_w(\mathbf{x}),
\;\; \psi^{\tau}_w\in \{\psi_o\rightarrow \tau\},
\\
{\cal \hat{\zeta}}_{\tau}^{\scscs (\eta)}
\psi^{\tau}_r(\mathbf{x})&=&
\xi_{r}^{\mbox{\tiny $\tau$}}
\psi^{\tau}_r(\mathbf{x}), \;\;
\psi^{\tau}_r\in \{\psi_u\rightarrow \tau\}.
\end{eqnarray}
\end{subequations}
Henceforth, the orbitals sets that satisfy Eqs.~(\ref{eF.noncan}) and
(\ref{eF.can}) are denoted by $\{\psi^{\tau}_o \leftarrow \tau, \mbox{\small
$\hat{\zeta}_{\tau}^{\scscs (\eta)}$}\}$ and $\{\psi^{\tau}_u \leftarrow \tau,
\mbox{\small $\hat{\zeta}_{\tau}^{\scscs (\eta)}$}\}$, indicating that they are
determined by $\tau$ and ${\cal \hat{\zeta}}_{\tau}^{\scscs (\eta)}$.  Since
theses orbitals, and their energies, can, perhaps, depend on $\eta$, it is more
precise to denote then by $\psi^{\tau\eta}_i$ and $\xi_{i}^{\mbox{\tiny
$\tau\eta$}}$, but we suppress the $\eta$ superscripts to keep the notation less
cluttered.

Substituting Eq.~(\ref{kernEfock}) into Eqs.~(\ref{eF.can}) gives generalized,
canonical Hartree--Fock Eqs:
\begin{eqnarray} \label{eF.canb} 
\left(
-\mbox{\small$\frac{1}{2}$}
\nabla_{\mbox{\tiny $1$}}^2
+ v(\mathbf{r}_1)
+
\int 
r_{12}^{-1}
\tau(\mathbf{x}_2,\mathbf{x}_2) 
\,d\mathbf{x}_2
+ \hat{\nu}_{\mathrm{xc}}^{\scscs \tau\eta v}(\mathbf{x}_1)
\right) 
\psi^{\tau}_i(\mathbf{x}_1)
=
\xi_{i}^{\mbox{\tiny $\tau$}}
\psi^{\tau}_i(\mathbf{x}_1).
\end{eqnarray}

\section{Treatment of the correlation-energy functionals 
$\bar{E}_{\mathrm{co}}^{\scscs (\eta)}$}

The Hamiltonian operator $H_v$ can be written in normal-ordered form
\cite{Cizek:66,Cizek:69,Lindgren:86,Paldus:75} with respect to the reference state
$|\gamma\rangle$: 
\begin{eqnarray} \label{Hv.norm} 
H_v = E_1[\gamma,v] + \{v\}_\gamma +
\{{-}\mbox{\footnotesize $\frac{1}{2}\nabla^2$}\}_\gamma +
\{r_{12}^{-1}\}_{\!\mbox{\tiny $1\!,\!2$}}^\gamma,
\end{eqnarray}
where $\{v\}_\gamma$, $\{{-}\mbox{\footnotesize $\frac{1}{2}\nabla^2$}\}_\gamma$,
and $\{r_{12}^{-1}\}_{\!\mbox{\tiny $1\!,\!2$}}^\gamma$ are terms from the external
potential, kinetic energy and electron-electron interactions:
\begin{eqnarray} \label{v.norm} 
\{v\}_\gamma &=& \sum_{i\sigma j\sigma} 
(i\sigma|v|j\sigma)
\{a_{i\sigma}^\dagger a_{j\sigma}\}_\gamma, \\
\label{kin.norm} 
\{{-}\mbox{\footnotesize $\frac{1}{2}\nabla^2$}\}_\gamma
&=& \sum_{i\sigma j\sigma} 
(i\sigma|{-}\mbox{\footnotesize $\frac{1}{2}\nabla^2$}|j\sigma)
\{a_{i\sigma}^\dagger a_{j\sigma}\}_\gamma,
\\
\label{el.norm} 
\{r_{12}^{-1}\}_{\!\mbox{\tiny $1\!,\!2$}}^\gamma &=&
\frac{1}{2} 
\sum_{i\sigma j\sigma}
\sum_{k\sigma^\prime l\sigma^\prime}
(i\sigma j\sigma|k\sigma^\prime l\sigma^\prime)
\{a_{i\sigma}^\dagger 
a_{k\sigma^\prime}^\dagger a_{l\sigma^\prime}  
a_{j\sigma}\}_\gamma + 
\sum_{i\sigma j\sigma} 
(i\sigma|\left(\hat{v}_{\mathrm{J}}^{\scscs \rho} + 
\hat{v}_{\mathrm{x}}^{\scscs \gamma}\right)|j\sigma)
\{a_{i\sigma}^\dagger a_{j\sigma}\}_\gamma, \nonumber \\
\label{nord.r12} 
\end{eqnarray}
Furthermore, the Coulomb $\hat{v}_{\mathrm{J}}^{\scscs \rho}$ operator
satisfies:
\begin{equation} \label{Coulomb} 
\hat{v}_{\mathrm{J}}^{\scscs \rho}\phi({\mathbf r}_1)
= \int r_{12}^{-1}\rho({\mathbf r}_2) \, 
\phi({\mathbf r}_1) \, 
d{\mathbf r}_2,
\end{equation}
and the exchange operator $\hat{v}_{\mathrm{x}}^{\scscs \gamma}$ is given by
Eq.~(\ref{exchop.hf}).

Substituting Eqs.~(\ref{Hv.norm}) into (\ref{Efuncts.var}) gives
\begin{equation} 
\bar{E}_{\mathrm{co}}^{\scscs (\eta)}[\gamma,v] =
V_{\mathrm{co}}^{\scscs (\eta)}[\gamma,v] +
T_{\mathrm{co}}^{\scscs (\eta)}[\gamma] +
U_{\mathrm{co}}^{\scscs (\eta)}[\gamma]
\end{equation}
where these terms are identified as the potential, kinetic, and
electron-electron-interaction contributions to the correlation-energy functionals:
\begin{eqnarray} \label{pot.co} 
V_{\mathrm{co}}^{\scscs (\eta)}[\gamma,v]&=& 
\langle \{v\}_\gamma \rangle_{\!\gamma\eta} =
\langle \{v\} \rangle_{\!\gamma\eta}
\\
\label{kin.co} 
T_{\mathrm{co}}^{\scscs (\eta)}[\gamma]&=&
\langle \{{-}\mbox{\footnotesize $\frac{1}{2}\nabla^2$}\}_\gamma\rangle_{\!\gamma\eta} 
=
\langle \{{-}\mbox{\footnotesize $\frac{1}{2}\nabla^2$}\} \rangle_{\!\gamma\eta},
\\
\label{ee.co} 
U_{\mathrm{co}}^{\scscs (\eta)}[\gamma]&=& 
\langle \{r_{12}^{-1}\}_{\!\mbox{\tiny $1\!,\!2$}}^\gamma \rangle_{\!\gamma\eta}
=
\langle \{r_{12}^{-1}\}_{\!\mbox{\tiny $1\!,\!2$}} \rangle_{\!\gamma\eta},
\end{eqnarray}
and where we have also introduced a more condensed notation where the vacuum state is
understood to agree with the trial wavefunction.

Similarly, substituting Eqs.~(\ref{Hv.norm}) into (\ref{targetEa}) gives
\begin{equation} \label{Eco.part} 
{\cal E}_{\mathrm{co}}[\gamma,v] =
V_{\mathrm{co}}[\gamma,v] +
T_{\mathrm{co}}[\gamma] +
U_{\mathrm{co}}[\gamma]
\end{equation}
where 
\begin{eqnarray}
V_{\mathrm{co}}[\gamma,v]&=& 
\langle \{v\} \rangle_{\!\gamma},\\
T_{\mathrm{co}}[\gamma]&=& 
\langle \{{-}\mbox{\footnotesize $\frac{1}{2}\nabla^2$}\} \rangle_{\!\gamma} \\
\label{U.exact} 
U_{\mathrm{co}}[\gamma]&=&
\langle \{r_{12}^{-1}\}_{\!\mbox{\tiny $1\!,\!2$}} \rangle_{\!\gamma}.
\end{eqnarray}

\section{Approximations}

If we know the exact correlation energy ${\cal E}_{\mathrm{co}}$ for some
Brueckner one-particle density-matrix, say $\tau^\prime$ with some external
potential, say $v^\prime$, where $\tau^\prime \rightarrow v^\prime$, then using
Eq.~(\ref{co=cof}), we obtain the following reasonable approximation:
\begin{eqnarray} \label{approx6b} 
\bar{E}_{\mathrm{co}}^{\scscs (\mathrm{III})}
[\gamma,v]&=&
{\cal E}_{\mathrm{co}}[\tau^\prime,v^\prime]_{(\tau^\prime=\gamma,v^\prime=v)},
\;\; \tau^\prime \rightarrow v^\prime,
\end{eqnarray}
where a similar approximation has been used previously in the $v$-dependent
approach \cite{Finley:bdmt.arxiv,Finley:temp,Finley:bdft.arxiv}, and we are are
assuming, as we have previously, that this approximation is most appropriate for
($\eta= \mbox{{\small III}}$). Similarly, a reasonable approximation for the
components of the correlation-energy functionals are given by the following
prescriptions:
\begin{eqnarray} \label{pot.coAP} 
V_{\mathrm{co}}^{\scscs (\mathrm{III})}[\gamma,v]&=&
V_{\mathrm{co}}[\tau^\prime,v^\prime]_{(\tau^\prime=\gamma,v^\prime=v)}, 
\;\; \tau^\prime \rightarrow v^\prime,
\\
\label{kin.coAP} 
T_{\mathrm{co}}^{\scscs (\mathrm{III})}[\gamma]&=&
T_{\mathrm{co}}
[\tau^\prime]_{(\tau^\prime=\gamma)} \\
\label{el.coAP} 
U_{\mathrm{co}}^{\scscs (\mathrm{III})}[\gamma]&=& 
U_{\mathrm{co}}[\tau^\prime]_{(\tau^\prime=\gamma)},
\end{eqnarray}
and the three functionals: $V_{\mathrm{co}}$, $T_{\mathrm{co}}$ and
$U_{\mathrm{co}}$, are known for a Brueckner one-particle density-matrix
$\tau^\prime$ that determines the external potential $v^\prime$, i.e.,
$\tau^\prime \rightarrow v^\prime$. Of course, we have many Brueckner one-particle
density-matrices $\tau$ coming from the same external potential $v$; presumable,
we have one $\tau$ from each $N$--electron sector of the Hilbert space, for
external potentials $v$ with nondegenerate ground states.

The correlation-energy functionals $\bar{E}_{\mathrm{co}}^{\scscs (\eta)}$
dependence on the external potential $v$ comes exclusively form the correlated
potential-energy-functional $V_{\mathrm{co}}^{\scscs (\mathrm{III})}$.  In many
cases it is reasonable to assume that the this functionals, and the kinetic energy
one $T_{\mathrm{co}}^{\scscs (\mathrm{III})}$, are small, since the potential- and
kinetic-energy contributions are treated well in first order. Therefore, the
following approximation seems reasonable:
\begin{eqnarray} \label{Eco=U} 
\bar{E}_{\mathrm{co}}^{\scscs (\mathrm{III})}[\gamma] \approx 
U_{\mathrm{co}}[\tau^\prime]_{(\tau^\prime=\gamma)},
\end{eqnarray}
and this approximation yields a universal functional in the Kohn--Sham sense --
the functional does not depend on $v$. However, even the potential energy
contribution to the correlation-energy functional, $V_{\mathrm{co}}^{\scscs
(\eta)}$, can be viewed -- in a more general sense -- as being universal, since if
this functional is known for an arbitrary external potential, than it is known for
other cases, since the manner in which this functional depends on the external
potential is the same for all systems. On the other hand, if we use a model system
to approximate the functionals, this will certainly not generate exact functionals
for real systems, only approximation ones.

Using the helium atom as a model system where $U_{\mathrm{co}}$ is presumable
known, the previous approximation becomes
\begin{eqnarray} \label{Eco=Uc} 
\bar{E}_{\mathrm{co}}^{\scscs (\mathrm{III})}[\gamma] \approx 
U_{\mathrm{co}}
[\tau_{\mbox{\tiny\textsc{h}e}}]
_{(\tau_{\mbox{\tiny\textsc{h}e}} = \gamma)},
\end{eqnarray}
where $\tau_{\mbox{\tiny\textsc{h}e}}$ is the Brueckner one-particle density
matrix from the helium atom. Assuming the Hartree--Fock one-particle
density-matrix, say $\tilde{\tau}_{\mbox{\tiny\textsc{h}e}}$, is approximately
equal to the Brueckner one, $\tau_{\mbox{\tiny\textsc{h}e}}$, we have
\begin{eqnarray} \label{Eco=Ucb} 
\bar{E}_{\mathrm{co}}^{\scscs (\mathrm{III})}[\gamma] \approx 
U_{\mathrm{co}}
[\tilde{\tau}_{\mbox{\tiny\textsc{h}e}}]
_{(\tilde{\tau}_{\mbox{\tiny\textsc{h}e}} = \gamma)},
\end{eqnarray}
and for a closed-shell systems that use spatially-restricted spin-orbitals, given
by Eq.~(\ref{spinorb.res}), we have
\begin{eqnarray} 
\bar{E}_{\mathrm{co}}^{\scscs (\mathrm{III})}[\rho_1] \approx 
U_{\mathrm{co}}
[\tilde{\varrho}_{\mbox{\tiny 1\hspace{-0.15ex}\textsc{h}e}}]
_{(\tilde{\varrho}_{\mbox{\tiny 1\hspace{-0.15ex}\textsc{h}e}} = \rho_1)},
\end{eqnarray}
where $\rho_1$ is the spinless one-particle density matrix
\cite{McWeeny:60,Parr:89}, given by Eq.~(\ref{gamma.sf}).  As demonstrated in
Appendix~\ref{CS.IDENT}, we can use the well known approximation for
$U_{\mathrm{co}}^{\mathrm{cs}} [\tilde{\varrho}_{\mbox{\tiny
1\hspace{-0.15ex}\textsc{h}e}}]$ given by the Colle and Salvetti functional
\cite{Colle:75,Lee:88}, giving
\begin{eqnarray} \label{U.CS} 
\bar{E}_{\mathrm{co}}^{\scscs (\mathrm{III})}[\rho_1] \approx
E_{\mathrm{co}}^{\mathrm{cs}}[\rho_1]
=
U_{\mathrm{co}}^{\mathrm{cs}}
[\tilde{\varrho}_{\mbox{\tiny 1\hspace{-0.15ex}\textsc{h}e}}]
_{(\tilde{\varrho}_{\mbox{\tiny 1\hspace{-0.15ex}\textsc{h}e}} = \rho_1)},
\end{eqnarray}
where this functional, denoted by $E_{\mathrm{co}}^{\mathrm{cs}}$, uses four
empirical parameters that are determined using data from the helium atom. In
addition, it is reaily verified that this approximate correlation-energy
functional neglects the potential- and kinetic-energy components,
$V_{\mathrm{co}}$ and $T_{\mathrm{co}}$, and these terms vanish when the {\em
exact} one-particle density matrix, say $\Gamma_1$, from the target state
$|\Psi_\gamma\rangle$, is equal to the reference state one, $\gamma$; these two
terms are considered to be small, or small enough to neglect, when
($\Gamma_1\approx\gamma_1$), as in the approach used when deriving the
Colle--Salvetti functional. (Note that $V_{\mathrm{co}}$ vanishes if the density
from the reference-state is the same as the density from the target state, as is
the case for the Kohn--Sham method.)

Using an identical derivation as in the $v$-dependent approach
\cite{Finley:bdft.arxiv}, it is readily verified that the well known
density-dependent approximation for the Colle--Salvetti functional \cite{Lee:88},
given by the LYP functional -- at least for closed shell ground states -- remains
valid for the current approach:
\begin{eqnarray} \label{CS.LYP} 
E_{\mathrm{co}}^{\mathrm{cs}}[\rho] \approx
E_{\mathrm{co}}^{\mathrm{lyp}}[\rho],
\end{eqnarray}
where the density $\rho$ dependence is associated with the reference state, and
not the target state.

An electron gas defined by an constant external potential is not a member of
$\{|\gamma\rangle_{v}\}$, as in the case for an electron gas with periodic
boundary conditions. However, let us assume that we can generalize the functional
$v(\rho)$ so that it yields the appropriate constant value, say $v_g$, from the
constant-density of an electron gas, say $\rho_g$; so, we have
($v(\rho_g)=v_g$). And if we denote the one-particle density-matrix of the
Brueckner reference state for an electron gas by $\tau_g$, we get
\begin{eqnarray}\label{Egas.approx} 
\bar{E}_{\mathrm{co}}^{\scscs (\mathrm{III})}[\gamma,v]&\approx& {\cal
E}_{\mathrm{co}}
[\tau_g,v_g]_{(\tau_{\mbox{\tiny$g$}}=\gamma,v_g=v)}.
\end{eqnarray}
However, since the correlation energy ${\cal E}_{\mathrm{co}}$ does not depend on
on the constant external potential $v_g$, we cannot make the substitution
($v_g=v$), and so we obtain an approximation that yields a universal functional:
\begin{eqnarray}\label{Egas.approxb} 
\bar{E}_{\mathrm{co}}^{\scscs (\mathrm{III})}[\gamma]&\approx& {\cal
E}_{\mathrm{co}}^{\scscs (\text{gas})}
[\tau_g]_{(\tau_{\mbox{\tiny$g$}}=\gamma)},
\end{eqnarray}
where ${\cal E}_{\mathrm{co}}^{\scscs (\text{gas})}$ is the correlation energy of
an electron gas, and this approximation is also identical to the approximation
used in the $v$-dependent approach \cite{Finley:bdmt.arxiv,Finley:temp}, but with
a slightly different interpretation and derivation. Furthermore, in order to
include $\tau_g$ in the set $\{|\gamma\rangle_{v}\}$, we only need to require
$v(\rho)$ to vanishes for any constant density: ($v(\rho_g)=0$), where we consider
two external-potentials that differ by a constant to be equivalent.

Starting with Eq.~(\ref{Egas.approxb}), except using a {\em uniform} electron gas,
the same correlation energy-function used in the local density approximation (LDA)
\cite{Kohn:65} was shown to be valid with the $v$-dependent approach, for closed
shell ground states \cite{Finley:bdft.arxiv}; however, in contrast to the
Kohn--Sham approach, the density dependence of the correlation-energy function is
only associated with the reference state, and not, in addition, the target
state. Furthermore, using an identical derivation as in the $v$-dependent
approach, it is easily verified that the correlation-energy functional from LDA
can also be used, as well, as an approximation for $\bar{E}_{\mathrm{co}}^{\scscs
(\mathrm{III})}$ within the current approach under consideration.  Furthermore,
since the exchange-energy functional in the current approach is identical with the
one from the $v$-dependent method, the exchange-energy functionals that are valid
in the $v$-dependent approach are also valid in the current approach, including the
Dirac exchange-functional, and the augmentation of this functional with the Becke
exchange correction \cite{Becke:88}. Hence, as in the $v$-dependent approach, the
LDA and the method known as BLYP are also valid in the current method. (At least
for closed shell ground states.) Furthermore, it is readily verified that the
B3LYP approach \cite{Becke:93,Stephens:94} -- that was demonstrated to be a
reasonable approximation within the $v$-dependent approach for closed-shell
ground-states \cite{Finley:bdft.arxiv} -- remains valid for current approach under
consideration.

While all functionals that have been shown, so far, to be valid approximations for
the $v$-dependent approach, are also valid in the current approach, the use of the
LYP and Colle--Salvetti functional appear more natural within the current
approach under consideration, since these functionals are universal ones that do
not have a dependence on the external potential. And since the BLYP and B3LYP
functionals contain the LYP functional, these approaches are also better suited with
the current approach.

\appendix

\section{Connection with Colle--Salvetti Functional} 
\label{CS.IDENT} 

In order to keep the discussion simple, we only consider closed-shell singlet
states that are well described by a single determinantal-state, where we use
spatially-restricted spin-orbitals, given by
\begin{equation} \label{spinorb.res} 
\psi_{j\sigma}(\mathbf{x})=
\chi_j(\bm{r})\sigma(\omega); \;\; \sigma=\alpha,\beta.
\end{equation}
By using these orbitals, it is easily demonstrated that the one-particle
density-matrix $\gamma$ is determined by the spinless one, as indicated by the
following relation:
\begin{equation} \label{gamma.sf} 
\gamma(\mathbf{x}_1,\mathbf{x}_2) = \frac12 \rho_1(\mathbf{r}_1,\mathbf{r}_2)
\delta_{\omega_1\omega_2}.
\end{equation}
Hence, any functional of $\gamma$ now becomes a functional of $\rho_1$; So, if we
use the Hartree--Fock spin-less one-particle density matrix, say
$\tilde{\varrho}_1$,  Eq.~(\ref{targetEa}), becomes
\begin{eqnarray} 
{\cal E}_{\mbox{\tiny $N$}v}  
=E_1[\tilde{\varrho}_1,v] +
{\cal E}_{\mathrm{co}}[\tilde{\varrho}_1,v].
\end{eqnarray}
Substituting Eq.~(\ref{Eco.part}) into this expression for ($\gamma_1=\varrho_1$),
and neglected the terms $V_{\mathrm{co}}$ and $T_{\mathrm{co}}$, we have
\begin{eqnarray} \label{E.hf.app} 
{\cal E}_{\mbox{\tiny $N$}v}  
\approx
E_1[\tilde{\varrho}_1,v] +
U_{\mathrm{co}}[\varrho_1],
\end{eqnarray}
and using Eq.~(\ref{first.eb}), we have
\begin{eqnarray} \label{E.hf} 
{\cal E}_{\mbox{\tiny $N$}v} 
\approx
\int \left[{-}\mbox{\small$\frac{1}{2}$}\nabla_{\mathbf{r}}^2\,
\tilde{\varrho}_1(\mathbf{r},\mathbf{r}^\prime) 
\right]_{\mbox{\tiny $\mathbf{r}^\prime\!\!=\!\!\mathbf{r}$}}
\!d\mathbf{r} + 
\int v(\mathbf{r}) 
\tilde{\varrho}(\mathbf{r})
\,d\mathbf{r} 
+ E_{\mathrm{J}}[\tilde{\varrho}] + E_{\mathrm{x}}[\tilde{\varrho}_1^\sigma] +
U_{\mathrm{co}}[\varrho_1].
\end{eqnarray}

Consider the total electron-electron potential energy, given as the expectation
value involving the target state $|\Psi_{\rho_1}\rangle$ and the electron-electron
repulsion energy operator $\{r_{12}^{-1}\}_0$:
\begin{eqnarray} 
\langle \{r_{12}^{-1}\}_0 \rangle_\gamma 
&=& 
\frac{\langle\Psi_{\rho_1}|
\{r_{12}^{-1}\}_0
|\Psi_{\rho_1}\rangle}
{\langle\Psi_{\rho_1}|\Psi_{\rho_1}\rangle} 
\end{eqnarray}
where the operator is given by
\begin{eqnarray}\label{el2A} 
\{r_{12}^{-1}\}_0 
&=&
\frac{1}{2} 
\sum_{i\sigma j\sigma}
\sum_{k\sigma^\prime l\sigma^\prime}
(i\sigma j\sigma|k\sigma^\prime l\sigma^\prime)
a_{i\sigma}^\dagger 
a_{k\sigma^\prime}^\dagger a_{l\sigma^\prime}  
a_{j\sigma},
\end{eqnarray}
and the $0$ subscript appended to $\{r_{12}^{-1}\}_0$ indicates normal-ordering
with respect to the true vacuum state,~$|\;\rangle$. Using the Hartree--Fock
closed-shell reference-state $|\tilde{\varrho}_1\rangle$, instead, as the
vacuum-state, it is readily demonstrated that we have
\begin{eqnarray} \label{ident.r12} 
\langle \{r_{12}^{-1}\}_0 \rangle_{\!\tilde{\varrho}_1}
&=& 
\langle \{r_{12}^{-1}\}_{\!\mbox{\tiny $1\!,\!2$}} \rangle_{\!\tilde{\varrho}_1}+
E_{\mathrm{J}}[\tilde{\varrho}] + E_{\mathrm{x}}[\tilde{\varrho}_1] =
U_{\mathrm{co}}[\tilde{\varrho}_1] +
E_{\mathrm{J}}[\tilde{\varrho}] + E_{\mathrm{x}}[\tilde{\varrho}_1],
\end{eqnarray}
where we have use Eq.~(\ref{U.exact}); in addition,
$\{r_{12}^{-1}\}_{\!\mbox{\tiny $1\!,\!2$}}$ is given by Eq.~(\ref{nord.r12}),
where the suppressed superscript, $\gamma$ -- the vacuum state -- is set to
$\tilde{\varrho}_1$.

It is well known that the total electron-electron potential energy can also be
expressed using the (diagonal portion of) the two-particle, spinless density-matrix
from the target state \cite{McWeeny:60,Parr:89}:
\begin{eqnarray}\label{ee.en} 
\langle \{r_{12}^{-1}\}_0 \rangle_{\!\tilde{\varrho}_1}
&=&
\int \int r_{12}^{-1 }\Gamma_2^{\tilde{\varrho}_1}(\mathbf{r}_1,\mathbf{r}_2) \,
d\mathbf{r}_1 d\mathbf{r}_2,
\end{eqnarray}
and we can use the approximate expression, involving a two-body function $\varphi$,
for the two-particle spinless-density-matrix -- valid for closed-shell systems --
that was derived by Colle and Salvetti \cite{Colle:75}:
\begin{eqnarray} \label{CS.2dm} 
\Gamma_2^{\tilde{\varrho}_1}
(\mathbf{r}_1,\mathbf{r}_2) =
\tilde{\varrho}_2(\mathbf{r}_1,\mathbf{r}_2) 
\left(
1 + \varphi^2(\mathbf{r}_1,\mathbf{r}_2) - 2\varphi(\mathbf{r}_1,\mathbf{r}_2) 
\right),
\end{eqnarray}
and the two-particle density-matrix, from the Hartree--Fock reference-state, is
given by
\begin{eqnarray} \label{2bod.sf.hf} 
\tilde{\varrho}_2(\mathbf{r}_1,\mathbf{r}_2) = 
\frac12
\tilde{\varrho}(\mathbf{r}_1)\tilde{\varrho}(\mathbf{r}_2) - 
\frac14
\tilde{\varrho}_1(\mathbf{r}_1,\mathbf{r}_2) 
\tilde{\varrho}_1(\mathbf{r}_2,\mathbf{r}_1),
\end{eqnarray}
where we also have
\begin{eqnarray} \label{coul.ex} 
\int \int r_{12}^{-1 }\tilde{\varrho}_2(\mathbf{r}_1,\mathbf{r}_2) \,
d\mathbf{r}_1 d\mathbf{r}_2
&=&
E_{\mathrm{J}}[\tilde{\varrho}] + E_{\mathrm{x}}[\tilde{\varrho}_1].
\end{eqnarray}
Substituting Eq.~(\ref{CS.2dm}) into (\ref{ee.en}) and using (\ref{coul.ex}), we
have
\begin{eqnarray}\label{ee.enb} 
\langle \{r_{12}^{-1}\}_0 \rangle_{\tilde{\varrho}_1}
&=&
E_{\mathrm{J}}[\tilde{\varrho}] + E_{\mathrm{x}}[\tilde{\varrho}_1] +
E_{\mathrm{co}}^{\mathrm{cs}}[\tilde{\varrho}_1],
\end{eqnarray}
where $E_{\mathrm{co}}^{\mathrm{cs}}$ is the Colle--Salvetti correlation-energy
functional \cite{Colle:75,Lee:88}:
\begin{eqnarray}
E_{\mathrm{co}}^{\mathrm{cs}}[\tilde{\varrho}_1]=
\int \int r_{12}^{-1}
\tilde{\varrho}_2(\mathbf{r}_1,\mathbf{r}_2) 
\left(
\varphi^2(\mathbf{r}_1,\mathbf{r}_2) - 2\varphi(\mathbf{r}_1,\mathbf{r}_2)
\right),
\end{eqnarray}
and this functional, after a series of approximations, is developed into one that
does depends explicitly on $\tilde{\varrho}_1$ \cite{Lee:88}. (Note that
$\tilde{\varrho}_2$ is determined by $\tilde{\varrho}_1$, as indicated by
Eq.~(\ref{2bod.sf.hf}).)

Comparing this Eqs.~(\ref{ident.r12}) and (\ref{ee.enb}), gives the desired
result:
\begin{eqnarray}
E_{\mathrm{co}}^{\mathrm{cs}}[\tilde{\varrho}_1] =
U_{\mathrm{co}}[\tilde{\varrho}_1].
\end{eqnarray}
and from Eq.~(\ref{E.hf.app}), we have
\begin{eqnarray}
\label{E.hf.app2} 
{\cal E}_{\mbox{\tiny $N$}v}  
=E_1[\tilde{\varrho}_1,v] +
E_{\mathrm{co}}^{\mathrm{cs}}[\tilde{\varrho}_1],
\end{eqnarray}
in agreement with the Colle--Salvetti electronic energy expresion used in their
derivation of $E_{\mathrm{co}}^{\mathrm{cs}}$ \cite{Colle:75}, where
$E_1[\tilde{\varrho}_1,v]$ is the Hartree--Fock energy.

\section{Derivation of E\lowercase{q}.~(\ref{vcopsi})} 
\label{prooft} 

Using the occupied and unoccupied Brueckner orbitals, given by
Eqs.~(\ref{Bocc.orbs}) and (\ref{Bunocc.orbs}), respectively, we can express the
one-body operator $\left(\hat{v}_{\mathrm{co}}^{\scscs \tau}\right)_{\mathrm{ex}}$
in the following manner:
\begin{eqnarray} \label{exp.vco} 
\left(\hat{v}_{\mathrm{co}}^{\scscs\tau}\right)_{\mathrm{ex}} = 
\sum_{w\sigma}  \sum_{r\sigma} 
\langle r\sigma|
\left(\hat{v}_{\mathrm{co}}^{\scscs\tau}\right)_{\mathrm{ex}}|w\sigma\rangle
a_{r\sigma}^\dagger a_{w\sigma},
\end{eqnarray}
where, as in $( H\chi_\tau)_1$, the vanishing terms involving the matrix elements
that do not preserve the spin state, , i.e., $\langle r\sigma^\prime|
\left(\hat{v}_{\mathrm{co}}^{\scscs\tau}\right)_{\mathrm{ex}}|w\sigma\rangle$ for
$\sigma\ne\sigma^\prime$, are omitted.  The matrix elements in the above
expression can be computed using the kernel from the operator:
\begin{eqnarray} \label{matel.vco} 
\langle r\sigma|
\left(\hat{v}_{\mathrm{co}}^{\scscs\tau}\right)_{\mathrm{ex}}|w\sigma\rangle =
\int \int \psi_{r\sigma}^{\scscs \tau}(\mathbf{x})
v_{\mathrm{co}}^{\scscs\tau\mathrm{ex}}(\mathbf{x},\mathbf{x}^\prime)
\psi_{w\sigma}^{\scscs \tau}(\mathbf{x}^\prime)
\, d \mathbf{x} \, d \mathbf{x}^\prime;
\end{eqnarray}
so, if we have the an expression for the matrix element on the left-hand side,
that has the general form given by the integral on the right-hand side, we should
by able to obtain the kernel
$v_{\mathrm{co}}^{\scscs\tau\mathrm{ex}}(\mathbf{x},\mathbf{x}^\prime)$ and,
therefore,
$(\hat{v}_{\mathrm{co}}^{\scscs\tau})_{\mathrm{ex}}\psi_{w\sigma}^{\scscs \tau}$,
from the following definition:
\begin{eqnarray} \label{def.vco} 
\left(\hat{v}_{\mathrm{co}}^{\scscs\tau} (\mathbf{x})\right)_{\mathrm{ex}}
\psi_{w\sigma}^{\scscs \tau}(\mathbf{x})  =
\int v_{\mathrm{co}}^{\scscs\tau\mathrm{ex}}(\mathbf{x},\mathbf{x}^\prime)
\psi_{w\sigma}^{\scscs \tau}(\mathbf{x}^\prime)
\, d \mathbf{x}^\prime.
\end{eqnarray}
In order to obtain an expression for the term on the left side of
Eq.~(\ref{also.nat2}), below we obtain the diagrammatic expansion of the open
portion of $\left( H\chi_\tau\right)_1$, where this operator gives the matrix
elements $\langle r\sigma^\prime| \hat{v}_{\mathrm{co}}^{\scscs\tau} |
w\sigma\rangle$ using Eqs.~(\ref{vco.def}) and (\ref{matel.vco}). We then use this
matrix element to determine
$\hat{v}_{\mathrm{co}}^{\scscs\tau}\psi_{w\sigma}^{\scscs \tau}$ and obtain the
identity given by Eq.~(\ref{vcopsi}).

It is well known that the correlation operator $\chi_\tau$ is given by a
linked-diagram expansion, where all disconnected pieces are open
\cite{Lindgren:86,Lindgren:74,Lindgren:78}. Since $\chi_\tau$ does not contain a
one-body portion, it is easily demonstrated that the open portion of $\left(
H\chi_\tau \right)_1$ is connected -- all disconnected pieces from $\chi_\tau$ are
connected by the Hamiltonian $H$.

Using the diagrammatic formalism presented in Appendix \ref{DIAG.CE} and elsewhere
\cite{Finley:bdmt.arxiv}, consider the following example of a diagram that
contributes to the open portion of $\left( H\chi_\tau\right)_1$:
\begin{eqnarray} \label{d1} 
\setlength{\unitlength}{0.00066667in}
\begingroup\makeatletter\ifx\SetFigFont\undefined%
\gdef\SetFigFont#1#2#3#4#5{%
  \reset@font\fontsize{#1}{#2pt}%
  \fontfamily{#3}\fontseries{#4}\fontshape{#5}%
  \selectfont}%
\fi\endgroup%
{\renewcommand{\dashlinestretch}{30}
\begin{picture}(1445,1124)(0,-10)
\put(2183.000,537.000){\arc{3750.000}{2.8578}{3.4254}}
\put(-1417.000,537.000){\arc{3750.000}{5.9994}{6.5670}}
\path(1433,1062)(1133,12)
\path(833,1062)(1133,12)
\path(442,282)(448,382)
\blacken\path(472.556,290.365)(448.000,382.000)(412.664,293.958)(444.227,319.113)(472.556,290.365)
\path(943,673)(916,769)
\blacken\path(969.247,690.484)(916.000,769.000)(911.488,674.239)(933.057,708.353)(969.247,690.484)
\path(1024,391)(997,487)
\blacken\path(1050.247,408.484)(997.000,487.000)(992.488,392.239)(1014.057,426.353)(1050.247,408.484)
\path(1327,697)(1300,601)
\blacken\path(1295.488,695.761)(1300.000,601.000)(1353.247,679.516)(1317.057,661.647)(1295.488,695.761)
\path(881,900)(854,996)
\blacken\path(907.247,917.484)(854.000,996.000)(849.488,901.239)(871.057,935.353)(907.247,917.484)
\path(1088,162)(1061,258)
\blacken\path(1114.247,179.484)(1061.000,258.000)(1056.488,163.239)(1078.057,197.353)(1114.247,179.484)
\path(311,564)(311,464)
\blacken\path(281.000,554.000)(311.000,464.000)(341.000,554.000)(311.000,527.000)(281.000,554.000)
\path(446,744)(434,844)
\blacken\path(474.509,758.215)(434.000,844.000)(414.937,751.067)(441.506,781.449)(474.509,758.215)
\put(1208,837){\blacken\ellipse{150}{80}}
\put(1208,837){\ellipse{150}{80}}
\put(758,312){\blacken\ellipse{150}{80}}
\put(758,312){\ellipse{150}{80}}
\put(83,1062){\blacken\ellipse{150}{80}}
\put(83,1062){\ellipse{150}{80}}
\dottedline{60}(463,537)(988,537)
\dottedline{60}(383,12)(1133,12)
\dottedline{60}(893,837)(1133,837)
\dottedline{60}(1048,312)(833,312)
\dottedline{60}(183,1062)(383,1062)
\end{picture}
}
\raisebox{2.8ex}{$\displaystyle  
\;\;\; = \;\;\;$} 
\raisebox{2.8ex}{$\displaystyle
\hat{F}_{\tau \mbox{\tiny $1$}}
\kappa_\tau({\mathbf x}_1,{\mathbf x}_2)
g({\mathbf x}_2,{\mathbf x}_{1^\prime})
\psi_{w\sigma}^{\scscs \tau}({\mathbf x}_{1^\prime})
\psi_{r\sigma}^{\scscs \tau*}({\mathbf x}_1)
a_{r\sigma}^\dagger a_{w\sigma}
,$} 
\end{eqnarray}
where the repeated indices, $w\sigma$ and $r\sigma$, are summed over and where the
two body function is given by
\begin{eqnarray} \label{d1b} 
g({\mathbf x}_2,{\mathbf x}_{1^\prime})
=\frac{1}{16}
\varepsilon_\tau^{-4}
\tau({\mathbf x}_4,{\mathbf x}_5)
\hat{F}_{\tau \mbox{\tiny $5$}}
\kappa_\tau({\mathbf x}_5,{\mathbf x}_6)
\kappa_\tau({\mathbf x}_6,{\mathbf x}_4)
r_{62}^{-1}
r_{41^\prime}^{-1}
\kappa_\tau({\mathbf x}_2,{\mathbf x}_3)
\hat{F}_{\tau \mbox{\tiny $3$}}
\kappa_\tau({\mathbf x}_3,{\mathbf x}_1^\prime);
\end{eqnarray}
furthermore, it is understood that there are no integrations over ${\mathbf x}_2 $
and ${\mathbf x}_{1^\prime}$ on the right side of Eq.~(\ref{d1b}), since these
variables are not repeated indices according to the following convention: When
determining which dummy indices are repeated indices, indices appearing within
operators are not counted. So, for example, the indices $\mathbf{x}_2$ and
$\mathbf{x}_1^\prime$ appear only once in the above Eq, and not two times, since
the dummy indices from the Coulombic operator, i.e., $\mathbf{x}_2$ (and
$\mathbf{x}_6$) from $r_{62}^{-1}$ and $\mathbf{x}_1^\prime$ (and $\mathbf{x}_4$)
from $r_{41^\prime}^{-1}$, are not counted.

The orbitals $\psi_{w\sigma}^{\scscs \tau}$ and $\psi_{r\sigma}^{\scscs \tau}$,
presented in Eq.~(\ref{d1}), can be any Brueckner orbitals, as defined by
Eqs.~(\ref{Bocc.orbs}) and (\ref{Bunocc.orbs}). However, for convenience we choose
the canonical orbitals that are eigenfunctions of ${\cal \hat{F}}_{\tau}$, as
defined by Eq.~(\ref{EHF.eq2}). Using these orbitals the Brueckner one-particle
density-matrix, $\tau$, and the orthogonal function, $\kappa_\tau$, are given by
\begin{eqnarray} \label{onepart.brB} 
\tau(\mathbf{x},\mathbf{x}^\prime)
= \sum_{x\sigma\in\{\psi_o \leftarrow \tau, \hat{F}_{\tau}\}}
\psi_{x\sigma}^{\scscs \tau}(\mathbf{x}) 
\left(\psi_{x\sigma}^{\scscs \tau}(\mathbf{x}^\prime)\right)^*, \\
 \label{orthpart.brB} 
\kappa_\tau(\mathbf{x},\mathbf{x}^\prime)
= \sum_{r\sigma\in\{\psi_u \leftarrow \tau, \hat{F}_{\tau}\}}
\psi_{r\sigma}^{\scscs \tau}(\mathbf{x}) 
\left(\psi_{r\sigma}^{\scscs \tau}(\mathbf{x}^\prime)\right)^*,
\end{eqnarray}
where these functions are also given by Eqs.~(\ref{onepart.br}) and (\ref{orthpart.br}).

According to Eq.~(\ref{vco.def}), the diagram from Eq.~(\ref{d1}) also contributes
to $\left(\hat{v}_{\mathrm{co}}^{\scscs \tau}\right)_{\mathrm{ex}}$. Comparing
Eqs.~(\ref{exp.vco}) and (\ref{d1}) we see that the following term:
\begin{eqnarray*}
\hat{F}_{\tau \mbox{\tiny $1$}} 
\kappa_\tau({\mathbf x}_1,{\mathbf x}_2) 
g({\mathbf x}_2,{\mathbf x}_{1^\prime}) 
\psi_{w\sigma}^{\scscs \tau}({\mathbf x}_{1^\prime})
\psi_{r\sigma}^{\scscs \tau*}({\mathbf x}_1),
\end{eqnarray*}
contributes to the matrix element $\langle r|
\left(\hat{v}_{\mathrm{co}}^{\scscs\tau}\right)_{\mathrm{ex}}|w\rangle$;
furthermore, and diagrammatically speaking, removing the incoming and outgoing
free-lines from the operator given by Eq.~(\ref{d1}), yields
\begin{eqnarray} \label{d2} 
\raisebox{-4ex}{
\setlength{\unitlength}{0.00066667in}
\begingroup\makeatletter\ifx\SetFigFont\undefined%
\gdef\SetFigFont#1#2#3#4#5{%
  \reset@font\fontsize{#1}{#2pt}%
  \fontfamily{#3}\fontseries{#4}\fontshape{#5}%
  \selectfont}%
\fi\endgroup%
{\renewcommand{\dashlinestretch}{30}
\begin{picture}(1291,1137)(0,-10)
\put(2183.000,550.000){\arc{3750.000}{2.8578}{3.4254}}
\put(-1417.000,550.000){\arc{3750.000}{5.9994}{6.5670}}
\path(908,850)(1133,25)
\path(442,295)(448,395)
\blacken\path(472.556,303.365)(448.000,395.000)(412.664,306.958)(444.227,332.113)(472.556,303.365)
\path(311,571)(311,471)
\blacken\path(281.000,561.000)(311.000,471.000)(341.000,561.000)(311.000,534.000)(281.000,561.000)
\path(446,760)(434,860)
\blacken\path(474.509,774.215)(434.000,860.000)(414.937,767.067)(441.506,797.449)(474.509,774.215)
\path(965,647)(937,743)
\blacken\path(991.000,665.000)(937.000,743.000)(933.400,648.200)(954.640,682.520)(991.000,665.000)
\path(1035,398)(1006,493)
\blacken\path(1060.970,415.680)(1006.000,493.000)(1003.584,398.162)(1024.394,432.745)(1060.970,415.680)
\path(1096,163)(1068,259)
\blacken\path(1122.000,181.000)(1068.000,259.000)(1064.400,164.200)(1085.640,198.520)(1122.000,181.000)
\put(83,1075){\blacken\ellipse{150}{80}}
\put(83,1075){\ellipse{150}{80}}
\put(758,325){\blacken\ellipse{150}{80}}
\put(758,325){\ellipse{150}{80}}
\put(1208,850){\blacken\ellipse{150}{80}}
\put(1208,850){\ellipse{150}{80}}
\dottedline{60}(383,25)(1133,25)
\dottedline{60}(463,550)(988,550)
\dottedline{60}(183,1075)(383,1075)
\dottedline{60}(1048,325)(833,325)
\dottedline{60}(908,850)(1133,850)
\put(798,810){\makebox(0,0)[lb]{\smash{{{\SetFigFont{5}{6.0}{\rmdefault}{\mddefault}{\updefault}$1$}}}}}
\put(1176,0){\makebox(0,0)[lb]{\smash{{{\SetFigFont{5}{6.0}{\rmdefault}{\mddefault}{\updefault}$1^\prime$}}}}}
\end{picture}
}
}
\raisebox{0.3ex}{$\displaystyle  
\;\;\; = \;\;\;$} 
\raisebox{0.3ex}{$\displaystyle
\hat{F}_{\tau \mbox{\tiny $1$}}
\kappa_\tau({\mathbf x}_1,{\mathbf x}_2)
g({\mathbf x}_2,{\mathbf x}_{1^\prime}),$} 
\end{eqnarray}
and this diagram contributes to
$v_{\mathrm{co}}^{\scscs\tau\mathrm{ex}}(\mathbf{x}_1,\mathbf{x}_{1^\prime})$, the
kernel of $\left(\hat{v}_{\mathrm{co}}^{\scscs \tau}\right)_{\mathrm{ex}}$, as
defined by Eq.~(\ref{matel.vco}); furthermore, we use the following diagrammatic
representation for this two-body function:
\begin{eqnarray} \label{vco.diag} 
\raisebox{0.3ex}{$\displaystyle
v_{\mathrm{co}}^{\scscs\tau\mathrm{ex}}
({\mathbf x}_1,{\mathbf x}_{1^\prime})$} 
\raisebox{0.3ex}{$\displaystyle \;\;\; = \;\;\;$}
\raisebox{-1ex}{
\setlength{\unitlength}{0.00066667in}
\begingroup\makeatletter\ifx\SetFigFont\undefined%
\gdef\SetFigFont#1#2#3#4#5{%
  \reset@font\fontsize{#1}{#2pt}%
  \fontfamily{#3}\fontseries{#4}\fontshape{#5}%
  \selectfont}%
\fi\endgroup%
{\renewcommand{\dashlinestretch}{30}
\begin{picture}(1250,370)(0,-10)
\path(780,145)(705,220)(630,145)
	(555,220)(480,145)
\path(155,145)(255,145)
\blacken\path(165.000,115.000)(255.000,145.000)(165.000,175.000)(192.000,145.000)(165.000,115.000)
\path(980,145)(1080,145)
\blacken\path(990.000,115.000)(1080.000,145.000)(990.000,175.000)(1017.000,145.000)(990.000,115.000)
\path(30,145)(480,145)
\path(780,145)(1230,145)
\put(0,0){\makebox(0,0)[lb]
{\smash{{{\SetFigFont{5}{6.0}{\rmdefault}{\mddefault}{\updefault}$1^\prime$}}}}}
\put(1204,0){\makebox(0,0)[lb]
{\smash{{{\SetFigFont{5}{6.0}{\rmdefault}{\mddefault}{\updefault}$1$}}}}}
\put(494,295){\makebox(0,0)[lb]
{\smash{{{\SetFigFont{6}{7.2}{\rmdefault}{\mddefault}{\updefault}\mbox{\small $v_{\mathrm{co}}$}}}}}}
\end{picture}
}
}
\end{eqnarray}
where, in addition, the non-dummy indices $\mathbf{x}_1$ and
$\mathbf{x}_{1^\prime}$ in the above two diagrams, by our convention, correspond to
the vertices of the {\em omitted} outgoing and incoming lines, respectively. Note
that we have labeled these indices in the diagrams above; however, we will often
omit these labels in similar (kernel) diagrams below.

As a slight variation of the diagram given within Eq.~(\ref{d1}), consider the
following diagram:
\begin{eqnarray} \label{d1mod} 
\raisebox{-4ex}{
\setlength{\unitlength}{0.00066667in}
\begingroup\makeatletter\ifx\SetFigFont\undefined%
\gdef\SetFigFont#1#2#3#4#5{%
  \reset@font\fontsize{#1}{#2pt}%
  \fontfamily{#3}\fontseries{#4}\fontshape{#5}%
  \selectfont}%
\fi\endgroup%
{\renewcommand{\dashlinestretch}{30}
\begin{picture}(1445,1124)(0,-10)
\put(2183.000,537.000){\arc{3750.000}{2.8578}{3.4254}}
\put(-1417.000,537.000){\arc{3750.000}{5.9994}{6.5670}}
\path(833,1062)(1133,12)
\path(442,282)(448,382)
\blacken\path(472.556,290.365)(448.000,382.000)(412.664,293.958)(444.227,319.113)(472.556,290.365)
\path(943,673)(916,769)
\blacken\path(969.247,690.484)(916.000,769.000)(911.488,674.239)(933.057,708.353)(969.247,690.484)
\path(1024,391)(997,487)
\blacken\path(1050.247,408.484)(997.000,487.000)(992.488,392.239)(1014.057,426.353)(1050.247,408.484)
\path(881,900)(854,996)
\blacken\path(907.247,917.484)(854.000,996.000)(849.488,901.239)(871.057,935.353)(907.247,917.484)
\path(1088,162)(1061,258)
\blacken\path(1114.247,179.484)(1061.000,258.000)(1056.488,163.239)(1078.057,197.353)(1114.247,179.484)
\path(311,564)(311,464)
\blacken\path(281.000,554.000)(311.000,464.000)(341.000,554.000)(311.000,527.000)(281.000,554.000)
\path(446,744)(434,844)
\blacken\path(474.509,758.215)(434.000,844.000)(414.937,751.067)(441.506,781.449)(474.509,758.215)
\path(1340,719)(1303,599)
\whiten\path(1289.860,760.020)(1303.000,599.000)(1404.533,724.662)(1333.938,699.339)(1289.860,760.020)
\path(1433,1062)(1133,12)
\put(1208,837){\blacken\ellipse{150}{80}}
\put(1208,837){\ellipse{150}{80}}
\put(758,312){\blacken\ellipse{150}{80}}
\put(758,312){\ellipse{150}{80}}
\put(83,1062){\blacken\ellipse{150}{80}}
\put(83,1062){\ellipse{150}{80}}
\dottedline{60}(463,537)(988,537)
\dottedline{60}(383,12)(1133,12)
\dottedline{60}(893,837)(1133,837)
\dottedline{60}(1048,312)(833,312)
\dottedline{60}(183,1062)(383,1062)
\end{picture}
}
}
\raisebox{0.3ex}{$\displaystyle  
\;\;\; = \;\;\;$} 
\raisebox{0.3ex}{$\displaystyle
\frac{1}{2}
\hat{F}_{\tau \mbox{\tiny $1$}}
\kappa_\tau({\mathbf x}_1,{\mathbf x}_2)
g({\mathbf x}_2,{\mathbf x}_{1^\prime})
\psi_{w}^{\scscs \tau}({\mathbf x}_{1^\prime})
\psi_{r}^{\scscs \tau*}({\mathbf x}_1)
a_{r}^\dagger a_{w},$} 
\end{eqnarray}
where the additional factor of $\frac{1}{2}$ comes from the diagonal term, given
by Eq.~(\ref{H01o.diag}). As in the diagram within Eq.~(\ref{d1}), this diagram
contributes to $\left(\hat{v}_{\mathrm{co}}^{\scscs \tau}\right)_{\mathrm{ex}}$;
the corresponding diagram that contributes to the kernel
$v_{\mathrm{co}}^{\scscs\tau\mathrm{ex}}(\mathbf{x}_1,\mathbf{x}_{1^\prime})$ can
be expressed in two alternative forms:
\begin{eqnarray} \label{d2mod} 
\raisebox{-4ex}{
\setlength{\unitlength}{0.00066667in}
\begingroup\makeatletter\ifx\SetFigFont\undefined%
\gdef\SetFigFont#1#2#3#4#5{%
  \reset@font\fontsize{#1}{#2pt}%
  \fontfamily{#3}\fontseries{#4}\fontshape{#5}%
  \selectfont}%
\fi\endgroup%
{\renewcommand{\dashlinestretch}{30}
\begin{picture}(1445,1137)(0,-10)
\put(2183.000,550.000){\arc{3750.000}{2.8578}{3.4254}}
\put(-1417.000,550.000){\arc{3750.000}{5.9994}{6.5670}}
\path(898,848)(1133,25)
\dottedline{90}(833,1075)(900,853)
\path(442,295)(448,395)
\blacken\path(472.556,303.365)(448.000,395.000)(412.664,306.958)(444.227,332.113)(472.556,303.365)
\path(311,571)(311,471)
\blacken\path(281.000,561.000)(311.000,471.000)(341.000,561.000)(311.000,534.000)(281.000,561.000)
\path(446,760)(434,860)
\blacken\path(474.509,774.215)(434.000,860.000)(414.937,767.067)(441.506,797.449)(474.509,774.215)
\path(943,686)(916,782)
\blacken\path(969.247,703.484)(916.000,782.000)(911.488,687.239)(933.057,721.353)(969.247,703.484)
\path(1024,404)(997,500)
\blacken\path(1050.247,421.484)(997.000,500.000)(992.488,405.239)(1014.057,439.353)(1050.247,421.484)
\path(1088,175)(1061,271)
\blacken\path(1114.247,192.484)(1061.000,271.000)(1056.488,176.239)(1078.057,210.353)(1114.247,192.484)
\path(1340,732)(1303,612)
\whiten\path(1289.860,773.020)(1303.000,612.000)(1404.533,737.662)(1333.938,712.339)(1289.860,773.020)
\dottedline{90}(1433,1075)(1133,25)
\put(83,1075){\blacken\ellipse{150}{80}}
\put(83,1075){\ellipse{150}{80}}
\put(758,325){\blacken\ellipse{150}{80}}
\put(758,325){\ellipse{150}{80}}
\put(1208,850){\blacken\ellipse{150}{80}}
\put(1208,850){\ellipse{150}{80}}
\dottedline{60}(383,25)(1133,25)
\dottedline{60}(183,1075)(383,1075)
\dottedline{60}(1048,325)(833,325)
\dottedline{60}(898,850)(1133,850)
\dottedline{60}(463,550)(988,550)
\put(798,810){\makebox(0,0)[lb]
{\smash{{{\SetFigFont{5}{6.0}{\rmdefault}{\mddefault}{\updefault}$1$}}}}}
\put(1176,0){\makebox(0,0)[lb]
{\smash{{{\SetFigFont{5}{6.0}{\rmdefault}{\mddefault}{\updefault}$1^\prime$}}}}}
\end{picture}
}
}
\raisebox{0.3ex}{$\displaystyle  
\;\;\; = \;\;\;$} 
\raisebox{-4ex}{
\setlength{\unitlength}{0.00066667in}
\begingroup\makeatletter\ifx\SetFigFont\undefined%
\gdef\SetFigFont#1#2#3#4#5{%
  \reset@font\fontsize{#1}{#2pt}%
  \fontfamily{#3}\fontseries{#4}\fontshape{#5}%
  \selectfont}%
\fi\endgroup%
{\renewcommand{\dashlinestretch}{30}
\begin{picture}(1431,1137)(0,-10)
\put(2183.000,550.000){\arc{3750.000}{2.8578}{3.4254}}
\put(-1417.000,550.000){\arc{3750.000}{5.9994}{6.5670}}
\path(898,848)(1133,25)
\path(442,295)(448,395)
\blacken\path(472.556,303.365)(448.000,395.000)(412.664,306.958)(444.227,332.113)(472.556,303.365)
\path(311,571)(311,471)
\blacken\path(281.000,561.000)(311.000,471.000)(341.000,561.000)(311.000,534.000)(281.000,561.000)
\path(446,760)(434,860)
\blacken\path(474.509,774.215)(434.000,860.000)(414.937,767.067)(441.506,797.449)(474.509,774.215)
\path(943,686)(916,782)
\blacken\path(969.247,703.484)(916.000,782.000)(911.488,687.239)(933.057,721.353)(969.247,703.484)
\path(1024,404)(997,500)
\blacken\path(1050.247,421.484)(997.000,500.000)(992.488,405.239)(1014.057,439.353)(1050.247,421.484)
\path(1088,175)(1061,271)
\blacken\path(1114.247,192.484)(1061.000,271.000)(1056.488,176.239)(1078.057,210.353)(1114.247,192.484)
\path(1340,732)(1303,612)
\whiten\path(1289.860,773.020)(1303.000,612.000)(1404.533,737.662)(1333.938,712.339)(1289.860,773.020)
\put(83,1075){\blacken\ellipse{150}{80}}
\put(83,1075){\ellipse{150}{80}}
\put(758,325){\blacken\ellipse{150}{80}}
\put(758,325){\ellipse{150}{80}}
\put(1208,850){\blacken\ellipse{150}{80}}
\put(1208,850){\ellipse{150}{80}}
\dottedline{60}(383,25)(1133,25)
\dottedline{60}(183,1075)(383,1075)
\dottedline{60}(1048,325)(833,325)
\dottedline{60}(898,850)(1133,850)
\dottedline{60}(463,550)(988,550)
\put(798,810){\makebox(0,0)[lb]
{\smash{{{\SetFigFont{5}{6.0}{\rmdefault}{\mddefault}{\updefault}$1$}}}}}
\put(1176,0){\makebox(0,0)[lb]
{\smash{{{\SetFigFont{5}{6.0}{\rmdefault}{\mddefault}{\updefault}$1^\prime$}}}}}
\end{picture}
}
}
\raisebox{0.3ex}{$\displaystyle  
\;\;\; = \;\;\;$} 
\raisebox{0.3ex}{$\displaystyle
\frac{1}{2}
\hat{F}_{\tau \mbox{\tiny $1$}}
\kappa_\tau({\mathbf x}_1,{\mathbf x}_2)
g({\mathbf x}_2,{\mathbf x}_{1^\prime}),$}
\end{eqnarray}
where the first diagram on the left side replaces the incoming and outgoing
omitted-lines with dotted lines; this form gives a visual aid in determining the
excitations involved and a psudo hole-line for the diagonal term to reside on.

The diagrams appearing in Eqs.~(\ref{d2}) and (\ref{d2mod}) are examples of one-body
kernel-diagrams where the omitted outgoing free-line is attached at a
$\hat{F}_{\tau \mbox{\tiny $1$}}$ vertex and the omitted incoming-line is attached
at the $\mathbf{x}_{1^\prime}$ vertex of a $r_{j1^\prime}^{-1}$ operator. (In this
particular case we have ($j=4$), according to Eq.~(\ref{d1b})).  Furthermore, note
that the Fock operator is acting upon excited orbitals, giving the $\hat{F}_{\tau
\mbox{\tiny $1$}} \kappa_\tau({\mathbf x}_1,{\mathbf x}_2)$ term.  Summing over
all diagrams of this type, we have
\begin{eqnarray} \label{kern.sum} 
\raisebox{0.3ex}{$\displaystyle
\hat{F}_{\tau \mbox{\tiny $1$}}
\kappa_\tau({\mathbf x}_1,{\mathbf x}_2)
G({\mathbf x}_2,{\mathbf x}_{1^\prime})
$} 
\hspace{65ex}\\ \nonumber \hspace{8ex}
\raisebox{0.3ex}{$\displaystyle  
\;\;\; = \;\;\;$} 
\raisebox{-4ex}{
\setlength{\unitlength}{0.00066667in}
\begingroup\makeatletter\ifx\SetFigFont\undefined%
\gdef\SetFigFont#1#2#3#4#5{%
  \reset@font\fontsize{#1}{#2pt}%
  \fontfamily{#3}\fontseries{#4}\fontshape{#5}%
  \selectfont}%
\fi\endgroup%
{\renewcommand{\dashlinestretch}{30}
\begin{picture}(1397,1124)(0,-10)
\put(1914.000,537.000){\arc{3750.000}{2.8578}{3.4254}}
\put(-1686.000,537.000){\arc{3750.000}{5.9994}{6.5670}}
\path(1014,462)(864,12)
\path(42,564)(42,464)
\blacken\path(12.000,554.000)(42.000,464.000)(72.000,554.000)(42.000,527.000)(12.000,554.000)
\path(189,499)(189,599)
\blacken\path(219.000,509.000)(189.000,599.000)(159.000,509.000)(189.000,536.000)(219.000,509.000)
\path(937,222)(970,316)
\blacken\path(968.494,221.144)(970.000,316.000)(911.882,241.018)(949.132,256.557)(968.494,221.144)
\put(1314,462){\blacken\ellipse{150}{80}}
\put(1314,462){\ellipse{150}{80}}
\put(414,1062){\blacken\ellipse{150}{80}}
\put(414,1062){\ellipse{150}{80}}
\dottedline{60}(114,12)(864,12)
\dottedline{60}(1014,462)(1264,462)
\dottedline{60}(114,1062)(364,1062)
\end{picture}
}
}
\raisebox{0.3ex}{$\displaystyle +$}
\raisebox{-4ex}{
\setlength{\unitlength}{0.00066667in}
\begingroup\makeatletter\ifx\SetFigFont\undefined%
\gdef\SetFigFont#1#2#3#4#5{%
  \reset@font\fontsize{#1}{#2pt}%
  \fontfamily{#3}\fontseries{#4}\fontshape{#5}%
  \selectfont}%
\fi\endgroup%
{\renewcommand{\dashlinestretch}{30}
\begin{picture}(1325,1124)(0,-10)
\path(942,462)(792,12)
\path(42,12)(42,1062)
\path(42,1062)(792,12)
\path(865,222)(898,316)
\blacken\path(896.494,221.144)(898.000,316.000)(839.882,241.018)(877.132,256.557)(896.494,221.144)
\path(42,420)(42,516)
\blacken\path(72.000,426.000)(42.000,516.000)(12.000,426.000)(42.000,453.000)(72.000,426.000)
\path(339,651)(397,561)
\blacken\path(323.030,620.400)(397.000,561.000)(373.464,652.902)(362.873,613.956)(323.030,620.400)
\put(1242,462){\blacken\ellipse{150}{80}}
\put(1242,462){\ellipse{150}{80}}
\put(342,1062){\blacken\ellipse{150}{80}}
\put(342,1062){\ellipse{150}{80}}
\dottedline{60}(42,12)(792,12)
\dottedline{60}(942,462)(1167,462)
\dottedline{60}(42,1062)(267,1062)
\end{picture}
}
}
\raisebox{0.3ex}{$\displaystyle +$}
\raisebox{-4ex}{
\setlength{\unitlength}{0.00066667in}
\begingroup\makeatletter\ifx\SetFigFont\undefined%
\gdef\SetFigFont#1#2#3#4#5{%
  \reset@font\fontsize{#1}{#2pt}%
  \fontfamily{#3}\fontseries{#4}\fontshape{#5}%
  \selectfont}%
\fi\endgroup%
{\renewcommand{\dashlinestretch}{30}
\begin{picture}(1397,1089)(0,-10)
\put(1914.000,537.000){\arc{3750.000}{2.8578}{3.4254}}
\put(-1686.000,537.000){\arc{3750.000}{5.9994}{6.5670}}
\path(564,1062)(864,12)
\path(1014,462)(864,12)
\path(42,564)(42,464)
\blacken\path(12.000,554.000)(42.000,464.000)(72.000,554.000)(42.000,527.000)(12.000,554.000)
\path(189,499)(189,599)
\blacken\path(219.000,509.000)(189.000,599.000)(159.000,509.000)(189.000,536.000)(219.000,509.000)
\path(703,571)(735,473)
\blacken\path(678.546,549.242)(735.000,473.000)(735.582,567.867)(715.445,532.888)(678.546,549.242)
\path(937,222)(970,316)
\blacken\path(968.494,221.144)(970.000,316.000)(911.882,241.018)(949.132,256.557)(968.494,221.144)
\put(1314,462){\blacken\ellipse{150}{80}}
\put(1314,462){\ellipse{150}{80}}
\dottedline{60}(114,12)(864,12)
\dottedline{60}(114,1062)(564,1062)
\dottedline{60}(1014,462)(1239,462)
\end{picture}
}
}
\raisebox{0.3ex}{$\displaystyle +$}
\raisebox{-4ex}{
\setlength{\unitlength}{0.00066667in}
\begingroup\makeatletter\ifx\SetFigFont\undefined%
\gdef\SetFigFont#1#2#3#4#5{%
  \reset@font\fontsize{#1}{#2pt}%
  \fontfamily{#3}\fontseries{#4}\fontshape{#5}%
  \selectfont}%
\fi\endgroup%
{\renewcommand{\dashlinestretch}{30}
\begin{picture}(1325,1089)(0,-10)
\path(492,1062)(792,12)
\path(942,462)(792,12)
\path(42,1062)(42,12)
\path(42,12)(492,1062)
\path(631,571)(663,473)
\blacken\path(606.546,549.242)(663.000,473.000)(663.582,567.867)(643.445,532.888)(606.546,549.242)
\path(865,222)(898,316)
\blacken\path(896.494,221.144)(898.000,316.000)(839.882,241.018)(877.132,256.557)(896.494,221.144)
\path(42,564)(42,464)
\blacken\path(12.000,554.000)(42.000,464.000)(72.000,554.000)(42.000,527.000)(12.000,554.000)
\path(255,494)(291,590)
\blacken\path(287.489,495.197)(291.000,590.000)(231.309,516.264)(268.879,531.011)(287.489,495.197)
\put(1242,462){\blacken\ellipse{150}{80}}
\put(1242,462){\ellipse{150}{80}}
\dottedline{60}(42,12)(792,12)
\dottedline{60}(42,1062)(492,1062)
\dottedline{60}(942,462)(1167,462)
\end{picture}
}
}
\raisebox{0.3ex}{$\displaystyle + \cdots$}
\raisebox{0.65ex}{,}
\end{eqnarray}
where the two-body function $G({\mathbf x}_2,{\mathbf x}_{1^\prime})$ is obtained
from the infinite-order expansion, and, for brevity, we have only displayed the
first four diagrams of the series. Introducing a diagrammatic symbol for
$G({\mathbf x}_2,{\mathbf x}_{1^\prime})$, we can represent the above expansion in
the following manner:
\begin{eqnarray} \label{kern.sumb} 
\raisebox{0.3ex}{$\displaystyle
\hat{F}_{\tau \mbox{\tiny $1$}}
\kappa_\tau({\mathbf x}_1,{\mathbf x}_2)
G({\mathbf x}_2,{\mathbf x}_{1^\prime})$} 
\raisebox{0.3ex}{$\displaystyle \;\;\; = \;\;\;$}
\raisebox{-2.5ex}{
\setlength{\unitlength}{0.00066667in}
\begingroup\makeatletter\ifx\SetFigFont\undefined%
\gdef\SetFigFont#1#2#3#4#5{%
  \reset@font\fontsize{#1}{#2pt}%
  \fontfamily{#3}\fontseries{#4}\fontshape{#5}%
  \selectfont}%
\fi\endgroup%
{\renewcommand{\dashlinestretch}{30}
\begin{picture}(1313,657)(0,-10)
\path(930,595)(780,145)
\path(780,145)(705,220)(630,145)
	(555,220)(480,145)
\path(853,355)(886,449)
\blacken\path(884.494,354.144)(886.000,449.000)(827.882,374.018)(865.132,389.557)(884.494,354.144)
\path(155,145)(255,145)
\blacken\path(165.000,115.000)(255.000,145.000)(165.000,175.000)(192.000,145.000)(165.000,115.000)
\put(1230,595){\blacken\ellipse{150}{80}}
\put(1230,595){\ellipse{150}{80}}
\dottedline{60}(930,595)(1180,595)
\dottedline{60}(30,145)(480,145)
\put(823,555){\makebox(0,0)[lb]
{\smash{{{\SetFigFont{5}{6.0}{\rmdefault}{\mddefault}{\updefault}$1$}}}}}
\put(755,0){\makebox(0,0)[lb]
{\smash{{{\SetFigFont{5}{6.0}{\rmdefault}{\mddefault}{\updefault}$2$}}}}}
\put(405,230){\makebox(0,0)[lb]
{\smash{{{\SetFigFont{6}{7.2}{\rmdefault}{\mddefault}{\updefault}G}}}}}
\put(0,0){\makebox(0,0)[lb]
{\smash{{{\SetFigFont{5}{6.0}{\rmdefault}{\mddefault}{\updefault}$1^\prime$}}}}}
\end{picture}
}
}
\end{eqnarray}

As a slight variation of the diagrams from the series appearing in
Eq.~(\ref{kern.sum}), we also have diagrams that have the Fock operator acting
upon occupied orbitals, e.g.,
\begin{eqnarray} \label{d2c} 
\raisebox{-4ex}{
\setlength{\unitlength}{0.00066667in}
\begingroup\makeatletter\ifx\SetFigFont\undefined%
\gdef\SetFigFont#1#2#3#4#5{%
  \reset@font\fontsize{#1}{#2pt}%
  \fontfamily{#3}\fontseries{#4}\fontshape{#5}%
  \selectfont}%
\fi\endgroup%
{\renewcommand{\dashlinestretch}{30}
\begin{picture}(1445,1124)(0,-10)
\put(-1417.000,537.000){\arc{3750.000}{5.9994}{6.5670}}
\path(983,612)(1208,837)
\path(1433,12)(1208,837)
\path(456,473)(459,573)
\blacken\path(486.288,482.141)(459.000,573.000)(426.315,483.940)(457.111,510.028)(486.288,482.141)
\path(423,885)(400,985)
\blacken\path(449.410,904.014)(400.000,985.000)(390.937,890.566)(414.121,923.603)(449.410,904.014)
\path(1372,230)(1344,327)
\blacken\path(1397.783,248.851)(1344.000,327.000)(1340.137,232.210)(1361.472,266.471)(1397.783,248.851)
\path(1270,612)(1242,709)
\blacken\path(1295.783,630.851)(1242.000,709.000)(1238.137,614.210)(1259.472,648.471)(1295.783,630.851)
\path(1110,741)(1045,670)
\blacken\path(1083.645,756.640)(1045.000,670.000)(1127.900,716.125)(1087.541,716.468)(1083.645,756.640)
\path(324,294)(345,169)
\whiten\path(260.977,306.986)(345.000,169.000)(379.319,326.868)(327.604,272.549)(260.977,306.986)
\put(2183.000,537.000){\arc{3750.000}{2.8578}{3.4254}}
\put(83,1062){\blacken\ellipse{150}{80}}
\put(83,1062){\ellipse{150}{80}}
\put(683,612){\blacken\ellipse{150}{80}}
\put(683,612){\ellipse{150}{80}}
\put(1012,462){\blacken\ellipse{150}{80}}
\put(1012,462){\ellipse{150}{80}}
\dottedline{60}(383,12)(1433,12)
\dottedline{60}(438,837)(1208,837)
\dottedline{60}(183,1062)(383,1062)
\dottedline{60}(973,612)(758,612)
\dottedline{60}(1302,462)(1087,462)
\end{picture}
}
}
\raisebox{0.3ex}{$\displaystyle  
\;\;\; = \;\;\;$} 
\raisebox{0.3ex}{$\displaystyle
\frac{1}{3}\hat{F}_{\tau \mbox{\tiny $1$}}
\tau({\mathbf x}_1,{\mathbf x}_2)
g({\mathbf x}_2,{\mathbf x}_{1^\prime})
,$} 
\end{eqnarray}
where $g({\mathbf x}_2,{\mathbf x}_{1^\prime})$ is given by
Eq.~(\ref{d1b}). Summing over all diagrams of this type, as in Eqs.~(\ref{kern.sum}) and
(\ref{kern.sumb}), we have
\begin{eqnarray} \label{kern.sumc} 
\raisebox{0.3ex}{$\displaystyle
\hat{F}_{\tau \mbox{\tiny $1$}}
\tau({\mathbf x}_1,{\mathbf x}_2)
I({\mathbf x}_2,{\mathbf x}_{1^\prime})$} 
\raisebox{0.3ex}{$\displaystyle \;\;\; = \;\;\;$}
\raisebox{-2.5ex}{
\setlength{\unitlength}{0.00066667in}
\begingroup\makeatletter\ifx\SetFigFont\undefined%
\gdef\SetFigFont#1#2#3#4#5{%
  \reset@font\fontsize{#1}{#2pt}%
  \fontfamily{#3}\fontseries{#4}\fontshape{#5}%
  \selectfont}%
\fi\endgroup%
{\renewcommand{\dashlinestretch}{30}
\begin{picture}(1313,674)(0,-10)
\path(930,49)(780,499)
\path(780,499)(705,574)(630,499)
	(555,574)(480,499)
\path(155,499)(255,499)
\blacken\path(165.000,469.000)(255.000,499.000)(165.000,529.000)(192.000,499.000)(165.000,469.000)
\path(853,274)(886,180)
\blacken\path(827.882,254.982)(886.000,180.000)(884.494,274.856)(865.132,239.443)(827.882,254.982)
\put(1230,49){\blacken\ellipse{150}{80}}
\put(1230,49){\ellipse{150}{80}}
\dottedline{60}(30,499)(480,499)
\dottedline{60}(930,49)(1180,49)
\put(405,584){\makebox(0,0)[lb]
{\smash{{{\SetFigFont{6}{7.2}{\rmdefault}{\mddefault}{\updefault}I}}}}}
\put(823,0){\makebox(0,0)[lb]
{\smash{{{\SetFigFont{5}{6.0}{\rmdefault}{\mddefault}{\updefault}$1$}}}}}
\put(807,534){\makebox(0,0)[lb]
{\smash{{{\SetFigFont{5}{6.0}{\rmdefault}{\mddefault}{\updefault}$2$}}}}}
\put(0,354){\makebox(0,0)[lb]
{\smash{{{\SetFigFont{5}{6.0}{\rmdefault}{\mddefault}{\updefault}$1^\prime$}}}}}
\end{picture}
}
}
\end{eqnarray}

The diagram sums represented by Eqs.~(\ref{kern.sumb}) and (\ref{kern.sumc}) include all
diagram where the omitted outgoing free-line is attached at a $\hat{F}_{\tau
\mbox{\tiny $1$}}$ vertex and the omitted incoming-line is attached at the
$\mathbf{x}_{1^\prime}$ vertex of a $r_{j1^\prime}^{-1}$ operator. Two examples
where both incoming and outgoing omitted-lines are connected to Fock operators are
given by the following two diagrams:
\begin{eqnarray} \label{d5} 
\raisebox{-4ex}{
\setlength{\unitlength}{0.00066667in}
\begingroup\makeatletter\ifx\SetFigFont\undefined%
\gdef\SetFigFont#1#2#3#4#5{%
  \reset@font\fontsize{#1}{#2pt}%
  \fontfamily{#3}\fontseries{#4}\fontshape{#5}%
  \selectfont}%
\fi\endgroup%
{\renewcommand{\dashlinestretch}{30}
\begin{picture}(1966,1124)(0,-10)
\path(1433,12)(1208,837)
\put(-1417.000,537.000){\arc{3750.000}{5.9994}{6.5670}}
\path(983,612)(1208,837)
\path(1583,312)(1433,12)
\path(1306,464)(1335,369)
\blacken\path(1280.030,446.320)(1335.000,369.000)(1337.416,463.838)(1316.606,429.255)(1280.030,446.320)
\path(456,473)(459,573)
\blacken\path(486.288,482.141)(459.000,573.000)(426.315,483.940)(457.111,510.028)(486.288,482.141)
\path(1071,700)(1136,771)
\blacken\path(1097.355,684.360)(1136.000,771.000)(1053.100,724.875)(1093.459,724.532)(1097.355,684.360)
\path(1498,144)(1542,233)
\blacken\path(1529.007,139.026)(1542.000,233.000)(1475.221,165.616)(1514.080,176.525)(1529.007,139.026)
\path(307,486)(318,361)
\whiten\path(245.082,505.163)(318.000,361.000)(364.620,515.682)(308.796,465.596)(245.082,505.163)
\path(340,210)(363,85)
\whiten\path(276.846,221.666)(363.000,85.000)(394.865,243.381)(343.999,188.266)(276.846,221.666)
\path(423,885)(400,985)
\blacken\path(449.410,904.014)(400.000,985.000)(390.937,890.566)(414.121,923.603)(449.410,904.014)
\put(2183.000,537.000){\arc{3750.000}{2.8578}{3.4254}}
\put(83,1062){\blacken\ellipse{150}{80}}
\put(83,1062){\ellipse{150}{80}}
\put(683,612){\blacken\ellipse{150}{80}}
\put(683,612){\ellipse{150}{80}}
\put(1883,312){\blacken\ellipse{150}{80}}
\put(1883,312){\ellipse{150}{80}}
\dottedline{60}(383,12)(1433,12)
\dottedline{60}(438,837)(1208,837)
\dottedline{60}(183,1062)(383,1062)
\dottedline{60}(973,612)(758,612)
\dottedline{45}(1583,312)(1808,312)
\end{picture}
}
}
\raisebox{0.3ex}{$\displaystyle  
\;\;\; = \;\;\;$} 
\raisebox{0.3ex}{$\displaystyle
\hat{F}_{\tau \mbox{\tiny $1$}}
\kappa_\tau({\mathbf x}_1,{\mathbf x}_2)
l({\mathbf x}_2,{\mathbf x}_{1^\prime})
\hat{F}_{\tau \mbox{\tiny $1^\prime$}},$} 
\end{eqnarray}
\begin{eqnarray} \label{d5b} 
\raisebox{-4ex}{
\setlength{\unitlength}{0.00066667in}
\begingroup\makeatletter\ifx\SetFigFont\undefined%
\gdef\SetFigFont#1#2#3#4#5{%
  \reset@font\fontsize{#1}{#2pt}%
  \fontfamily{#3}\fontseries{#4}\fontshape{#5}%
  \selectfont}%
\fi\endgroup%
{\renewcommand{\dashlinestretch}{30}
\begin{picture}(1966,1124)(0,-10)
\put(-1417.000,537.000){\arc{3750.000}{5.9994}{6.5670}}
\path(983,612)(1208,837)
\path(1583,312)(1433,12)
\path(1433,12)(1208,837)
\path(456,473)(459,573)
\blacken\path(486.288,482.141)(459.000,573.000)(426.315,483.940)(457.111,510.028)(486.288,482.141)
\path(307,486)(318,361)
\whiten\path(245.082,505.163)(318.000,361.000)(364.620,515.682)(308.796,465.596)(245.082,505.163)
\path(340,210)(363,85)
\whiten\path(276.846,221.666)(363.000,85.000)(394.865,243.381)(343.999,188.266)(276.846,221.666)
\path(423,885)(400,985)
\blacken\path(449.410,904.014)(400.000,985.000)(390.937,890.566)(414.121,923.603)(449.410,904.014)
\path(1101,731)(1036,660)
\blacken\path(1074.645,746.640)(1036.000,660.000)(1118.900,706.125)(1078.541,706.468)(1074.645,746.640)
\path(1309,483)(1280,578)
\blacken\path(1334.970,500.680)(1280.000,578.000)(1277.584,483.162)(1298.394,517.745)(1334.970,500.680)
\path(1537,213)(1493,124)
\blacken\path(1505.993,217.974)(1493.000,124.000)(1559.779,191.384)(1520.920,180.475)(1505.993,217.974)
\put(2183.000,537.000){\arc{3750.000}{2.8578}{3.4254}}
\put(83,1062){\blacken\ellipse{150}{80}}
\put(83,1062){\ellipse{150}{80}}
\put(683,612){\blacken\ellipse{150}{80}}
\put(683,612){\ellipse{150}{80}}
\put(1883,312){\blacken\ellipse{150}{80}}
\put(1883,312){\ellipse{150}{80}}
\dottedline{60}(383,12)(1433,12)
\dottedline{60}(183,1062)(383,1062)
\dottedline{60}(973,612)(758,612)
\dottedline{45}(1583,312)(1808,312)
\dottedline{60}(438,837)(1208,837)
\end{picture}
}
}
\raisebox{0.3ex}{$\displaystyle  
\;\;\; = \;\;\;$} 
\raisebox{0.3ex}{$\displaystyle
\hat{F}_{\tau \mbox{\tiny $1$}}
\tau({\mathbf x}_1,{\mathbf x}_2)
m({\mathbf x}_2,{\mathbf x}_{1^\prime})
\hat{F}_{\tau \mbox{\tiny $1^\prime$}},$} 
\end{eqnarray}
where, in these examples, we have
\begin{eqnarray} \label{d5c} 
l({\mathbf x}_2,{\mathbf x}_{1^\prime})&=&
\frac{1}{96} \varepsilon_\tau^{-4}
\tau({\mathbf x}_4,{\mathbf x}_5)
\hat{F}_{\tau \mbox{\tiny $5$}}
\kappa_\tau({\mathbf x}_5,{\mathbf x}_6)
\kappa_\tau({\mathbf x}_6,{\mathbf x}_4)
r_{42}^{-1} r_{63}^{-1}
\tau({\mathbf x}_2,{\mathbf x}_3)
\kappa_\tau({\mathbf x}_3,{\mathbf x}_{1^\prime}),
\\
\label{d5d} 
m({\mathbf x}_2,{\mathbf x}_{1^\prime})&=&
\frac{1}{96} \varepsilon_\tau^{-4}
\tau({\mathbf x}_4,{\mathbf x}_5)
\hat{F}_{\tau \mbox{\tiny $5$}}
\kappa_\tau({\mathbf x}_5,{\mathbf x}_6)
\kappa_\tau({\mathbf x}_6,{\mathbf x}_4)
r_{43}^{-1} r_{62}^{-1}
\kappa_\tau({\mathbf x}_2,{\mathbf x}_3)
\tau({\mathbf x}_3,{\mathbf x}_{1^\prime}),
\end{eqnarray}
and again, we can sum over all diagrams of these types:
\begin{eqnarray} \label{kern.sumd} 
\raisebox{0.3ex}{$\displaystyle
\hat{F}_{\tau \mbox{\tiny $1$}}
\kappa_\tau({\mathbf x}_1,{\mathbf x}_2)
L({\mathbf x}_2,{\mathbf x}_{1^\prime})
\hat{F}_{\tau \mbox{\tiny $1^\prime$}}
$} 
\raisebox{0.3ex}{$\displaystyle \;\;\; = \;\;\;$}
\raisebox{-2.5ex}{
\setlength{\unitlength}{0.00066667in}
\begingroup\makeatletter\ifx\SetFigFont\undefined%
\gdef\SetFigFont#1#2#3#4#5{%
  \reset@font\fontsize{#1}{#2pt}%
  \fontfamily{#3}\fontseries{#4}\fontshape{#5}%
  \selectfont}%
\fi\endgroup%
{\renewcommand{\dashlinestretch}{30}
\begin{picture}(1666,657)(0,-10)
\path(1283,595)(1133,145)
\path(1133,145)(1058,220)(983,145)
	(908,220)(833,145)
\path(1206,355)(1239,449)
\blacken\path(1237.494,354.144)(1239.000,449.000)(1180.882,374.018)(1218.132,389.557)(1237.494,354.144)
\path(508,145)(608,145)
\blacken\path(518.000,115.000)(608.000,145.000)(518.000,175.000)(545.000,145.000)(518.000,115.000)
\put(83,145){\blacken\ellipse{150}{80}}
\put(83,145){\ellipse{150}{80}}
\put(1583,595){\blacken\ellipse{150}{80}}
\put(1583,595){\ellipse{150}{80}}
\path(383,145)(833,145)
\dottedline{60}(1283,595)(1533,595)
\dottedline{60}(383,145)(133,145)
\put(1176,555){\makebox(0,0)[lb]
{\smash{{{\SetFigFont{5}{6.0}{\rmdefault}{\mddefault}{\updefault}$1$}}}}}
\put(1108,0){\makebox(0,0)[lb]
{\smash{{{\SetFigFont{5}{6.0}{\rmdefault}{\mddefault}{\updefault}$2$}}}}}
\put(758,230){\makebox(0,0)[lb]
{\smash{{{\SetFigFont{6}{7.2}{\rmdefault}{\mddefault}{\updefault}L}}}}}
\put(353,0){\makebox(0,0)[lb]
{\smash{{{\SetFigFont{5}{6.0}{\rmdefault}{\mddefault}{\updefault}$1^\prime$}}}}}
\end{picture}
}
} \\
 \label{kern.sume} 
\raisebox{0.3ex}{$\displaystyle
\hat{F}_{\tau \mbox{\tiny $1$}}
\tau({\mathbf x}_1,{\mathbf x}_2)
M({\mathbf x}_2,{\mathbf x}_{1^\prime})
\hat{F}_{\tau \mbox{\tiny $1^\prime$}}
$} 
\raisebox{0.3ex}{$\displaystyle \;\;\; = \;\;\;$}
\raisebox{-2.5ex}{
\setlength{\unitlength}{0.00066667in}
\begingroup\makeatletter\ifx\SetFigFont\undefined%
\gdef\SetFigFont#1#2#3#4#5{%
  \reset@font\fontsize{#1}{#2pt}%
  \fontfamily{#3}\fontseries{#4}\fontshape{#5}%
  \selectfont}%
\fi\endgroup%
{\renewcommand{\dashlinestretch}{30}
\begin{picture}(1666,674)(0,-10)
\path(1283,49)(1133,499)
\path(1133,499)(1058,574)(983,499)
	(908,574)(833,499)
\path(508,499)(608,499)
\blacken\path(518.000,469.000)(608.000,499.000)(518.000,529.000)(545.000,499.000)(518.000,469.000)
\path(1206,274)(1239,180)
\blacken\path(1180.882,254.982)(1239.000,180.000)(1237.494,274.856)(1218.132,239.443)(1180.882,254.982)
\put(1583,49){\blacken\ellipse{150}{80}}
\put(1583,49){\ellipse{150}{80}}
\put(83,499){\blacken\ellipse{150}{80}}
\put(83,499){\ellipse{150}{80}}
\path(383,499)(833,499)
\dottedline{60}(1283,49)(1533,49)
\dottedline{60}(383,499)(133,499)
\put(758,584){\makebox(0,0)[lb]
{\smash{{{\SetFigFont{6}{7.2}{\rmdefault}{\mddefault}{\updefault}M}}}}}
\put(1176,0){\makebox(0,0)[lb]
{\smash{{{\SetFigFont{5}{6.0}{\rmdefault}{\mddefault}{\updefault}$1$}}}}}
\put(1160,534){\makebox(0,0)[lb]
{\smash{{{\SetFigFont{5}{6.0}{\rmdefault}{\mddefault}{\updefault}$2$}}}}}
\put(353,354){\makebox(0,0)[lb]
{\smash{{{\SetFigFont{5}{6.0}{\rmdefault}{\mddefault}{\updefault}$1^\prime$}}}}}
\end{picture}
}
}
\end{eqnarray}

The other two cases of interest involve diagrams where both incoming and outgoing
omitted lines are attached to the two-body part of the Hamiltonian, e.g.,
$r^{-1}_{ij}$, and diagrams where the omitted incoming free-line is attached at a
$\hat{F}_{\tau \mbox{\tiny $1^\prime$}}$ vertex and the omitted outgoing-line is
attached at the $\mathbf{x}_{1}$ vertex of a $r_{j1}^{-1}$ operator. Examples of
these two case are given by the following two diagrams:
\begin{eqnarray} \label{d4} 
\lefteqn{\raisebox{-4ex}{
\setlength{\unitlength}{0.00066667in}
\begingroup\makeatletter\ifx\SetFigFont\undefined%
\gdef\SetFigFont#1#2#3#4#5{%
  \reset@font\fontsize{#1}{#2pt}%
  \fontfamily{#3}\fontseries{#4}\fontshape{#5}%
  \selectfont}%
\fi\endgroup%
{\renewcommand{\dashlinestretch}{30}
\begin{picture}(1670,1174)(0,-10)
\put(-1417.000,537.000){\arc{3750.000}{5.9994}{6.5670}}
\path(1133,12)(908,837)
\dottedline{90}(1133,12)(1658,1062)
\path(442,282)(448,382)
\blacken\path(472.556,290.365)(448.000,382.000)(412.664,293.958)(444.227,319.113)(472.556,290.365)
\path(446,747)(434,847)
\blacken\path(474.509,761.215)(434.000,847.000)(414.937,754.067)(441.506,784.449)(474.509,761.215)
\path(1076,215)(1104,119)
\blacken\path(1050.000,197.000)(1104.000,119.000)(1107.600,213.800)(1086.360,179.480)(1050.000,197.000)
\path(1006,464)(1035,369)
\blacken\path(980.030,446.320)(1035.000,369.000)(1037.416,463.838)(1016.606,429.255)(980.030,446.320)
\path(945,699)(973,603)
\blacken\path(919.000,681.000)(973.000,603.000)(976.600,697.800)(955.360,663.480)(919.000,681.000)
\path(852,1060)(887,938)
\whiten\path(787.962,1065.638)(887.000,938.000)(903.309,1098.730)(858.045,1038.929)(787.962,1065.638)
\put(2183.000,537.000){\arc{3750.000}{2.8578}{3.4254}}
\put(758,312){\blacken\ellipse{150}{80}}
\put(758,312){\ellipse{150}{80}}
\put(83,1062){\blacken\ellipse{150}{80}}
\put(83,1062){\ellipse{150}{80}}
\put(1283,537){\blacken\ellipse{150}{80}}
\put(1283,537){\ellipse{150}{80}}
\dottedline{60}(383,12)(1133,12)
\dottedline{60}(1048,312)(833,312)
\dottedline{60}(183,1062)(383,1062)
\dottedline{60}(438,837)(903,837)
\dottedline{45}(983,537)(1208,537)
\dottedline{60}(914,836)(853,1065)
\end{picture}
}
}} \hspace{15ex}
&\raisebox{0.3ex}{$\displaystyle  \;\;\;= \;\;\;$}&
\raisebox{0.3ex}{$\displaystyle
p({\mathbf x}_1,{\mathbf x}_{1^\prime}),$} 
\\
\label{d3} 
\lefteqn{\raisebox{-4ex}{
\setlength{\unitlength}{0.00066667in}
\begingroup\makeatletter\ifx\SetFigFont\undefined%
\gdef\SetFigFont#1#2#3#4#5{%
  \reset@font\fontsize{#1}{#2pt}%
  \fontfamily{#3}\fontseries{#4}\fontshape{#5}%
  \selectfont}%
\fi\endgroup%
{\renewcommand{\dashlinestretch}{30}
\begin{picture}(1291,1124)(0,-10)
\path(1133,12)(908,837)
\put(2183.000,537.000){\arc{3750.000}{2.8578}{3.4254}}
\put(-1417.000,537.000){\arc{3750.000}{5.9994}{6.5670}}
\path(1076,215)(1104,119)
\blacken\path(1050.000,197.000)(1104.000,119.000)(1107.600,213.800)(1086.360,179.480)(1050.000,197.000)
\path(1006,464)(1035,369)
\blacken\path(980.030,446.320)(1035.000,369.000)(1037.416,463.838)(1016.606,429.255)(980.030,446.320)
\path(945,699)(973,603)
\blacken\path(919.000,681.000)(973.000,603.000)(976.600,697.800)(955.360,663.480)(919.000,681.000)
\path(442,282)(448,382)
\blacken\path(472.556,290.365)(448.000,382.000)(412.664,293.958)(444.227,319.113)(472.556,290.365)
\path(311,558)(311,458)
\blacken\path(281.000,548.000)(311.000,458.000)(341.000,548.000)(311.000,521.000)(281.000,548.000)
\path(446,747)(434,847)
\blacken\path(474.509,761.215)(434.000,847.000)(414.937,754.067)(441.506,784.449)(474.509,761.215)
\put(758,312){\blacken\ellipse{150}{80}}
\put(758,312){\ellipse{150}{80}}
\put(1208,837){\blacken\ellipse{150}{80}}
\put(1208,837){\ellipse{150}{80}}
\put(83,1062){\blacken\ellipse{150}{80}}
\put(83,1062){\ellipse{150}{80}}
\dottedline{60}(383,12)(1133,12)
\dottedline{60}(463,537)(988,537)
\dottedline{60}(1048,312)(833,312)
\dottedline{45}(908,837)(1133,837)
\dottedline{60}(183,1062)(383,1062)
\end{picture}
}
}}  \hspace{15ex}
&\raisebox{0.3ex}{$\displaystyle  \;\;\;\; = \;\;\;$}&
\raisebox{0.3ex}{$\displaystyle
n({\mathbf x}_1,{\mathbf x}_{1^\prime})
\hat{F}_{\tau \mbox{\tiny $1^\prime$}}
,$} 
\end{eqnarray}
where, for these examples, we have
\begin{eqnarray}
\label{p.kern} 
p({\mathbf x}_1,{\mathbf x}_{1^\prime})\!&=& \!
\mbox{$\frac{1}{128}$}
\varepsilon_\tau^{-4}
\tau({\mathbf x}_4,{\mathbf x}_5)
\hat{F}_{\tau \mbox{\tiny $5$}}
\kappa_\tau({\mathbf x}_5,{\mathbf x}_6)
\kappa_\tau({\mathbf x}_6,{\mathbf x}_4)
r_{61^\prime}^{-1} r_{41}^{-1}
\tau({\mathbf x}_1,{\mathbf x}_2)
\hat{F}_{\tau \mbox{\tiny $2$}}
\tau({\mathbf x}_2,{\mathbf x}_3)
\hat{F}_{\tau \mbox{\tiny $3$}}
\tau({\mathbf x}_3,{\mathbf x}_{1^\prime}).
\nonumber \\ \\
n({\mathbf x}_1,{\mathbf x}_{1^\prime})&=&
\mbox{$\frac{1}{16}$}
\varepsilon_\tau^{-4}
\tau({\mathbf x}_4,{\mathbf x}_5)
\hat{F}_{\tau \mbox{\tiny $5$}}
\kappa_\tau({\mathbf x}_5,{\mathbf x}_6)
\kappa_\tau({\mathbf x}_6,{\mathbf x}_4)
r_{63}^{-1} r_{41}^{-1}
\tau({\mathbf x}_1,{\mathbf x}_2)
\hat{F}_{\tau \mbox{\tiny $2$}}
\tau({\mathbf x}_2,{\mathbf x}_3)
\tau({\mathbf x}_3,{\mathbf x}_{1^\prime})
\nonumber \\
\end{eqnarray}
Summing over all diagrams of these types, gives
\begin{eqnarray} \label{kern.sumf} 
\raisebox{0.3ex}{$\displaystyle
P({\mathbf x}_1,{\mathbf x}_{1^\prime})
$} 
&\raisebox{0.3ex}{$\displaystyle \;\;\; = \;\;\;$}&
\raisebox{-1ex}{
\setlength{\unitlength}{0.00066667in}
\begingroup\makeatletter\ifx\SetFigFont\undefined%
\gdef\SetFigFont#1#2#3#4#5{%
  \reset@font\fontsize{#1}{#2pt}%
  \fontfamily{#3}\fontseries{#4}\fontshape{#5}%
  \selectfont}%
\fi\endgroup%
{\renewcommand{\dashlinestretch}{30}
\begin{picture}(1250,335)(0,-10)
\path(780,145)(705,220)(630,145)
	(555,220)(480,145)
\path(155,145)(255,145)
\blacken\path(165.000,115.000)(255.000,145.000)(165.000,175.000)(192.000,145.000)(165.000,115.000)
\path(980,145)(1080,145)
\blacken\path(990.000,115.000)(1080.000,145.000)(990.000,175.000)(1017.000,145.000)(990.000,115.000)
\dottedline{60}(30,145)(480,145)
\dottedline{60}(780,145)(1230,145)
\put(0,0){\makebox(0,0)[lb]
{\smash{{{\SetFigFont{5}{6.0}{\rmdefault}{\mddefault}{\updefault}$1^\prime$}}}}}
\put(604,245){\makebox(0,0)[lb]
{\smash{{{\SetFigFont{6}{7.2}{\rmdefault}{\mddefault}{\updefault}P}}}}}
\put(1204,0){\makebox(0,0)[lb]
{\smash{{{\SetFigFont{5}{6.0}{\rmdefault}{\mddefault}{\updefault}$1$}}}}}
\end{picture}
}
} \\
\label{kern.sumg} 
\raisebox{0.3ex}{$\displaystyle
N({\mathbf x}_1,{\mathbf x}_{1^\prime})
\hat{F}_{\tau \mbox{\tiny $1^\prime$}}
$} 
&\raisebox{0.3ex}{$\displaystyle \;\;\; = \;\;\;$}&
\hspace{-1.5ex}
\raisebox{-1ex}{
\setlength{\unitlength}{0.00066667in}
\begingroup\makeatletter\ifx\SetFigFont\undefined%
\gdef\SetFigFont#1#2#3#4#5{%
  \reset@font\fontsize{#1}{#2pt}%
  \fontfamily{#3}\fontseries{#4}\fontshape{#5}%
  \selectfont}%
\fi\endgroup%
{\renewcommand{\dashlinestretch}{30}
\begin{picture}(1603,320)(0,-10)
\path(1133,145)(1058,220)(983,145)
	(908,220)(833,145)
\path(508,145)(608,145)
\blacken\path(518.000,115.000)(608.000,145.000)(518.000,175.000)(545.000,145.000)(518.000,115.000)
\path(1333,145)(1433,145)
\blacken\path(1343.000,115.000)(1433.000,145.000)(1343.000,175.000)(1370.000,145.000)(1343.000,115.000)
\put(83,145){\blacken\ellipse{150}{80}}
\put(83,145){\ellipse{150}{80}}
\path(383,145)(833,145)
\dottedline{60}(383,145)(133,145)
\dottedline{60}(1133,145)(1583,145)
\put(758,230){\makebox(0,0)[lb]
{\smash{{{\SetFigFont{6}{7.2}{\rmdefault}{\mddefault}{\updefault}N}}}}}
\put(353,0){\makebox(0,0)[lb]
{\smash{{{\SetFigFont{5}{6.0}{\rmdefault}{\mddefault}{\updefault}$1^\prime$}}}}}
\put(1557,0){\makebox(0,0)[lb]
{\smash{{{\SetFigFont{5}{6.0}{\rmdefault}{\mddefault}{\updefault}$1$}}}}}
\end{picture}
}
}
\end{eqnarray}

Since the diagrams represented in Eq.~(\ref{kern.sumb}), (\ref{kern.sumc}),
(\ref{kern.sumd}), (\ref{kern.sume}), (\ref{kern.sumf}), and (\ref{kern.sumg})
include all possible diagrams that can contribute to
$v_{\mathrm{co}}^{\scscs\tau\mathrm{ex}}(\mathbf{x}_1,\mathbf{x}_{1^\prime})$,
using these expression and Eq.~(\ref{vco.diag}), we have the following
diagrammatic and algebraic relations:
\begin{eqnarray} \label{vco.diagb} 
\raisebox{-1ex}{
\setlength{\unitlength}{0.00066667in}
\begingroup\makeatletter\ifx\SetFigFont\undefined%
\gdef\SetFigFont#1#2#3#4#5{%
  \reset@font\fontsize{#1}{#2pt}%
  \fontfamily{#3}\fontseries{#4}\fontshape{#5}%
  \selectfont}%
\fi\endgroup%
{\renewcommand{\dashlinestretch}{30}
\begin{picture}(1250,370)(0,-10)
\path(780,145)(705,220)(630,145)
	(555,220)(480,145)
\path(155,145)(255,145)
\blacken\path(165.000,115.000)(255.000,145.000)(165.000,175.000)(192.000,145.000)(165.000,115.000)
\path(980,145)(1080,145)
\blacken\path(990.000,115.000)(1080.000,145.000)(990.000,175.000)(1017.000,145.000)(990.000,115.000)
\path(30,145)(480,145)
\path(780,145)(1230,145)
\put(0,0){\makebox(0,0)[lb]
{\smash{{{\SetFigFont{5}{6.0}{\rmdefault}{\mddefault}{\updefault}$1^\prime$}}}}}
\put(1204,0){\makebox(0,0)[lb]
{\smash{{{\SetFigFont{5}{6.0}{\rmdefault}{\mddefault}{\updefault}$1$}}}}}
\put(494,295){\makebox(0,0)[lb]
{\smash{{{\SetFigFont{6}{7.2}{\rmdefault}{\mddefault}{\updefault}\mbox{\small $v_{\mathrm{co}}$}}}}}}
\end{picture}
}
} 
\raisebox{0.3ex}{$\displaystyle \;\;\; = \;\;\;$}
\raisebox{-2.5ex}{
\setlength{\unitlength}{0.00066667in}
\begingroup\makeatletter\ifx\SetFigFont\undefined%
\gdef\SetFigFont#1#2#3#4#5{%
  \reset@font\fontsize{#1}{#2pt}%
  \fontfamily{#3}\fontseries{#4}\fontshape{#5}%
  \selectfont}%
\fi\endgroup%
{\renewcommand{\dashlinestretch}{30}
\begin{picture}(1313,657)(0,-10)
\path(930,595)(780,145)
\path(780,145)(705,220)(630,145)
	(555,220)(480,145)
\path(853,355)(886,449)
\blacken\path(884.494,354.144)(886.000,449.000)(827.882,374.018)(865.132,389.557)(884.494,354.144)
\path(155,145)(255,145)
\blacken\path(165.000,115.000)(255.000,145.000)(165.000,175.000)(192.000,145.000)(165.000,115.000)
\put(1230,595){\blacken\ellipse{150}{80}}
\put(1230,595){\ellipse{150}{80}}
\dottedline{60}(930,595)(1180,595)
\dottedline{60}(30,145)(480,145)
\put(823,555){\makebox(0,0)[lb]
{\smash{{{\SetFigFont{5}{6.0}{\rmdefault}{\mddefault}{\updefault}$1$}}}}}
\put(755,0){\makebox(0,0)[lb]
{\smash{{{\SetFigFont{5}{6.0}{\rmdefault}{\mddefault}{\updefault}$2$}}}}}
\put(405,230){\makebox(0,0)[lb]
{\smash{{{\SetFigFont{6}{7.2}{\rmdefault}{\mddefault}{\updefault}G}}}}}
\put(0,0){\makebox(0,0)[lb]
{\smash{{{\SetFigFont{5}{6.0}{\rmdefault}{\mddefault}{\updefault}$1^\prime$}}}}}
\end{picture}
}
} +
\raisebox{-2.5ex}{
\setlength{\unitlength}{0.00066667in}
\begingroup\makeatletter\ifx\SetFigFont\undefined%
\gdef\SetFigFont#1#2#3#4#5{%
  \reset@font\fontsize{#1}{#2pt}%
  \fontfamily{#3}\fontseries{#4}\fontshape{#5}%
  \selectfont}%
\fi\endgroup%
{\renewcommand{\dashlinestretch}{30}
\begin{picture}(1313,674)(0,-10)
\path(930,49)(780,499)
\path(780,499)(705,574)(630,499)
	(555,574)(480,499)
\path(155,499)(255,499)
\blacken\path(165.000,469.000)(255.000,499.000)(165.000,529.000)(192.000,499.000)(165.000,469.000)
\path(853,274)(886,180)
\blacken\path(827.882,254.982)(886.000,180.000)(884.494,274.856)(865.132,239.443)(827.882,254.982)
\put(1230,49){\blacken\ellipse{150}{80}}
\put(1230,49){\ellipse{150}{80}}
\dottedline{60}(30,499)(480,499)
\dottedline{60}(930,49)(1180,49)
\put(405,584){\makebox(0,0)[lb]
{\smash{{{\SetFigFont{6}{7.2}{\rmdefault}{\mddefault}{\updefault}I}}}}}
\put(823,0){\makebox(0,0)[lb]
{\smash{{{\SetFigFont{5}{6.0}{\rmdefault}{\mddefault}{\updefault}$1$}}}}}
\put(807,534){\makebox(0,0)[lb]
{\smash{{{\SetFigFont{5}{6.0}{\rmdefault}{\mddefault}{\updefault}$2$}}}}}
\put(0,354){\makebox(0,0)[lb]
{\smash{{{\SetFigFont{5}{6.0}{\rmdefault}{\mddefault}{\updefault}$1^\prime$}}}}}
\end{picture}
}
} +
\raisebox{-2.5ex}{
\setlength{\unitlength}{0.00066667in}
\begingroup\makeatletter\ifx\SetFigFont\undefined%
\gdef\SetFigFont#1#2#3#4#5{%
  \reset@font\fontsize{#1}{#2pt}%
  \fontfamily{#3}\fontseries{#4}\fontshape{#5}%
  \selectfont}%
\fi\endgroup%
{\renewcommand{\dashlinestretch}{30}
\begin{picture}(1666,657)(0,-10)
\path(1283,595)(1133,145)
\path(1133,145)(1058,220)(983,145)
	(908,220)(833,145)
\path(1206,355)(1239,449)
\blacken\path(1237.494,354.144)(1239.000,449.000)(1180.882,374.018)(1218.132,389.557)(1237.494,354.144)
\path(508,145)(608,145)
\blacken\path(518.000,115.000)(608.000,145.000)(518.000,175.000)(545.000,145.000)(518.000,115.000)
\put(83,145){\blacken\ellipse{150}{80}}
\put(83,145){\ellipse{150}{80}}
\put(1583,595){\blacken\ellipse{150}{80}}
\put(1583,595){\ellipse{150}{80}}
\path(383,145)(833,145)
\dottedline{60}(1283,595)(1533,595)
\dottedline{60}(383,145)(133,145)
\put(1176,555){\makebox(0,0)[lb]
{\smash{{{\SetFigFont{5}{6.0}{\rmdefault}{\mddefault}{\updefault}$1$}}}}}
\put(1108,0){\makebox(0,0)[lb]
{\smash{{{\SetFigFont{5}{6.0}{\rmdefault}{\mddefault}{\updefault}$2$}}}}}
\put(758,230){\makebox(0,0)[lb]
{\smash{{{\SetFigFont{6}{7.2}{\rmdefault}{\mddefault}{\updefault}L}}}}}
\put(353,0){\makebox(0,0)[lb]
{\smash{{{\SetFigFont{5}{6.0}{\rmdefault}{\mddefault}{\updefault}$1^\prime$}}}}}
\end{picture}
}
} 
\hspace{12ex} \\ \nonumber \hspace{28ex} 
+
\raisebox{-2.5ex}{
\setlength{\unitlength}{0.00066667in}
\begingroup\makeatletter\ifx\SetFigFont\undefined%
\gdef\SetFigFont#1#2#3#4#5{%
  \reset@font\fontsize{#1}{#2pt}%
  \fontfamily{#3}\fontseries{#4}\fontshape{#5}%
  \selectfont}%
\fi\endgroup%
{\renewcommand{\dashlinestretch}{30}
\begin{picture}(1666,674)(0,-10)
\path(1283,49)(1133,499)
\path(1133,499)(1058,574)(983,499)
	(908,574)(833,499)
\path(508,499)(608,499)
\blacken\path(518.000,469.000)(608.000,499.000)(518.000,529.000)(545.000,499.000)(518.000,469.000)
\path(1206,274)(1239,180)
\blacken\path(1180.882,254.982)(1239.000,180.000)(1237.494,274.856)(1218.132,239.443)(1180.882,254.982)
\put(1583,49){\blacken\ellipse{150}{80}}
\put(1583,49){\ellipse{150}{80}}
\put(83,499){\blacken\ellipse{150}{80}}
\put(83,499){\ellipse{150}{80}}
\path(383,499)(833,499)
\dottedline{60}(1283,49)(1533,49)
\dottedline{60}(383,499)(133,499)
\put(758,584){\makebox(0,0)[lb]
{\smash{{{\SetFigFont{6}{7.2}{\rmdefault}{\mddefault}{\updefault}M}}}}}
\put(1176,0){\makebox(0,0)[lb]
{\smash{{{\SetFigFont{5}{6.0}{\rmdefault}{\mddefault}{\updefault}$1$}}}}}
\put(1160,534){\makebox(0,0)[lb]
{\smash{{{\SetFigFont{5}{6.0}{\rmdefault}{\mddefault}{\updefault}$2$}}}}}
\put(353,354){\makebox(0,0)[lb]
{\smash{{{\SetFigFont{5}{6.0}{\rmdefault}{\mddefault}{\updefault}$1^\prime$}}}}}
\end{picture}
}
} +
\raisebox{-1ex}{
\setlength{\unitlength}{0.00066667in}
\begingroup\makeatletter\ifx\SetFigFont\undefined%
\gdef\SetFigFont#1#2#3#4#5{%
  \reset@font\fontsize{#1}{#2pt}%
  \fontfamily{#3}\fontseries{#4}\fontshape{#5}%
  \selectfont}%
\fi\endgroup%
{\renewcommand{\dashlinestretch}{30}
\begin{picture}(1250,335)(0,-10)
\path(780,145)(705,220)(630,145)
	(555,220)(480,145)
\path(155,145)(255,145)
\blacken\path(165.000,115.000)(255.000,145.000)(165.000,175.000)(192.000,145.000)(165.000,115.000)
\path(980,145)(1080,145)
\blacken\path(990.000,115.000)(1080.000,145.000)(990.000,175.000)(1017.000,145.000)(990.000,115.000)
\dottedline{60}(30,145)(480,145)
\dottedline{60}(780,145)(1230,145)
\put(0,0){\makebox(0,0)[lb]
{\smash{{{\SetFigFont{5}{6.0}{\rmdefault}{\mddefault}{\updefault}$1^\prime$}}}}}
\put(604,245){\makebox(0,0)[lb]
{\smash{{{\SetFigFont{6}{7.2}{\rmdefault}{\mddefault}{\updefault}P}}}}}
\put(1204,0){\makebox(0,0)[lb]
{\smash{{{\SetFigFont{5}{6.0}{\rmdefault}{\mddefault}{\updefault}$1$}}}}}
\end{picture}
}
} +
\raisebox{-1ex}{
\setlength{\unitlength}{0.00066667in}
\begingroup\makeatletter\ifx\SetFigFont\undefined%
\gdef\SetFigFont#1#2#3#4#5{%
  \reset@font\fontsize{#1}{#2pt}%
  \fontfamily{#3}\fontseries{#4}\fontshape{#5}%
  \selectfont}%
\fi\endgroup%
{\renewcommand{\dashlinestretch}{30}
\begin{picture}(1603,320)(0,-10)
\path(1133,145)(1058,220)(983,145)
	(908,220)(833,145)
\path(508,145)(608,145)
\blacken\path(518.000,115.000)(608.000,145.000)(518.000,175.000)(545.000,145.000)(518.000,115.000)
\path(1333,145)(1433,145)
\blacken\path(1343.000,115.000)(1433.000,145.000)(1343.000,175.000)(1370.000,145.000)(1343.000,115.000)
\put(83,145){\blacken\ellipse{150}{80}}
\put(83,145){\ellipse{150}{80}}
\path(383,145)(833,145)
\dottedline{60}(383,145)(133,145)
\dottedline{60}(1133,145)(1583,145)
\put(758,230){\makebox(0,0)[lb]
{\smash{{{\SetFigFont{6}{7.2}{\rmdefault}{\mddefault}{\updefault}N}}}}}
\put(353,0){\makebox(0,0)[lb]
{\smash{{{\SetFigFont{5}{6.0}{\rmdefault}{\mddefault}{\updefault}$1^\prime$}}}}}
\put(1557,0){\makebox(0,0)[lb]
{\smash{{{\SetFigFont{5}{6.0}{\rmdefault}{\mddefault}{\updefault}$1$}}}}}
\end{picture}
}
},
\end{eqnarray}
\begin{eqnarray} \!\!
v_{\mathrm{co}}^{\scscs\tau\mathrm{ex}}({\mathbf x}_1,{\mathbf x}_{1^\prime}) \!=\!
\hat{F}_{\tau \mbox{\tiny $1$}}
\kappa_\tau({\mathbf x}_1,{\mathbf x}_2)
G({\mathbf x}_2,{\mathbf x}_{1^\prime}) +
\hat{F}_{\tau \mbox{\tiny $1$}}
\tau({\mathbf x}_1,{\mathbf x}_2)
I({\mathbf x}_2,{\mathbf x}_{1^\prime}) +
\hat{F}_{\tau \mbox{\tiny $1$}}
\kappa_\tau({\mathbf x}_1,{\mathbf x}_2)
L({\mathbf x}_2,{\mathbf x}_{1^\prime})
\hat{F}_{\tau \mbox{\tiny $1^\prime$}}
\nonumber \\
\mbox{}+ 
\hat{F}_{\tau \mbox{\tiny $1$}}
\tau({\mathbf x}_1,{\mathbf x}_2)
M({\mathbf x}_2,{\mathbf x}_{1^\prime})
\hat{F}_{\tau \mbox{\tiny $1^\prime$}} +
P({\mathbf x}_1,{\mathbf x}_{1^\prime}) +
N({\mathbf x}_1,{\mathbf x}_{1^\prime})
\hat{F}_{\tau \mbox{\tiny $1^\prime$}}.
\hspace{4ex}
\end{eqnarray}
Substituting this expression into Eq.~(\ref{def.vco}), and reordering terms, we have
\begin{eqnarray} \label{vco.psi} 
\hspace{-1ex}
\left(
\hat{v}_{\mathrm{co}}^{\scscs\tau} (\mathbf{x}_1)\right)_{\mathrm{ex}}
\psi_{w\sigma}^{\scscs \tau}(\mathbf{x}_1) &&\hspace{-1.5ex}=
P({\mathbf x}_1,{\mathbf x}_{1^\prime}) 
\psi_{w\sigma}^{\scscs \tau}(\mathbf{x}_{1^\prime})
+
N({\mathbf x}_1,{\mathbf x}_{1^\prime})
\hat{F}_{\tau \mbox{\tiny $1^\prime$}}
\psi_{w\sigma}^{\scscs \tau}(\mathbf{x}_{1^\prime})
\\ \nonumber 
\mbox{}+&&\hspace{-2ex}
\hat{F}_{\tau \mbox{\tiny $1$}}
\kappa_\tau({\mathbf x}_1,{\mathbf x}_2)
G({\mathbf x}_2,{\mathbf x}_{1^\prime})
\psi_{w\sigma}^{\scscs \tau}(\mathbf{x}_{1^\prime})
+
\hat{F}_{\tau \mbox{\tiny $1$}}
\tau({\mathbf x}_1,{\mathbf x}_2)
I({\mathbf x}_2,{\mathbf x}_{1^\prime}) 
\psi_{w\sigma}^{\scscs \tau}(\mathbf{x}_{1^\prime})
\\ \nonumber 
\mbox{} 
+&&\hspace{-2ex}
\hat{F}_{\tau \mbox{\tiny $1$}}
\kappa_\tau({\mathbf x}_1,{\mathbf x}_2)
L({\mathbf x}_2,{\mathbf x}_{1^\prime})
\hat{F}_{\tau \mbox{\tiny $1^\prime$}}
\psi_{w\sigma}^{\scscs \tau}(\mathbf{x}_{1^\prime})
+ 
\hat{F}_{\tau \mbox{\tiny $1$}}
\tau({\mathbf x}_1,{\mathbf x}_2)
M({\mathbf x}_2,{\mathbf x}_{1^\prime})
\hat{F}_{\tau \mbox{\tiny $1^\prime$}} 
\psi_{w\sigma}^{\scscs \tau}(\mathbf{x}_{1^\prime})
\nonumber 
\end{eqnarray}

If we substitute the above expression into Eq.~(\ref{wehf.prop2}), it is easily
demonstrated that the first two terms from the above expression vanish, e.g., for
the first term, we have
\begin{eqnarray} 
\lim_{\mathbf{r}_1\rightarrow\mathbf{R}} 
|\mathbf{r}_1-\mathbf{R}| 
P({\mathbf x}_1,{\mathbf x}_{1^\prime})
\psi_{w\sigma}^{\scscs \tau}(\mathbf{x}_{1^\prime}) 
= 0,
\;\;\; \mbox{for all $\mathbf{R}$},
\end{eqnarray}
where the variable ${\mathbf x}_1$ within $P({\mathbf x}_1,{\mathbf x}_{2}) $ is
the independent variable for functions of the general form
$r_{j1}^{-1}\tau({\mathbf x}_1,{\mathbf x}_i)$ or $r_{j1}^{-1}\kappa_\tau({\mathbf
x}_1,{\mathbf x}_i)$, and the dummy indices in these function: ${\mathbf x}_i$ and
${\mathbf x}_j$, are integrated over; furthermore, and in general, all other
variable that appear in diagrams that contribute to $P({\mathbf x}_1,{\mathbf
x}_{1^\prime})$ -- e.g., ${\mathbf x}_1$, ${\mathbf x}_2$, ${\mathbf x}_3$,
${\mathbf x}_4$, ${\mathbf x}_4$, and ${\mathbf x}_6$ for $p({\mathbf
x}_1,{\mathbf x}_{1^\prime})$ as presented in Eq.~(\ref{p.kern}) -- are also dummy
integration variables.  Hence, since the functions $r_{j1}^{-1}\tau({\mathbf x}_1$
${\mathbf x}_i)$ and $r_{j1}^{-1}\kappa_\tau({\mathbf x}_1,{\mathbf x}_i)$ do {\em
not} and contain a laplacian term -- i.e.,
$\mbox{$-\frac{1}{2}$}\nabla^2_{\mathbf{r}_1}\tau({\mathbf x}_1,{\mathbf x}_i)$ or
$\mbox{$-\frac12$}\nabla^2_{\mathbf{r}_1}\kappa_\tau({\mathbf x}_1,{\mathbf x}_i)$
-- or a singularity -- i.e., $|\mathbf{r}_1-\mathbf{R}_m|^{-1}$ -- the above
identity holds. 

Using a similar analysis, we also obtain the following identity:
\begin{eqnarray} 
\lim_{\mathbf{r}_1\rightarrow\mathbf{R}} 
|\mathbf{r}_1-\mathbf{R}| 
N({\mathbf x}_1,{\mathbf x}_{1^\prime})
\hat{F}_{\tau \mbox{\tiny $1^\prime$}}
\psi_{w\sigma}^{\scscs \tau}(\mathbf{x}_{1^\prime}) 
= 0.
\end{eqnarray}

Substituting Eq.~(\ref{vco.psi}) into (\ref{wehf.prop2}), and using the above two
identities, we get
\begin{eqnarray} 
\lim_{\mathbf{r}_1\rightarrow\mathbf{R}} 
|\mathbf{r}_1-\mathbf{R}|
\left[
\hat{v}_{\mathrm{co}}^{\scscs\tau} (\mathbf{x}_1)
\right]_{\mathrm{ex}}
\psi_{w\sigma}^{\scscs \tau}(\mathbf{x}_1) =
\hspace{53ex}
\\ \nonumber
\hspace{10ex}
\lim_{\mathbf{r}_1\rightarrow\mathbf{R}} 
|\mathbf{r}_1-\mathbf{R}| \left[
\hat{F}_{\tau \mbox{\tiny $1$}}
\kappa_\tau({\mathbf x}_1,{\mathbf x}_2)
G({\mathbf x}_2,{\mathbf x}_{1^\prime})
\psi_{w\sigma}^{\scscs \tau}(\mathbf{x}_{1^\prime})
+
\hat{F}_{\tau \mbox{\tiny $1$}}
\tau({\mathbf x}_1,{\mathbf x}_2)
I({\mathbf x}_2,{\mathbf x}_{1^\prime}) 
\psi_{w\sigma}^{\scscs \tau}(\mathbf{x}_{1^\prime})
\right.
\\ 
\hspace{11ex}
\nonumber 
\left.
\mbox{} +
\hat{F}_{\tau \mbox{\tiny $1$}}
\kappa_\tau({\mathbf x}_1,{\mathbf x}_2)
L({\mathbf x}_2,{\mathbf x}_{1^\prime})
\hat{F}_{\tau \mbox{\tiny $1^\prime$}}
\psi_{w\sigma}^{\scscs \tau}(\mathbf{x}_{1^\prime})
+ 
\hat{F}_{\tau \mbox{\tiny $1$}}
\tau({\mathbf x}_1,{\mathbf x}_2)
M({\mathbf x}_2,{\mathbf x}_{1^\prime})
\hat{F}_{\tau \mbox{\tiny $1^\prime$}} 
\psi_{w\sigma}^{\scscs \tau}(\mathbf{x}_{1^\prime})\right],
\nonumber 
\end{eqnarray}
and this expression can be written as
\begin{eqnarray} \label{intexprA} 
\!\!\lim_{\mathbf{r}_1\rightarrow\mathbf{R}} 
|\mathbf{r}_1-\mathbf{R}|
\left[
\hat{v}_{\mathrm{co}}^{\scscs\tau} (\mathbf{x}_1)
\right]_{\mathrm{ex}}
\psi_{w\sigma}^{\scscs \tau}(\mathbf{x}_1) = 
\hspace{54ex} \\ \hspace{28ex}
\lim_{\mathbf{r}_1\rightarrow\mathbf{R}} 
|\mathbf{r}_1-\mathbf{R}| \left[
\hat{F}_{\tau \mbox{\tiny $1$}}
\tau({\mathbf x}_1,{\mathbf x}_2)
A_{w\sigma}({\mathbf x}_2)
\!+\!
\hat{F}_{\tau \mbox{\tiny $1$}}
\kappa_\tau({\mathbf x}_1,{\mathbf x}_2)
B_{w\sigma}({\mathbf x}_2)
\right], 
\nonumber
\end{eqnarray}
where
\begin{eqnarray}
A_{w\sigma}({\mathbf x}_2)&=& I({\mathbf x}_2,{\mathbf x}_{1^\prime}) 
\psi_{w\sigma}^{\scscs \tau}(\mathbf{x}_{1^\prime})
+
M({\mathbf x}_2,{\mathbf x}_{1^\prime})
\hat{F}_{\tau \mbox{\tiny $1^\prime$}} 
\psi_{w\sigma}^{\scscs \tau}(\mathbf{x}_{1^\prime}),
\\
B_{w\sigma}({\mathbf x}_2)&=&
G({\mathbf x}_2,{\mathbf x}_{1^\prime})
\psi_{w\sigma}^{\scscs \tau}(\mathbf{x}_{1^\prime})
+
L({\mathbf x}_2,{\mathbf x}_{1^\prime})
\hat{F}_{\tau \mbox{\tiny $1^\prime$}}
\psi_{w\sigma}^{\scscs \tau}(\mathbf{x}_{1^\prime}).
\end{eqnarray}
Substituting into Eq.~(\ref{intexprA}) the Fock operator, Eq.~(\ref{Fockop}), it is
easily seen that the terms involving the Coulomb and exchange operator vanish, so
we have
\begin{eqnarray} 
\lim_{\mathbf{r}_1\rightarrow\mathbf{R}} 
|\mathbf{r}_1-\mathbf{R}|
\left[\hat{v}_{\mathrm{co}}^{\scscs\tau} (\mathbf{x}_1)
\right]_{\mathrm{ex}}
\psi_{w\sigma}^{\scscs \tau}(\mathbf{x}_1) 
= 
\hspace{53ex} \\ \nonumber \hspace{28ex}
\lim_{\mathbf{r}_1\rightarrow\mathbf{R}} 
|\mathbf{r}_1-\mathbf{R}| \left[
\hat{h}_{v \mbox{\tiny $1$}}
\tau({\mathbf x}_1,{\mathbf x}_2)
A_{w\sigma}({\mathbf x}_2)
+
\hat{h}_{v \mbox{\tiny $1$}}
\kappa_\tau({\mathbf x}_1,{\mathbf x}_2)
B_{w\sigma}({\mathbf x}_2)
\right], 
\nonumber
\end{eqnarray}
where $\hat{h}_{v \mbox{\tiny $1$}}$ is given by Eq.~(\ref{hv1}). Using
Eqs.~(\ref{onepart.br}) and (\ref{orthpart.br}), we obtain Eq.~(\ref{vcopsi}),
where
\begin{eqnarray} \label{Cwx} 
C_{w\sigma}^{x\sigma} =A_{w\sigma}({\mathbf x}_2) 
\left(\psi_{x\sigma}^{\scscs \tau}(\mathbf{x}_2)\right)^* \\
\label{Dwr} 
D_{w\sigma}^{r\sigma} =B_{w\sigma}({\mathbf x}_2)
\left(\psi_{r\sigma}^{\scscs \tau}(\mathbf{x}_2)\right)^*
\end{eqnarray}
and the terms that do not preserve the spin state, e.g.,
$D_{w\sigma}^{x\sigma^\prime}$, ore omitted, since these terms vanish;
furthermore, there are summations over the repeated indices $x\sigma$ and
$r\sigma$ for the orbital sets $\{\psi_o \leftarrow \tau, \hat{F}_{\tau}\}$ and
$\{\psi_u \leftarrow \tau, \hat{F}_{\tau}\}$, respectively.

\section{Diagrammatic formalism for the correlation energy ${\mathcal E}_{\mathrm{co}}$} 
\label{DIAG.CE} 

In order to keep the notation less cluttered, for this section we use a combined
spin-spatial notation for spin-orbital indices; for example,
$\psi_{w\sigma}^{\scscs \tau}$ is now denoted by $\psi_{w}^{\scscs \tau}$.

A diagrammatic expansion for correlation-energy ${\mathcal E}_{\mathrm{co}}$, or
for $\chi_\tau$ using Lindgren's formalism
\cite{Lindgren:86,Lindgren:74,Lindgren:78}), is easier to obtain when all
operators involved are written in normal-ordered form
\cite{Bogoliubov:59,Paldus:75,Lindgren:86,Cizek:69}. For example, the Hamiltonian,
given by Eq.~(\ref{H}), can be written as
\begin{eqnarray} \label{H.norm} 
H = E_1[\tau] + \{\hat{F}_\tau\} + \{r_{12}^{-1}\}_\tau,
\end{eqnarray}
where
\begin{eqnarray}
E_1[\tau]&=& 
\langle \tau |H| \tau \rangle \nonumber \\
&=&\sum_{w} [w|\hat{h}|w] +  \frac12 
\sum_{wx}\left([w w|x x]-[w x|x w] \right), \\
\label{H1F} 
\{\hat{F}_\tau\} &=&  \sum_{ij} \left[[i|\hat{h}|j]+  \sum_{w} 
\left([w w|i j]- [w i |j w]\right)\right] \{a_{i}^\dagger a_{j}\}_\tau,
\\
\label{H2R12} 
\{r_{12}^{-1}\}_\tau &=&
\frac{1}{2} \sum_{ijkl}
[i j|k l]\{a_{i}^\dagger a_{k}^\dagger a_{l}  a_{j} \}_\tau,
\end{eqnarray}
and the integrals are now spin-dependent as indicated by the square brackets
$[\cdots]$ \cite{Szabo:82}.  Denoting the one-body portion of $H$ by
$\{\hat{F}_\tau\}$, is appropriate, since this term is the Fock-operator, except
that the second quantized operators are normal-ordered with respect to the
$|\tau\rangle$ vacuum state, instead of the true vacuum $|\,\rangle$, as in
$\hat{F}_\tau$. The two body portion of $H$ is denoted by $\{r_{12}^{-1}\}_\tau$
emphasizing that this operator is determined by $r_{12}^{-1}$ and the vacuum state
$|\tau\rangle$; furthermore, except for the shifted vacuum, the two-body portion
of $H$ is $r_{12}^{-1}$, when this operator is expressed in second quantization.

For a perturbative treatment, we partition the Hamiltonian into a zeroth-order
Hamiltonian $H_0$ and a perturbation~$V$:
\begin{equation} \label{Hpart} 
H = H_0 + V,
\end{equation}
where we require the reference state
$|\tau\rangle$ to be an eigenfunction of $H_0$, a one-body operator:
\begin{eqnarray} 
\label{h0.eigen} 
H_0 |\tau\rangle&=&E_0 |\tau\rangle, \\
\label{H0.onebod} 
H_0&=&\sum_{ij} \epsilon_{ij} a^\dagger_i a_j,
\end{eqnarray}
and the zeroth-order Hamiltonian is defined by its matrix elements; we choose
them by requiring the following relation to be satisfied:
\begin{subequations} \label{H0.matel} 
\begin{equation}
\epsilon_{ij}= \epsilon_{ji}= \epsilon_{ij}^{\mbox{\tiny $\tau$}}, \; 
\end{equation}
where
\begin{eqnarray} 
\label{H0.wr} 
\epsilon_{wr}^{\mbox{\tiny $\tau$}}&=& 0, \; 
\\ \label{H0.wx} 
\epsilon_{wx}^{\mbox{\tiny $\tau$}}&=&
\langle \psi_w^{\scscs \tau}|\hat{f}_o^{\scscs \tau}|\psi_x^{\scscs \tau} \rangle,
\\ \label{H0.rs} 
\epsilon_{rs}^{\mbox{\tiny $\tau$}}&=&
\langle \psi_r^{\scscs \tau}|\hat{f}_u^{\scscs \tau}|\psi_s^{\scscs \tau} \rangle,
\end{eqnarray}
\end{subequations}
and the one-body operators, $\hat{f}_o^{\scscs \tau}$ and $\hat{f}_u^{\scscs
\tau}$, are determined by the reference state $|\tau\rangle$, but the dependence
of $\hat{f}_o^{\scscs \tau}$ and $\hat{f}_u^{\scscs \tau}$ upon $|\tau\rangle$ is
at our disposal; the orbital subspaces are, again, defined by
Eqs.~(\ref{Bocc.orbs}) and (\ref{Bunocc.orbs}).

Using the above choice, our zeroth-order Hamiltonian becomes
\begin{equation} \label{H0.nd} 
H_{0}^{\tau} = \sum_{w,x\in \{\psi_o\rightarrow \tau\}}
\epsilon_{wx}^{\mbox{\tiny $\tau$}} a^\dagger_w a_x + \sum_{r,s\in \{\psi_u\rightarrow \tau\}}
\epsilon_{rs}^{\mbox{\tiny $\tau$}} a^\dagger_r a_s,
\end{equation}
where the appended ${\tau}$ superscript indicates that $H_0^{\scscs \tau}$ now
depends on the reference state~$|\tau\rangle$.

A linked diagram expansion for $\chi_{\tau}$ and ${\cal E}_{\mathrm{co}} [\tau]$
is known to exist for a zeroth-order Hamiltonian that is a diagonal, one-body,
operator
\cite{Goldstone:57,Hugenholtz:57,Sanders:69,Raimes:72,Paldus:75,Lindgren:74,Wilson:85,Lindgren:86,Harris:92}. A
diagonal form for our one-body operator, $H_0^\tau$, is obtained when we choose its
orbital sets -- $\{\psi_o\mbox{\small $\rightarrow\tau$}\}$ and
$\{\psi_u\mbox{\small $\rightarrow\tau$}\}$ -- to satisfy the following
conditions:
\begin{subequations} \label{vphi.matel} 
\begin{eqnarray} \label{vphi.wx} 
\langle \psi_w^{\scscs \tau}|\hat{f}_o^{\scscs \tau}|\psi_x^{\scscs \tau} \rangle&=& 
\delta_{wx} \epsilon_w^{\mbox{\tiny $\tau$}},
\\ \label{vphi.rs} 
\langle \psi_r^{\scscs \tau}|\hat{f}_u^{\scscs \tau}|\psi_s^{\scscs \tau} \rangle&=&
\delta_{rs} \epsilon_r^{\mbox{\tiny $\tau$}},
\end{eqnarray}
\end{subequations}
where we denote these particular sets of orbitals by $\{\psi_o \mbox{\small
$\leftarrow\tau,\hat{f}_o^{\tau}$}\}$ and $\{\psi_u \mbox{\small
$\leftarrow\tau,\hat{f}_u^{\tau}$}\}$, indicating that they are uniquely
determined by $|\tau\rangle$ and their one-particle operator, $\hat{f}_o^{\scscs
\tau}$ or $\hat{f}_u^{\scscs \tau}$.

Using these orbitals, $H_0^{\tau}$ can be written as
\begin{equation} \label{h0.part} 
H_0^\tau=\hat{\mbox{\sc \Large $o$}}_\tau + \hat{\mbox{\sc \large $u$}}_\tau,
\end{equation}
where these terms -- $\hat{\mbox{\sc \Large $o$}}_\tau$ and $\hat{\mbox{\sc
\large $u$}}_\tau$ -- are the occupied and unoccupied portions of $H_0^\tau$
-- $(H_0^\tau)_{\text{oc}}$ and $(H_0^\tau)_{\text{un}}$ -- and are given by
the following:
\begin{subequations} \label{h0.partB} 
\begin{eqnarray}
\label{H01o} 
\hat{\mbox{\sc \Large $o$}}_\tau &=& 
\!\!\!\!\!
\sum_{w\in\{\psi_o \leftarrow \tau, \hat{f}_o^{\tau}\}}
\!\!\!\!\!
\epsilon_{w}^{\mbox{\tiny $\tau$}} \, a_{w}^\dagger a_{w},
\\  \label{H01u} 
\hat{\mbox{\sc \large $u$}}_\tau &=& 
\!\!\!\!\!
\sum_{r\in \{\psi_u\leftarrow \tau,\hat{f}_u^{\tau}\}} 
\!\!\!\!\!
\epsilon_r^{\mbox{\tiny $\tau$}} \, a_r^\dagger a_r,
\end{eqnarray}
\end{subequations}
where our partitioning can be written as
\begin{equation} \label{Hpart.phi} 
H = H_0^{\tau} + V_\tau.
\end{equation}

Using the above notation, our zeroth-order Hamiltonian in normal-ordered form can
be written as
\begin{equation} \label{h0.norm} 
H_0^\tau=E_0[\tau] + 
\{\hat{\mbox{\sc \Large $o$}}_\tau\} + \hat{\mbox{\sc \large $u$}}_\tau,
\end{equation}
where $\hat{\mbox{\sc \large $u$}}_\tau$ is already normal-ordered; the constant term
$E_0[\tau]$ is the zeroth-order energy of $|\tau\rangle$:
\begin{equation}
H_0^\tau|\tau\rangle= E_0[\tau]\,|\tau\rangle,
\end{equation}
and is given by
\begin{equation}
E_0[\tau]\;=
\sum_{w\in \{\psi_o\leftarrow \tau,\hat{f}_o^{\tau}\}} 
\!\!\!\!\!\!\!
\epsilon_w^{\mbox{\tiny $\tau$}}.
\end{equation}
Note that the first-order and the correlation energies, $E_1[\tau]$ and
${\mathcal E}_{\mathrm{co}}$, do {\em not} depend the zeroth-order energy
$E_0[\tau]$,

The perturbation $V_\tau$, defined by Eqs.~(\ref{Hpart.phi}), can also be written
in normal-ordered form:
\begin{equation}
V_\tau=V_c^{\tau} + V_1^{\tau} + V_2^{\tau},
\end{equation}
where, from Eqs.~(\ref{H.norm}), and (\ref{h0.norm}), the individual terms are
given by the following expressions:
\begin{subequations}
\begin{eqnarray} \label{Vc} 
V_c^{\tau}&=& E_1[\tau] - E_0[\tau],
\\  \label{V1} 
V_1^{\tau}&=& \{\hat{F}_\tau\} - \{\hat{\mbox{\sc \Large $o$}}_\tau\} - \hat{\mbox{\sc \large $u$}}_\tau,
\\ \label{V2} 
V_2^{\tau}&=&\{r_{12}^{-1}\}_\tau.
\end{eqnarray}
\end{subequations}

The one- and two-body parts of $H$, $\{\hat{F}_\tau\}$ and $\{r_{12}^{-1}\}_\tau$,
are given by Eqs.~(\ref{H1F}) and (\ref{H2R12}), respectively. The Goldstone
diagrammatic representation of these operators can be written in the following
manner
\cite{Goldstone:57,Hugenholtz:57,Sanders:69,Raimes:72,Paldus:75,Lindgren:74,Wilson:85,Lindgren:86}:
\begin{subequations}  
\label{h.diag} 
\begin{eqnarray}  \label{h1.diag} 
\raisebox{3.6ex}{$\{\hat{F}_\tau\}$} &\raisebox{3.6ex}{$\;=\;\;$}& 
\setlength{\unitlength}{0.00062500in}
\begingroup\makeatletter\ifx\SetFigFont\undefined%
\gdef\SetFigFont#1#2#3#4#5{%
  \reset@font\fontsize{#1}{#2pt}%
  \fontfamily{#3}\fontseries{#4}\fontshape{#5}%
  \selectfont}%
\fi\endgroup%
{\renewcommand{\dashlinestretch}{30}
\begin{picture}(470,939)(0,-10)
\path(458,912)(458,12)
\put(83,462){\blacken\ellipse{150}{80}}
\put(83,462){\ellipse{150}{80}}
\dottedline{45}(158,462)(428,462)
\end{picture}
}
\raisebox{4.0ex}{\, ,} \\ \label{h2.diag} 
\raisebox{3.6ex}{$\{r_{12}^{-1}\}_\tau$} &\raisebox{3.6ex}{$\;=\;\;$}& 
\setlength{\unitlength}{0.00062500in}
\begingroup\makeatletter\ifx\SetFigFont\undefined%
\gdef\SetFigFont#1#2#3#4#5{%
  \reset@font\fontsize{#1}{#2pt}%
  \fontfamily{#3}\fontseries{#4}\fontshape{#5}%
  \selectfont}%
\fi\endgroup%
{\renewcommand{\dashlinestretch}{30}
\begin{picture}(474,939)(0,-10)
\path(462,912)(462,12)
\path(12,912)(12,12)
\dottedline{45}(12,462)(462,462)
\end{picture}
}
\raisebox{4.0ex}{\, .}
\end{eqnarray}
\end{subequations} 

The one-body part of the perturbation $V_1^{\tau}$ is usually represented by a
{\em single} diagrammatic operator. However, for our purposes, it is convenient to
use separate diagrammatic operators for the three terms on the right side of
Eq.~(\ref{V1}), where $\{\hat{F}_\tau\}$ is presented by
Eq.~(\ref{h1.diag}). Since the other two terms are diagonal, it is appropriate to
simply represent them as unfilled arrows:
\begin{subequations}
\label{H01.diag} 
\begin{eqnarray}
\label{H01o.diag} 
\raisebox{3.6ex}{$-\{\hat{\mbox{\sc \Large $o$}}_\tau\}$} &\raisebox{3.6ex}{$\;=\;\;\;$}& 
\setlength{\unitlength}{0.00062500in}
\begingroup\makeatletter\ifx\SetFigFont\undefined%
\gdef\SetFigFont#1#2#3#4#5{%
  \reset@font\fontsize{#1}{#2pt}%
  \fontfamily{#3}\fontseries{#4}\fontshape{#5}%
  \selectfont}%
\fi\endgroup%
{\renewcommand{\dashlinestretch}{30}
\begin{picture}(159,939)(0,-10)
\path(80,387)(80,386)
\whiten\path(35.000,521.000)(80.000,386.000)(125.000,521.000)(80.000,480.500)(35.000,521.000)
\path(80,387)(80,386)
\whiten\path(12.500,551.000)(80.000,386.000)(147.500,551.000)(80.000,501.500)(12.500,551.000)
\path(80,912)(80,12)
\end{picture}
}
\raisebox{4.0ex}{\, ,} \\
\label{H01u.diag} 
\raisebox{3.6ex}{$- \hat{\mbox{\sc \large $u$}}_\tau$} &\raisebox{3.6ex}{$\;=\;\;\;$}& 
\setlength{\unitlength}{0.00062500in}
\begingroup\makeatletter\ifx\SetFigFont\undefined%
\gdef\SetFigFont#1#2#3#4#5{%
  \reset@font\fontsize{#1}{#2pt}%
  \fontfamily{#3}\fontseries{#4}\fontshape{#5}%
  \selectfont}%
\fi\endgroup%
{\renewcommand{\dashlinestretch}{30}
\begin{picture}(159,939)(0,-10)
\path(80,612)(80,613)
\whiten\path(125.000,478.000)(80.000,613.000)(35.000,478.000)(80.000,518.500)(125.000,478.000)
\path(80,612)(80,613)
\whiten\path(147.500,448.000)(80.000,613.000)(12.500,448.000)(80.000,497.500)(147.500,448.000)
\path(80,12)(80,912)
\end{picture}
}
\raisebox{4.0ex}{\, .} 
\end{eqnarray}
\end{subequations}
In contrast, hole- and particle-lines, by themselves, are represented by
filled arrows: \put(-4,-1){
\setlength{\unitlength}{0.00062500in}
\begingroup\makeatletter\ifx\SetFigFont\undefined%
\gdef\SetFigFont#1#2#3#4#5{%
  \reset@font\fontsize{#1}{#2pt}%
  \fontfamily{#3}\fontseries{#4}\fontshape{#5}%
  \selectfont}%
\fi\endgroup%
{\renewcommand{\dashlinestretch}{30}
\begin{picture}(114,223)(0,-10)
\path(57,13)(57,12)
\whiten\path(12.000,147.000)(57.000,12.000)(102.000,147.000)(57.000,106.500)(12.000,147.000)
\path(57,13)(57,12)
\blacken\path(12.000,147.000)(57.000,12.000)(102.000,147.000)(57.000,106.500)(12.000,147.000)
\end{picture}
}
} \hspace{1.5ex} and 
\put(-3,-3){
\setlength{\unitlength}{0.00062500in}
\begingroup\makeatletter\ifx\SetFigFont\undefined%
\gdef\SetFigFont#1#2#3#4#5{%
  \reset@font\fontsize{#1}{#2pt}%
  \fontfamily{#3}\fontseries{#4}\fontshape{#5}%
  \selectfont}%
\fi\endgroup%
{\renewcommand{\dashlinestretch}{30}
\begin{picture}(114,222)(0,-10)
\path(57,194)(57,195)
\blacken\path(102.000,60.000)(57.000,195.000)(12.000,60.000)(57.000,100.500)(102.000,60.000)
\end{picture}
}
} \,\,\, .

As a slight alternative to the usual approach to evaluate the diagrams of the
correlation energy ${\mathcal E}_{\mathrm{co}}$ and the correlation operator
$\chi_\tau$
\cite{Goldstone:57,Hugenholtz:57,Sanders:69,Raimes:72,Paldus:75,Lindgren:74,Szabo:82,Wilson:85,Lindgren:86},
we associate an internal hole-line corresponding to a $w$-occupied orbital with a
$\psi_w({\mathbf x}_1)\psi_w^*({\mathbf x}_2)$ factor; we associate a particle
line corresponding to an $r$-unoccupied orbital with a $\psi_r({\mathbf
x}_2)\psi_r^*({\mathbf x}_1)$ factor, where ${\mathbf x}_1$ and ${\mathbf x}_2$
denote the dummy integration variables that arise from the vertices.  Using this
convention, the sole diagram involving the Fock operator $\hat{F}_\tau$ from
second-order perturbation theory can be evaluated in the following manner:
\begin{widetext}
\begin{equation} \label{secondfa} 
\raisebox{-4ex}{
\setlength{\unitlength}{0.00062500in}
\begingroup\makeatletter\ifx\SetFigFont\undefined%
\gdef\SetFigFont#1#2#3#4#5{%
  \reset@font\fontsize{#1}{#2pt}%
  \fontfamily{#3}\fontseries{#4}\fontshape{#5}%
  \selectfont}%
\fi\endgroup%
{\renewcommand{\dashlinestretch}{30}
\begin{picture}(890,1009)(0,-10)
\path(833,422)(833,421)
\blacken\path(788.000,556.000)(833.000,421.000)(878.000,556.000)(833.000,515.500)(788.000,556.000)
\path(533,572)(533,573)
\blacken\path(578.000,438.000)(533.000,573.000)(488.000,438.000)(533.000,478.500)(578.000,438.000)
\put(83.000,497.000){\arc{1500.000}{5.6397}{6.9267}}
\put(1283.000,497.000){\arc{1500.000}{2.4981}{3.7851}}
\put(83,47){\blacken\ellipse{150}{80}}
\put(83,47){\ellipse{150}{80}}
\put(83,947){\blacken\ellipse{150}{80}}
\put(83,947){\ellipse{150}{80}}
\dottedline{60}(158,947)(633,947)
\dottedline{60}(158,47)(633,47)
\end{picture}
}
} \raisebox{0.5ex}{$ \displaystyle  
\;\;\; = \; 
(\varepsilon_{rw}^{\mbox{\tiny $\tau$}})^{-1}
\int 
d\,{\mathbf x}_1 \,d\,{\mathbf x}_2 \,
\left(\hat{F}_{\tau \mbox{\tiny $1$}}
\tau_w({\mathbf x}_1,{\mathbf x}_2)
\right) \cdot
\hat{F}_{\tau \mbox{\tiny $2$}}
\tau_r({\mathbf x}_2,{\mathbf x}_1),$}
\end{equation}
\end{widetext}
where 
\begin{equation}
\varepsilon_{rw}^{\mbox{\tiny $\tau$}}=
\epsilon_w^{\mbox{\tiny $\tau$}} - \epsilon_r^{\mbox{\tiny $\tau$}},
\end{equation}
and the repeated indices -- $r$ and $w$ -- are summed over;
$\hat{F}_{\tau \mbox{\tiny $i$}}$ denotes the Fock operator $\hat{F}_\tau$ --
given by Eq.~(\ref{Fockop}) -- acting upon $({\mathbf x}_i)$; the term
$(\hat{F}_{\tau \mbox{\tiny $i$}}\cdots)\cdot$ indicates that $\hat{F}_{\tau
\mbox{\tiny $i$}}$ {\em exclusively} acts within the brackets; furthermore, the
$w$th component of the (one-particle) density-matrix $\tau$ is denoted by
\begin{subequations}
\label{tauwr} 
\begin{equation} \label{tauw} 
\tau_w({\mathbf x}_1,{\mathbf x}_2)=\psi_w({\mathbf x}_1)\psi_w^*({\mathbf x}_2);
\end{equation}
the $r$th orthogonal-component of $\tau$ is denoted by
\begin{equation} \label{taur} 
\tau_r({\mathbf x}_1,{\mathbf x}_2)=\psi_r({\mathbf x}_1)\psi_r^*({\mathbf
x}_2),
\end{equation}
\end{subequations}
where, for a complete set of orbital states, we have \cite{Raimes:72}
\begin{equation} \label{Comp} 
\delta({\mathbf x}_1 - {\mathbf x}_2) = \sum_w\tau_w({\mathbf x}_1,{\mathbf x}_2) 
+ \sum_r \tau_r({\mathbf x}_1,{\mathbf x}_2),
\end{equation}
which is a shorthand notations for
\begin{equation} 
\delta({\mathbf x}_1 - {\mathbf x}_2) = 
\delta({\mathbf r}_1 - {\mathbf r}_2)\delta_{\omega_1\omega_2}.
\end{equation}

If we remove the top interaction from the diagram given by Eq.~(\ref{secondfa}),
we see that this is a first-order diagram that contributes to the one-body portion
of the correlation operator $\chi_\tau$
\cite{Lindgren:86,Lindgren:74,Lindgren:78}. Since the infinite-order sum of all
one-body diagrams for $\chi_\tau$ must vanish for a Bruckner orbital description,
this diagram can be omitted from the expansion for the correlation energy
${\mathcal E}_{\mathrm{co}}$ However, we will still consider it as a simple
example to illustrate our approach and notation.

In order to further compress our notation, we use the convention that all repeated
dummy indices are integrated over and restrict the Fock operator $\hat{F}_{\tau
\mbox{\tiny $i$}}$ to exclusively act upon the first variable of any two-body
function, i.e., ($\hat{F}_{\tau \mbox{\tiny $i$}} \alpha^\prime({\mathbf
x}_j,{\mathbf x}_i) \alpha({\mathbf x}_i,{\mathbf x}_j) = \alpha^\prime({\mathbf
x}_j,{\mathbf x}_i) \hat{F}_{\tau \mbox{\tiny $i$}} \alpha({\mathbf x}_i,{\mathbf
x}_j) $); Eq.~(\ref{secondfa}) can then be written as
\begin{subequations}
\begin{equation} \label{secondfb} 
\raisebox{-4ex}{
\setlength{\unitlength}{0.00062500in}
\begingroup\makeatletter\ifx\SetFigFont\undefined%
\gdef\SetFigFont#1#2#3#4#5{%
  \reset@font\fontsize{#1}{#2pt}%
  \fontfamily{#3}\fontseries{#4}\fontshape{#5}%
  \selectfont}%
\fi\endgroup%
{\renewcommand{\dashlinestretch}{30}
\begin{picture}(890,1009)(0,-10)
\path(833,422)(833,421)
\blacken\path(788.000,556.000)(833.000,421.000)(878.000,556.000)(833.000,515.500)(788.000,556.000)
\path(533,572)(533,573)
\blacken\path(578.000,438.000)(533.000,573.000)(488.000,438.000)(533.000,478.500)(578.000,438.000)
\put(83.000,497.000){\arc{1500.000}{5.6397}{6.9267}}
\put(1283.000,497.000){\arc{1500.000}{2.4981}{3.7851}}
\put(83,47){\blacken\ellipse{150}{80}}
\put(83,47){\ellipse{150}{80}}
\put(83,947){\blacken\ellipse{150}{80}}
\put(83,947){\ellipse{150}{80}}
\dottedline{60}(158,947)(633,947)
\dottedline{60}(158,47)(633,47)
\end{picture}
}
} \raisebox{0.5ex}{$ \displaystyle  
\;\;\; = \; 
(\varepsilon_{rw}^{\mbox{\tiny $\tau$}})^{-1}
\hat{F}_{\tau \mbox{\tiny $1$}}
\tau_w({\mathbf x}_1,{\mathbf x}_2)
\hat{F}_{\tau \mbox{\tiny $2$}}
\tau_r({\mathbf x}_2,{\mathbf x}_1) ,$}
\end{equation}
and the other two diagrams from second-order perturbation theory
have the following forms:
\begin{widetext}
\begin{equation} \label{secondj.d} 
\raisebox{-4ex}{
\setlength{\unitlength}{0.00062500in}
\begingroup\makeatletter\ifx\SetFigFont\undefined%
\gdef\SetFigFont#1#2#3#4#5{%
  \reset@font\fontsize{#1}{#2pt}%
  \fontfamily{#3}\fontseries{#4}\fontshape{#5}%
  \selectfont}%
\fi\endgroup%
{\renewcommand{\dashlinestretch}{30}
\begin{picture}(1164,939)(0,-10)
\path(57,537)(57,538)
\blacken\path(102.000,403.000)(57.000,538.000)(12.000,403.000)(57.000,443.500)(102.000,403.000)
\path(357,387)(357,386)
\blacken\path(312.000,521.000)(357.000,386.000)(402.000,521.000)(357.000,480.500)(312.000,521.000)
\path(807,537)(807,538)
\blacken\path(852.000,403.000)(807.000,538.000)(762.000,403.000)(807.000,443.500)(852.000,403.000)
\path(1107,387)(1107,386)
\blacken\path(1062.000,521.000)(1107.000,386.000)(1152.000,521.000)(1107.000,480.500)(1062.000,521.000)
\put(-393.000,462.000){\arc{1500.000}{5.6397}{6.9267}}
\put(807.000,462.000){\arc{1500.000}{2.4981}{3.7851}}
\put(357.000,462.000){\arc{1500.000}{5.6397}{6.9267}}
\put(1557.000,462.000){\arc{1500.000}{2.4981}{3.7851}}
\dottedline{60}(207,912)(957,912)
\dottedline{60}(207,12)(957,12)
\end{picture}
}
}
\raisebox{0.3ex}{$\displaystyle  
\;\;\; 
= \frac12 
(\varepsilon_{rwsx}^{\mbox{\tiny $\tau$}})^{-1}
r_{12}^{-1}r_{34}^{-1}
\tau_w({\mathbf x}_1,{\mathbf x}_3)
\tau_r({\mathbf x}_3,{\mathbf x}_1)
\tau_x({\mathbf x}_2,{\mathbf x}_4)
\tau_s({\mathbf x}_4,{\mathbf x}_2),$} 
\end{equation}
\begin{equation} \label{secondk.d} 
\raisebox{-4ex}{
\setlength{\unitlength}{0.00062500in}
\begingroup\makeatletter\ifx\SetFigFont\undefined%
\gdef\SetFigFont#1#2#3#4#5{%
  \reset@font\fontsize{#1}{#2pt}%
  \fontfamily{#3}\fontseries{#4}\fontshape{#5}%
  \selectfont}%
\fi\endgroup%
{\renewcommand{\dashlinestretch}{30}
\begin{picture}(1314,943)(0,-10)
\path(57,389)(57,388)
\blacken\path(12.000,523.000)(57.000,388.000)(102.000,523.000)(57.000,482.500)(12.000,523.000)
\path(319,127)(531,339)
\blacken\path(467.360,211.721)(531.000,339.000)(403.721,275.360)(464.178,272.178)(467.360,211.721)
\path(1019,102)(807,314)
\blacken\path(934.279,250.360)(807.000,314.000)(870.640,186.721)(873.822,247.178)(934.279,250.360)
\path(1257,389)(1257,388)
\blacken\path(1212.000,523.000)(1257.000,388.000)(1302.000,523.000)(1257.000,482.500)(1212.000,523.000)
\put(807.000,464.000){\arc{1500.000}{2.4981}{3.7851}}
\put(507.000,464.000){\arc{1500.000}{5.6397}{6.9267}}
\path(207,914)(1109,12)
\path(207,14)(1109,916)
\dottedline{60}(207,914)(1107,914)
\dottedline{60}(207,14)(1107,14)
\end{picture}
}
}
\raisebox{0.3ex}{$ \displaystyle  \;\;\;
=-\frac12 
(\varepsilon_{rwsx}^{\mbox{\tiny $\tau$}})^{-1}
r_{12}^{-1}r_{34}^{-1}
\tau_w({\mathbf x}_1,{\mathbf x}_3)
\tau_r({\mathbf x}_3,{\mathbf x}_2)
\tau_x({\mathbf x}_2,{\mathbf x}_4)
\tau_s({\mathbf x}_4,{\mathbf x}_1),$}
\end{equation}
\end{widetext}
\end{subequations}
where
\begin{equation}
\varepsilon_{rwsx}^{\mbox{\tiny $\tau$}}=
\varepsilon_{rw}^{\mbox{\tiny $\tau$}}+
\varepsilon_{sx}^{\mbox{\tiny $\tau$}}.
\end{equation}
Let us also mention that when determining which dummy indices are repeated
indices, it is not necessary to count indices appearing within operators. So, for
example, the indices $\mathbf{x}_1$ and $\mathbf{x}_2$ appear twice in
Eq.~(\ref{secondfb}), and not three times, since the dummy indices from the Fock
operators, i.e., $\hat{F}_{\tau \mbox{\tiny $1$}}$ and $\hat{F}_{\tau \mbox{\tiny
$2$}}$, are not counted.

The diagonal terms arising from the zeroth-order Hamiltonian, given by
$-\{\hat{\mbox{\sc \Large $o$}}_\tau\}$ and $-\hat{\mbox{\sc \large $u$}}_\tau$,
and represented by Eqs.~(\ref{H01.diag}), first appear in third order. For
example, the following two diagrams can be obtained by inserting $-
\{\hat{\mbox{\sc \Large $o$}}_\tau\}$ and $- \hat{\mbox{\sc \large $u$}}_\tau$
into the diagram on the left side of Eq.~(\ref{secondfb}):
\begin{subequations}
\begin{eqnarray} \label{thirdo} 
\raisebox{-4ex}{
\setlength{\unitlength}{0.00062500in}
\begingroup\makeatletter\ifx\SetFigFont\undefined%
\gdef\SetFigFont#1#2#3#4#5{%
  \reset@font\fontsize{#1}{#2pt}%
  \fontfamily{#3}\fontseries{#4}\fontshape{#5}%
  \selectfont}%
\fi\endgroup%
{\renewcommand{\dashlinestretch}{30}
\begin{picture}(912,1009)(0,-10)
\path(533,572)(533,573)
\blacken\path(578.000,438.000)(533.000,573.000)(488.000,438.000)(533.000,478.500)(578.000,438.000)
\path(833,422)(833,421)
\whiten\path(765.500,586.000)(833.000,421.000)(900.500,586.000)(833.000,536.500)(765.500,586.000)
\put(83.000,497.000){\arc{1500.000}{5.6397}{6.9267}}
\put(1283.000,497.000){\arc{1500.000}{2.4981}{3.7851}}
\put(83,47){\blacken\ellipse{150}{80}}
\put(83,47){\ellipse{150}{80}}
\put(83,947){\blacken\ellipse{150}{80}}
\put(83,947){\ellipse{150}{80}}
\dottedline{60}(158,947)(633,947)
\dottedline{60}(158,47)(633,47)
\end{picture}
}
} \hspace{-4ex} &&\raisebox{0.5ex}{$ \displaystyle  
\;\;\; = \; 
-\frac{(-\epsilon_w)}{(\varepsilon_{rw}^{\mbox{\tiny $\tau$}})^{2}}
\hat{F}_{\tau \mbox{\tiny $1$}}
\tau_w({\mathbf x}_1,{\mathbf x}_2)
\hat{F}_{\tau \mbox{\tiny $2$}}
\tau_r({\mathbf x}_2,{\mathbf x}_1) ,$} 
\nonumber \\ \\
\label{thirdu} 
\raisebox{-4ex}{
\setlength{\unitlength}{0.00062500in}
\begingroup\makeatletter\ifx\SetFigFont\undefined%
\gdef\SetFigFont#1#2#3#4#5{%
  \reset@font\fontsize{#1}{#2pt}%
  \fontfamily{#3}\fontseries{#4}\fontshape{#5}%
  \selectfont}%
\fi\endgroup%
{\renewcommand{\dashlinestretch}{30}
\begin{picture}(890,1009)(0,-10)
\path(833,422)(833,421)
\blacken\path(788.000,556.000)(833.000,421.000)(878.000,556.000)(833.000,515.500)(788.000,556.000)
\path(533,572)(533,573)
\whiten\path(600.500,408.000)(533.000,573.000)(465.500,408.000)(533.000,457.500)(600.500,408.000)
\put(83.000,497.000){\arc{1500.000}{5.6397}{6.9267}}
\put(1283.000,497.000){\arc{1500.000}{2.4981}{3.7851}}
\put(83,47){\blacken\ellipse{150}{80}}
\put(83,47){\ellipse{150}{80}}
\put(83,947){\blacken\ellipse{150}{80}}
\put(83,947){\ellipse{150}{80}}
\dottedline{60}(158,947)(633,947)
\dottedline{60}(158,47)(633,47)
\end{picture}
}
} \hspace{-4ex} &&\raisebox{0.5ex}{$ \displaystyle  
\;\;\; = \; 
\frac{(-\epsilon_r)}{(\varepsilon_{rw}^{\mbox{\tiny $\tau$}})^{2}}
\hat{F}_{\tau \mbox{\tiny $1$}}
\tau_w({\mathbf x}_1,{\mathbf x}_2)
\hat{F}_{\tau \mbox{\tiny $2$}}
\tau_r({\mathbf x}_2,{\mathbf x}_1).$}
\nonumber \\ 
\end{eqnarray}
\end{subequations}
The hole-line operator $\{\hat{\mbox{\sc \Large $o$}}_\tau\}$ generates an
additional hole line when inserted into a diagram and, therefore, a factor of $-1$
is included when diagram (\ref{thirdo}) is evaluated, where this factor cancels
the $-1$ factor from $-\epsilon_w$. Since this type of cancellation always occurs,
as an alternative, we associate a factor of $\epsilon_w$ for $\{\hat{\mbox{\sc
\Large $o$}}_\tau\}$ insertions, and treat $\{\hat{\mbox{\sc \Large $o$}}_\tau\}$
vertices as ones that do not generate additional hole lines; $\hat{\mbox{\sc
\large $u$}}_\tau$ is associated with a $-\epsilon_r$ factor.  Keep in mind, also,
that these operators generate an additional energy-denominator factor, e.g.,
$\varepsilon_{rw}^{\mbox{\tiny $\tau$}}$, when inserted into a diagram.

The individual diagrams depend, in part, on each of the $\tau_w$ components, given
by Eq.~(\ref{tauw}), and the orthogonal components $\tau_r$, given by
Eq.~(\ref{taur}).  In addition, each diagram depends on the set of orbital
energies $\{\epsilon^{\mbox{\tiny $\tau$}}\}$, which are at our disposal. In order
to make each diagram an explicit functional of the one-particle density-matrix
$\tau$, given by
\begin{eqnarray} \label{onepart} 
\tau({\mathbf x}_1,{\mathbf x}_2) =
\sum_w \tau_w({\mathbf x}_1,{\mathbf x}_2),
\end{eqnarray}
and its orthogonal component, $\kappa_\tau$, given by
\begin{eqnarray} 
\label{onepart.virt} 
\kappa_\tau({\mathbf x}_1,{\mathbf x}_2) &=&
\sum_r \tau_r({\mathbf x}_1,{\mathbf x}_2),
\end{eqnarray}
where 
$\kappa_\tau$ depends, explicitly, on $\tau$:
\begin{equation} \label{delta} 
\delta({\mathbf x}_1 - {\mathbf x}_2) =
\tau({\mathbf x}_1,{\mathbf x}_2) + \kappa_\tau({\mathbf x}_1,{\mathbf x}_2),
\end{equation}
we choose all occupied orbitals to be degenerate, with energy
$\epsilon_o^\tau$; also, we choose all unoccupied orbitals to be degenerate,
with energy $\epsilon_u^\tau$. With these choices, the zeroth-order Hamiltonian,
given by Eqs.~(\ref{h0.part}) and (\ref{h0.partB}), becomes
\begin{equation} \label{H0.deg} 
H_0^{\tau} = 
\epsilon_o^\tau \!\!\!\!\!\!\!\!
\sum_{w\in \{\psi_o\leftarrow \tau,\hat{f}_o^{\tau}\}} 
a^\dagger_w a_w 
+ \epsilon_u^\tau \!\!\!\!\!\!\!\!
\sum_{r\in \{\psi_u\leftarrow \tau,\hat{f}_u^{\tau}\}} 
a^\dagger_r a_r,
\end{equation}
and since this operator is invariant to a unitary transformation of occupied or
unoccupied orbitals, it no longer depends on $\hat{f}_o^{\tau}$ and
$\hat{f}_u^{\tau}$ -- any set of orbitals defining $\tau$ is appropriate -- so
we can write
\begin{equation} \label{H0.degb} 
H_0^\tau =
\epsilon_o^\tau \!\!\!\!\!\!
\sum_{w\in \{\psi_o\rightarrow \tau\}} 
a^\dagger_w a_w 
+ \epsilon_u^\tau \!\!\!\!\!\!
\sum_{r\in \{\psi_u\rightarrow \tau\}}
a^\dagger_r a_r.
\end{equation}
It is easily proven that all perturbative orders, except for the zeroth-order,
depend on the orbital-energy difference $\varepsilon_\tau$, given by
\begin{equation} \label{veps.tau} 
\varepsilon_\tau=
\epsilon_o^\tau - \epsilon_u^\tau,
\end{equation}
and not on the individual orbital-energies, $\epsilon_o^\tau$ and
$\epsilon_u^\tau$.  Therefore, we can choose ($\epsilon_u^\tau=0$), and so our
only parameter is $\varepsilon_\tau$. With this choice we have
\begin{equation} \label{H0.tau} 
H_0^\tau = \varepsilon_\tau \hat{N}_\tau,
\end{equation}
where $\hat{N}_\tau$ is the number operator for the occupied orbitals,
\begin{equation} \label{Nocc} 
\hat{N}_\tau = \sum_{w\in \{\psi_o\rightarrow \tau\}} a^\dagger_w a_w,
\end{equation}
and it gives the total number of occupied orbitals when acting on a single
determinant. In the {\em one-particle} Hilbert space, this operator is the
projector for the occupied subspace -- spanned by $\{\psi_o\mbox{\small
$\rightarrow\tau$}\}$ -- or, the one-particle density-matrix operator:
\begin{equation} \label{Nocc.dm} 
\hat{N}_\tau =
\sum_{w\in \{\psi_o\rightarrow \tau\}} |\psi_w\rangle\langle\psi_w| = 
\hat{\tau}.
\end{equation}
Using the above two expressions, let us generalize the definition of $\hat{\tau}$:
\begin{equation} \label{gen.gam} 
\hat{\tau}= \sum_{w\in \{\psi_o\rightarrow \tau\}} a^\dagger_w a_w,
\end{equation}
and write the zeroth-order Hamiltonian in a simplified form, given by
\begin{equation} \label{H0.taub} 
H_0^\tau = 
\varepsilon_\tau\,\hat{\tau}.
\end{equation}
By normal-ordering this expression, we have
\begin{equation} \label{H0.tau.n} 
H_0^\tau = \varepsilon_\tau N_\tau + 
\varepsilon_\tau\{\hat{\tau}\},
\end{equation}
where $N_\tau$ is the number of particles within $|\tau\rangle$, and
from Eq.~(\ref{h0.norm}), we get the following identities:
\begin{eqnarray} 
E_0[\tau] &=&\varepsilon_\tau N_\tau, \\
\label{ident.oc} 
\{\hat{\mbox{\sc \Large $o$}}_\tau\}&=&\varepsilon_\tau\{\hat{\tau}\},\\
\label{ident.un} 
\hat{\mbox{\sc \large $u$}}_\tau &=&0;
\end{eqnarray}
furthermore, our zero- and one-body portion of the perturbation, Eqs.~(\ref{Vc})
and (\ref{V1}), have the following modified forms:
\begin{subequations}
\begin{eqnarray} \label{Vcb} 
V_c^{\tau}&=& E_1[\tau] - \varepsilon_\tau N_\tau,
\\  \label{V1b} 
V_1^{\tau}&=& \{\hat{F}_\tau\} - \varepsilon_\tau\{\hat{\tau}\}.
\end{eqnarray}
\end{subequations}

Eq.~(\ref{ident.un}) indicates that the unoccupied operator, $\hat{\mbox{\sc \large
$u$}}_\tau$, represented by Eq.~(\ref{H01u.diag}), does not appear in the
expansion of the correlation-energy ${\mathcal E}_{\mathrm{co}}$;
$\{\hat{\mbox{\sc \Large $o$}}_\tau\}$, represented by Eq.~(\ref{H01o.diag}) and
given by $\varepsilon_\tau\{\hat{\tau}\}$, is associated with a factor of
$\varepsilon_\tau$.  Each diagram now becomes an explicit functional of $\tau$
and $\kappa_\tau$. For example, the second-order diagrams can be written in the
following manner:
\begin{widetext}
\begin{equation}
\label{secondf} 
\setlength{\unitlength}{0.00062500in}
\begingroup\makeatletter\ifx\SetFigFont\undefined%
\gdef\SetFigFont#1#2#3#4#5{%
  \reset@font\fontsize{#1}{#2pt}%
  \fontfamily{#3}\fontseries{#4}\fontshape{#5}%
  \selectfont}%
\fi\endgroup%
{\renewcommand{\dashlinestretch}{30}
\begin{picture}(890,1009)(0,-10)
\path(833,422)(833,421)
\blacken\path(788.000,556.000)(833.000,421.000)(878.000,556.000)(833.000,515.500)(788.000,556.000)
\path(533,572)(533,573)
\blacken\path(578.000,438.000)(533.000,573.000)(488.000,438.000)(533.000,478.500)(578.000,438.000)
\put(83.000,497.000){\arc{1500.000}{5.6397}{6.9267}}
\put(1283.000,497.000){\arc{1500.000}{2.4981}{3.7851}}
\put(83,47){\blacken\ellipse{150}{80}}
\put(83,47){\ellipse{150}{80}}
\put(83,947){\blacken\ellipse{150}{80}}
\put(83,947){\ellipse{150}{80}}
\dottedline{60}(158,947)(633,947)
\dottedline{60}(158,47)(633,47)
\end{picture}
}
\raisebox{3.7ex}{$ \displaystyle  
\;\;\; = \varepsilon_\tau^{-1}
\hat{F}_{\tau \mbox{\tiny $1$}}
\tau({\mathbf x}_1,{\mathbf x}_2)
\hat{F}_{\tau \mbox{\tiny $2$}}
\kappa_\tau({\mathbf x}_2,{\mathbf x}_1), $}
\end{equation}
\begin{equation} \label{secondj} 
\setlength{\unitlength}{0.00062500in}
\begingroup\makeatletter\ifx\SetFigFont\undefined%
\gdef\SetFigFont#1#2#3#4#5{%
  \reset@font\fontsize{#1}{#2pt}%
  \fontfamily{#3}\fontseries{#4}\fontshape{#5}%
  \selectfont}%
\fi\endgroup%
{\renewcommand{\dashlinestretch}{30}
\begin{picture}(1164,939)(0,-10)
\path(57,537)(57,538)
\blacken\path(102.000,403.000)(57.000,538.000)(12.000,403.000)(57.000,443.500)(102.000,403.000)
\path(357,387)(357,386)
\blacken\path(312.000,521.000)(357.000,386.000)(402.000,521.000)(357.000,480.500)(312.000,521.000)
\path(807,537)(807,538)
\blacken\path(852.000,403.000)(807.000,538.000)(762.000,403.000)(807.000,443.500)(852.000,403.000)
\path(1107,387)(1107,386)
\blacken\path(1062.000,521.000)(1107.000,386.000)(1152.000,521.000)(1107.000,480.500)(1062.000,521.000)
\put(-393.000,462.000){\arc{1500.000}{5.6397}{6.9267}}
\put(807.000,462.000){\arc{1500.000}{2.4981}{3.7851}}
\put(357.000,462.000){\arc{1500.000}{5.6397}{6.9267}}
\put(1557.000,462.000){\arc{1500.000}{2.4981}{3.7851}}
\dottedline{60}(207,912)(957,912)
\dottedline{60}(207,12)(957,12)
\end{picture}
}
\raisebox{3.7ex}{$ \displaystyle  
\;\;\; = \frac{1}{4} \varepsilon_\tau^{-1}
r_{12}^{-1}r_{34}^{-1}
\tau({\mathbf x}_1,{\mathbf x}_3)
\kappa_\tau({\mathbf x}_3,{\mathbf x}_1)
\tau({\mathbf x}_2,{\mathbf x}_4)
\kappa_\tau({\mathbf x}_4,{\mathbf x}_2),$}
\end{equation}
\begin{equation} \label{secondk} 
\setlength{\unitlength}{0.00062500in}
\begingroup\makeatletter\ifx\SetFigFont\undefined%
\gdef\SetFigFont#1#2#3#4#5{%
  \reset@font\fontsize{#1}{#2pt}%
  \fontfamily{#3}\fontseries{#4}\fontshape{#5}%
  \selectfont}%
\fi\endgroup%
{\renewcommand{\dashlinestretch}{30}
\begin{picture}(1314,943)(0,-10)
\path(57,389)(57,388)
\blacken\path(12.000,523.000)(57.000,388.000)(102.000,523.000)(57.000,482.500)(12.000,523.000)
\path(319,127)(531,339)
\blacken\path(467.360,211.721)(531.000,339.000)(403.721,275.360)(464.178,272.178)(467.360,211.721)
\path(1019,102)(807,314)
\blacken\path(934.279,250.360)(807.000,314.000)(870.640,186.721)(873.822,247.178)(934.279,250.360)
\path(1257,389)(1257,388)
\blacken\path(1212.000,523.000)(1257.000,388.000)(1302.000,523.000)(1257.000,482.500)(1212.000,523.000)
\put(807.000,464.000){\arc{1500.000}{2.4981}{3.7851}}
\put(507.000,464.000){\arc{1500.000}{5.6397}{6.9267}}
\path(207,914)(1109,12)
\path(207,14)(1109,916)
\dottedline{60}(207,914)(1107,914)
\dottedline{60}(207,14)(1107,14)
\end{picture}
}
\raisebox{3.7ex}{$ \displaystyle  
\;\;\; = - \frac{1}{4} 
\varepsilon_\tau^{-1}
r_{12}^{-1}r_{34}^{-1}
\tau({\mathbf x}_1,{\mathbf x}_3)
\kappa_\tau({\mathbf x}_3,{\mathbf x}_2)
\tau({\mathbf x}_2,{\mathbf x}_4)
\kappa_\tau({\mathbf x}_4,{\mathbf x}_1),$}
\end{equation}
\end{widetext}
where $\kappa_\tau$ is given by Eq.~(\ref{delta}).  Higher order examples are
presented elsewhere \cite{Finley:bdmt.arxiv} and, in addition, a method that
yields diagrams for the correlation-energy ${\mathcal E}_{\mathrm{co}}$ that
explicitly depend on the one particle density-matrix, $\tau$.

It is well known that the correlation operator $\chi_\tau$ is given by a
linked-diagram expansion, where all disconnected pieces are open
\cite{Lindgren:86,Lindgren:74,Lindgren:78}. Using our approach here, these
diagrams can be evaluated in an identical manner as the diagrams for the
correlation energy ${\mathcal E}_{\mathrm{co}}$; Eq.~(\ref{d1.chi})
gives an example of a fifth-order one-body $\chi_\tau$ diagram:
\begin{eqnarray} \label{d1.chi} 
\raisebox{-4ex}{
\setlength{\unitlength}{0.00066667in}
\begingroup\makeatletter\ifx\SetFigFont\undefined%
\gdef\SetFigFont#1#2#3#4#5{%
  \reset@font\fontsize{#1}{#2pt}%
  \fontfamily{#3}\fontseries{#4}\fontshape{#5}%
  \selectfont}%
\fi\endgroup%
{\renewcommand{\dashlinestretch}{30}
\begin{picture}(1445,1124)(0,-10)
\put(2183.000,537.000){\arc{3750.000}{2.8578}{3.4254}}
\put(-1417.000,537.000){\arc{3750.000}{5.9994}{6.5670}}
\path(1433,1062)(1133,12)
\path(833,1062)(1133,12)
\path(442,282)(448,382)
\blacken\path(472.556,290.365)(448.000,382.000)(412.664,293.958)(444.227,319.113)(472.556,290.365)
\path(943,673)(916,769)
\blacken\path(969.247,690.484)(916.000,769.000)(911.488,674.239)(933.057,708.353)(969.247,690.484)
\path(1024,391)(997,487)
\blacken\path(1050.247,408.484)(997.000,487.000)(992.488,392.239)(1014.057,426.353)(1050.247,408.484)
\path(1327,697)(1300,601)
\blacken\path(1295.488,695.761)(1300.000,601.000)(1353.247,679.516)(1317.057,661.647)(1295.488,695.761)
\path(881,900)(854,996)
\blacken\path(907.247,917.484)(854.000,996.000)(849.488,901.239)(871.057,935.353)(907.247,917.484)
\path(1088,162)(1061,258)
\blacken\path(1114.247,179.484)(1061.000,258.000)(1056.488,163.239)(1078.057,197.353)(1114.247,179.484)
\path(311,564)(311,464)
\blacken\path(281.000,554.000)(311.000,464.000)(341.000,554.000)(311.000,527.000)(281.000,554.000)
\path(446,744)(434,844)
\blacken\path(474.509,758.215)(434.000,844.000)(414.937,751.067)(441.506,781.449)(474.509,758.215)
\put(1208,837){\blacken\ellipse{150}{80}}
\put(1208,837){\ellipse{150}{80}}
\put(758,312){\blacken\ellipse{150}{80}}
\put(758,312){\ellipse{150}{80}}
\put(83,1062){\blacken\ellipse{150}{80}}
\put(83,1062){\ellipse{150}{80}}
\dottedline{60}(463,537)(988,537)
\dottedline{60}(383,12)(1133,12)
\dottedline{60}(893,837)(1133,837)
\dottedline{60}(1048,312)(833,312)
\dottedline{60}(183,1062)(383,1062)
\end{picture}
}

}
\raisebox{0.3ex}{$\displaystyle  
\;\;\; = \;\;\;$} 
\raisebox{0.3ex}{$\displaystyle
\varepsilon_\tau^{-1}
\hat{F}_{\tau \mbox{\tiny $1$}}
\kappa_\tau({\mathbf x}_1,{\mathbf x}_2)
g({\mathbf x}_2,{\mathbf x}_{1^\prime})
\psi_{w}^{\scscs \tau}({\mathbf x}_{1^\prime})
\psi_{r}^{\scscs \tau*}({\mathbf x}_1)
a_{r}^\dagger a_{w}
,$} 
\end{eqnarray}
where the repeated indices, $w$ and $r$, are summed over and where the two body
function is given by Eq.~(\ref{d1b}). (This diagram is identical to the $\left(
H\chi_\tau\right)_1$ diagram appearing in Eq.~(\ref{d1}), but that diagram is
evaluated slighly different, since there is no energy denominator associated with
$H$.)


\bibliography{ref}
\end{document}